\documentclass[aps,rmp,twocolumn,showpacs]{revtex4}
\usepackage{graphicx,amsmath,amssymb,float,bm,times}
\usepackage[breaklinks]{hyperref}

\newcommand{\Lambdak}{\bm{\Lambda}}

\newcommand{\ek}{\epsilon_{\mathbf{k}}}

\newcommand{\Ek}{E_{\mathbf{k}}}

\newcommand{\phik}{\varphi_{\mathbf{k}}}

\newcommand{\mb}[1]{{\mathbf{#1}}}
\newcommand{\sumk}{\sum_{\mathbf{k}}}

\newcommand{\uk}{u_{\mathbf{k}}}
\newcommand{\vk}{v_{\mathbf{k}}}
\newcommand{\createa}[1]{a^\dagger_{#1}}
\newcommand{\destroya}[1]{a^{\phantom \dagger}_{#1}}
\newcommand{\createb}[1]{b^\dagger_{#1}}
\newcommand{\destroyb}[1]{b^{\phantom\dagger}_{#1}}
\newcommand{\vect}[1] {\mathbf{#1}}

\newcommand{\realpart} {\mathrm{Re}}

\newcommand{\xt} {\vect{x},t}
\newcommand{\kw} {\vect{k},\omega}
\newcommand{\dif} {\mathrm{d}}
\newcommand{\evolve} {\frac{\partial}{\partial t}}
\newcommand{\InvariantEnergy} {\biggl(i\evolve-e^*\phi(\xt)\biggr)}
\newcommand{\InvariantMomentum}{\bigl(-i\nabla-e^*\vect{A}(\xt)\bigr)}

\newcommand{\measure} { \frac{\dif^3k}{(2\pi)^3}\frac{\dif\omega}{2\pi} }

\begin{document}

\title{BCS-BEC Crossover: From High Temperature Superconductors to
Ultracold Superfluids}

\author{Qijin Chen $^1$} 
\altaffiliation{Current address: James Franck
  Institute and Department of Physics, University of Chicago, Chicago,
  Illinois 60637} 

\author{Jelena Stajic$^2$}
\altaffiliation{Current address: Los Alamos National Laboratory, Los
  Alamos, NM 87545} 

\author{Shina Tan$^2$}
\author{K.  Levin$^2$}

\affiliation{$^1$ Department of Physics and Astronomy, Johns Hopkins
University, Baltimore, Maryland 21218}

\affiliation{$^2$ James Franck Institute and Department of Physics,
  University of Chicago, Chicago, Illinois 60637}

\begin{abstract}
  We review the BCS to Bose Einstein condensation (BEC) crossover
  scenario which is based on the well known crossover generalization of
  the BCS ground state wavefunction $\Psi_0$. While this ground state
  has been summarized extensively in the literature, this Review is
  devoted to less widely discussed issues: understanding the effects of
  finite temperature, primarily below $T_c$, in a manner consistent with
  $\Psi_0$. Our emphasis is on the intersection of two important
  problems: high $T_c$ superconductivity and superfluidity in ultracold
  fermionic atomic gases. We address the ``pseudogap state" in the
  copper oxide superconductors from the vantage point of a BCS-BEC
  crossover scenario, although there is no consensus on the
  applicability of this scheme to high $T_c$. We argue that it also
  provides a useful basis for studying atomic gases near the unitary
  scattering regime; they are most likely in the counterpart pseudogap
  phase.  That is, superconductivity takes place out of a non-Fermi
  liquid state where preformed, metastable fermion pairs are present at
  the onset of their Bose condensation.  As a microscopic basis for this
  work, we summarize a variety of $T$-matrix approaches, and assess their
  theoretical consistency. A close connection with conventional
  superconducting fluctuation theories is emphasized and exploited. 
\vskip -6.5mm \ %
\end{abstract}
\pacs{03.75.-b, 
 74.20.-z 
\hfill \textsf{Journal Ref: Physics Reports \textbf{412}, 1-88 (2005).}\hspace*{2cm}\ 
}


\maketitle
\tableofcontents

\section{\textbf{Introduction to Qualitative Crossover Picture}}
\label{sec:1}

This review addresses an extended form of the BCS theory of
superfluidity (or superconductivity) known as ``BCS-Bose Einstein
condensation (BEC) crossover theory". Here we contemplate a
generalization of the standard mean field description of superfluids now
modified so that one allows attractive interactions of arbitrary
strength. In this way one describes how the system smoothly goes from
being a superfluid system of the BCS type (where the attraction is
arbitrarily weak) to a Bose Einstein condensate of diatomic molecules
(where the attraction is arbitrarily strong).  One, thereby, sees that
BCS theory is intimately connected to BEC.  Attractive interactions
between fermions are needed to form ``bosonic-like" molecules (called
Cooper pairs) which then are driven statistically to Bose condense.  BCS
theory is a special case in which the condensation and pair formation
temperatures coincide.

The importance of obtaining a generalization of BCS theory which
addresses the crossover from BCS to BEC at general temperatures $ T \leq
T_c$ cannot be overestimated.  BCS theory as originally postulated can
be viewed as a paradigm among theories of condensed matter systems; it
is complete, in many ways generic and model independent, and well
verified experimentally.  The observation that a BCS-like approach
goes beyond strict BCS theory, suggests that there is
a larger mean field theory to be addressed.  Equally exciting is the
possibility that this mean field theory can be discovered and
simultaneously tested in a very controlled fashion using ultracold
fermionic atoms. We presume here that it may also have
applicability to other short coherence length materials, such as the
high temperature superconductors.

This Review is written in order to convey important new developments
from one subfield of physics to another.  Here it is hoped that we will
communicate the recent excitement felt by the cold atom community to
condensed matter physicists. Conversely we wish to communicate a
comparable excitement in studies of high temperature superconductivity
to atomic physicists.  There has been a revolution in our understanding
of ultracold fermions in traps in the last few years.  The milestones in
this research were the creation of a degenerate Fermi gas (1999), the
formation of dimers of fermions (2003), Bose-Einstein condensation (BEC)
of these dimers (late 2003) and finally superfluidity of fermionic pairs
(2004).

In high temperature superconductors the BCS-BEC crossover picture has
been investigated for many years now. It leads to a particular
interpretation of a fascinating, but not well understood phase, known as
the ``pseudogap state".  In the cold atom system, this crossover
description is not just a scenario, but has been realized in the
laboratory.  This is because one has the ability (via Feshbach
resonances) to apply magnetic fields in a controlled way to tune the
strength of the attraction.  We note, finally, that research in this
field has not been limited exclusively to the two communities. One has
seen the application of these crossover ideas to studies of excitons in
solids \cite{Lai}, to nuclear physics \cite{Baldo,Heiselberg3} and to
particle physics \cite{Heinz,Volovik,Itakura} as well.  There are not
many problems in physics which have as great an overlap with different
subfield communities as this ``BCS-BEC crossover problem".

\begin{figure}
\centerline{\includegraphics{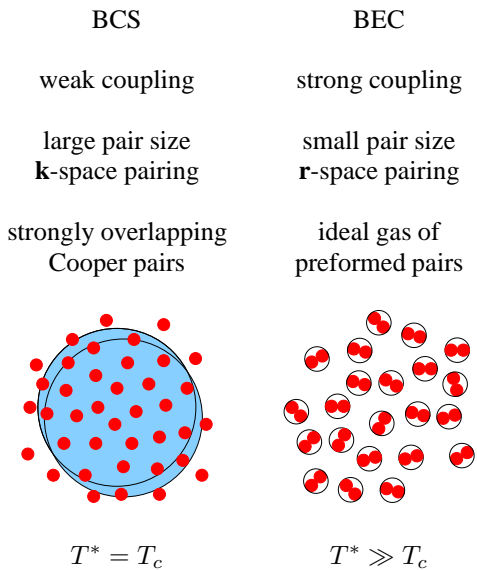}}
 \caption{Contrast between BCS- and BEC-based superfluids.
 Here $T^*$ represents the temperature at which pairs form,
 while $T_c$ is that at which they condense.}
 \label{fig:1}
\end{figure}

In this Review, we begin with an introduction to the qualitative picture
of the BCS-BEC crossover scenario which is represented schematically in
Fig.~\ref{fig:1}. We discuss this picture in the context of both high
$T_c$ superconductivity and ultracold atomic superfluidity, summarizing
experiments in both.  In Section II, we present a more quantitative
study of this crossover primarily at and below the superfluid transition
temperature $T_c$.  Section III contains a detailed discussion of the
superfluid density and gauge invariance issues, in large part to
emphasize self consistency checks for different theoretical approaches
to crossover. These alternative theories are summarized and compared in
Section IV.  Sections V and VI, respectively, present the physical
implications of crossover physics, in one particular rendition, in the
context of ultracold fermionic superfluids and high temperature
superconductors.  Our conclusions are presented in Section VII and a
number of Appendices are included for clarity.  A reader interested in
only the broad overview may choose to omit Sections II-IV.

The convention and notation as well as a list of abbreviations used in
this Review are listed in Appendix \ref{sec:Convention} at the end.

\subsection{Fermionic Pseudogaps and Meta-stable
Pairs: Two Sides of the Same Coin}
\label{sec:1a}

A number of years ago Eagles \cite{Eagles} and Leggett \cite{Leggett}
independently noted\footnote{Through private correspondence it appears
  that F. Dyson may have made similar observations even earlier.}  that
the BCS ground state wavefunction
\begin{equation}
\Psi_0=\Pi_{\bf k}(\uk+\vk c_{k,\uparrow}^{\dagger}
c_{-k,\downarrow}^{\dagger})|0\rangle   
\label{eq:1a}
\end{equation}
had a greater applicability than had been appreciated at the time of its
original proposal by Bardeen, Cooper and Schrieffer (BCS).  As the
strength of the attractive pairing interaction $U$ ($<0$) between
fermions is increased, this wavefunction is also capable of describing a
continuous evolution from BCS like behavior to a form of Bose Einstein
condensation.  What is essential is that the chemical potential $\mu$ of
the fermions be self consistently computed as $U$ varies.
 
The variational parameters $\vk$ and $u_{\bf k}$ which appear in
Eq.~\ref{eq:1a} are usually represented by the two more directly
accessible parameters $\Delta_{sc}(0)$ and $\mu$, which characterize the
fermionic system. Here $\Delta_{sc}(0)$ is the zero temperature
superconducting order parameter. These fermionic parameters are uniquely
determined in terms of $U$ and the fermionic density $n$. The
variationally determined self consistency conditions are given by two
BCS-like equations which we refer to as the ``gap" and ``number"
equations respectively.
\begin{eqnarray}
\label{eq:2}
\Delta_{sc}(0)&=&-U \sum_{\bf k} \Delta_{sc}(0) \frac{1}{2 \Ek}   \\
n&=& \sum_{\bf k} \left[ 1 -\frac{\ek - \mu}{\Ek} 
 \right]  
\label{eq:02}
\end{eqnarray}
where \footnote{In general, the order parameter $\Delta_{sc}$ and the
  excitation gap may have a {\bf k}-dependence, $\Delta _k = \Delta
  \phik$, except for a contact potential. To keep the mathematical
  expressions simpler, however, we neglect $\phik$ for the $s$-wave
  pairing under investigation of this Review. It can be easily
  reinserted back when needed in actual calculations. In particular, we
  use $\phik = \exp(-k^2/2k_c^2)$ with a large cutoff $k_c$ in most of
  our numerical calculations.}
\begin{equation}
\Ek \equiv \sqrt{ (\ek -\mu)^2 + \Delta_{sc}^2 (0) }
\label{eq:dispersion}
\end{equation}
and $\ek=k^2/2 m$ is the fermion energy dispersion.\footnote{Throughout
  this Review, we set $\hbar=1$ and $k_B=1$.}  Within this ground state
there have been extensive studies of collective modes
\cite{Micnas,randeriareview,Cote}, effects of two dimensionality
\cite{randeriareview}, and, more recently, extensions to atomic gases
\cite{Stringari,Strinati}.  Nozi\'eres and Schmitt-Rink were the first
\cite{NSR} to address nonzero $T$. We will outline some of their
conclusions later in this Review.  Randeria and co-workers reformulated
the approach of Nozi\'eres and Schmitt-Rink (NSR) and moreover, raised
the interesting possibility that crossover physics might be relevant to
high temperature superconductors \cite{randeriareview}.  Subsequently
other workers have applied this picture to the high $T_c$ cuprates
\cite{Chen2,Micnas1,Ranninger} and ultracold fermions
\cite{Milstein,Griffin} as well as formulated alternative schemes
\cite{Griffin2,Strinati2} for addressing $ T \neq 0$.  Importantly, a
number of experimentalists, most notably Uemura \cite{Uemura}, have
claimed evidence in support \cite{Renner,Deutscher,Junod} of the BCS-BEC
crossover picture for high $T_c$ materials.

Compared to work on the ground state, considerably less has been written
on crossover effects at nonzero temperature based on Eq.~(\ref{eq:1a}).
Because our understanding has increased substantially since the
pioneering work of NSR, and because they are the most interesting, this
review is focussed on these finite $T$ effects, at the level
of a simple mean field theory.
Mean field approaches are always approximate.  We can ascribe the
simplicity and precision of BCS theory to the fact that in conventional
superconductors the coherence length $\xi$ is extremely long. As a
result, the kind of averaging procedure implicit in mean field theory
becomes nearly exact. Once $\xi$ becomes small, BCS is not expected to
work at the same level of precision.  Nevertheless even when they are
not exact, mean field approaches are excellent ways of building up
intuition.  And further progress is not likely to be made without
investigating first the simplest of mean field approaches, associated
with Eq.~(\ref{eq:1a}).

\begin{figure}
\includegraphics[width=2.3in,clip]{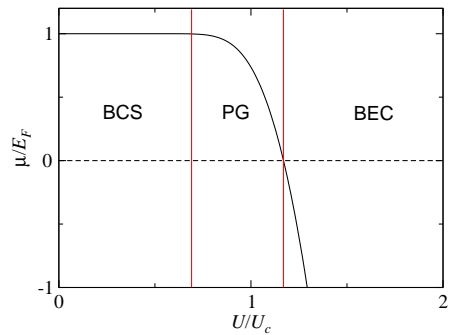}
\caption{Typical behavior of the $T=0$ chemical potential in the three
  regimes. As the pairing strength increases from 0, the chemical
  potential $\mu$ starts to decrease and then becomes negative. The
  character of the system changes from fermionic ($\mu>0$) to bosonic
  ($\mu < 0$).  The PG (pseudogap) case corresponds to non-Fermi liquid
  based superconductivity, and $U_c$ corresponds to critical coupling
  for forming a two fermion bound state in vacuum, as in 
  Eq.~(\ref{eq:Uc}) below.}
\label{fig:2a}
\end{figure}

The effects of BEC-BCS crossover are most directly reflected in the
behavior of the fermionic chemical potential $\mu$.  We plot the
behavior of $\mu$ in Fig.~\ref{fig:2a}, which indicates the BCS and
BEC regimes.  In the weak coupling regime $\mu = E_F$ and ordinary BCS
theory results.  However at sufficiently strong coupling, $\mu$ begins
to decrease, eventually crossing zero and then ultimately becoming
negative in the BEC regime, with increasing $|U|$.  We generally view $\mu
= 0 $ as a crossing point.  For positive $\mu$ the system has a remnant
Fermi surface, and we say that it is ``fermionic".  For negative
$\mu$, the Fermi surface is gone and the material is ``bosonic".

\textit{The new and largely unexplored physics of this problem lies in
  the fact that once outside the BCS regime, but before BEC,
  superconductivity or superfluidity emerge out of a very exotic,
  non-Fermi liquid normal state}.  Indeed, it should be clear that Fermi
liquid theory must break down somewhere in the intermediate regime, as
one goes continuously from fermionic to bosonic statistics. This
intermediate case has many names: ``resonant superfluidity", the
``unitary regime" or, as we prefer (because it is more descriptive of
the many body physics) the ``pseudogap (PG) regime". The PG regime
plotted in Fig.~\ref{fig:2a} is drawn so that its boundary on the left
coincides with the onset of an appreciable excitation gap at $T_c$,
called $\Delta(T_c)$; its boundary on the right coincides with where
$\mu =0$. This non-Fermi liquid based superconductivity has been
extensively studied\cite{Chen2,Chen3,Janko,Maly1}.

Fermi liquid theory is inappropriate here principally because there is a
gap for creating fermionic excitations.  The non-Fermi liquid like
behavior is reflected in that of the fermionic spectral function
\cite{Maly1} which has a two-peak structure in the normal state,
broadened somewhat, but rather similar to its counterpart in BCS theory.
Importantly, the onset of superconductivity occurs in the presence of
fermion pairs, which reflect this excitation gap. Unlike their
counterparts in the BEC limit, these pairs are not infinitely long
lived.  Their presence is apparent even in the normal state where an
energy must be applied to create fermionic excitations.  This energy
cost derives from the breaking of the metastable pairs.  Thus we say
that there is a ``pseudogap" at and above $T_c$. But pseudogap effects
are not restricted to this normal state.  Pre-formed pairs have a
natural counterpart in the superfluid phase, as noncondensed pairs.

It will be stressed throughout this Review that gaps in the fermionic
spectrum and bosonic degrees of freedom are two sides of the same coin.
A particularly important observation to make is that the starting point
for crossover physics is based on the fermionic degrees of freedom.
Bosonic degrees of freedom are deduced from these; they are not primary.
A nonzero value of the excitation gap $\Delta$ is equivalent to the
presence of metastable or stable fermion pairs. And it is only in this
indirect fashion that we can probe the presence of these ``bosons",
within the framework of Eq.~(\ref{eq:1a}).

In many ways this crossover theory appears to represent a more generic
form of superfluidity.  Without doing any calculations we can anticipate
some of the effects of finite temperature. Except for very weak
coupling, \textit{pairs form and condense at different temperatures}.
The BCS limit might be viewed as the anomaly.  Because the attractive
interaction is presumed to be arbitrarily weak, in BCS the normal state
is unaffected by $U$ and superfluidity appears precipitously, that is
without warning at $T_c$.  More generally, in the presence of a
moderately strong attractive interaction it pays energetically to take
some advantage and to form pairs (say roughly at temperature $T^*$)
within the normal state. Then, for statistical reasons these bosonic
degrees of freedom ultimately are driven to condense at $T_c < T^*$,
just as in BEC.

\begin{figure}
\includegraphics[width=2.5in,clip]{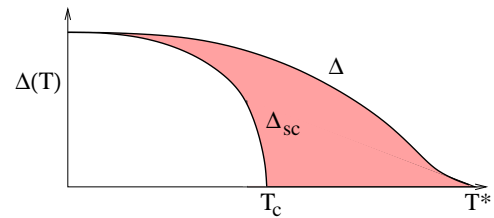}
\caption{Contrasting behavior of the excitation gap $\Delta(T)$ 
  and superfluid order parameter $\Delta_{sc}(T)$ versus temperature.
  The height of the shaded region roughly reflects the density of
  noncondensed pairs at each temperature.}
\label{fig:Delta_Deltasc}
\end{figure}

Just as there is a distinction between $T_c$ and $T^*$, \textit{there
  must be a distinction between the superconducting order parameter
  $\Delta_{sc}$ and the excitation gap $\Delta$}.  The presence of a
normal state excitation gap or pseudogap for fermions is inextricably
connected to this generalized BCS wavefunction.  In Figure
\ref{fig:Delta_Deltasc} we present a schematic plot of these two energy
parameters.  It may be seen that the order parameter vanishes at $T_c$,
as in a second order phase transition, while the excitation gap turns on
smoothly below $T^*$.  It should also be stressed that there is only one
gap energy scale in the ground state \cite{Leggett} of Eq.~(\ref{eq:1a}).
Thus at zero temperature
\begin{equation}
\Delta_{sc}(0) = \Delta(0)\;.
\label{eq:3ac}
\end{equation}

\begin{figure}
\includegraphics[width=3.4in,clip]{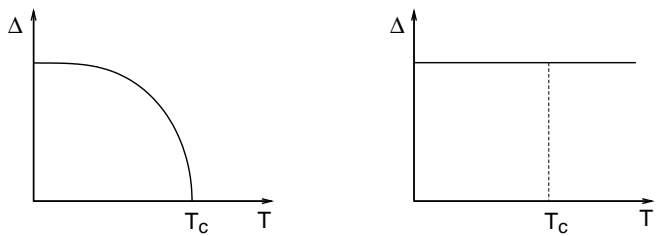}
\caption{Comparison of typical temperature dependences of the excitation
  gaps in the BCS  (left) and BEC (right) limits. For the former, the
  gap is small and vanishes at $T_c$; whereas for the latter, the gap is
  very large and essentially temperature independent.}
\label{fig:2}
\end{figure}

In addition to the distinction between $\Delta$ and $\Delta_{sc}$,
another important way in which \textit{bosonic degrees of freedom are
  revealed is indirectly through the temperature dependence of $\Delta$.
  In the BEC regime where fermion pairs are pre-formed, $\Delta$ is
  essentially constant for all $ T \leq T_c$ (as is $\mu$)}. By contrast
in the BCS regime it exhibits the well known temperature dependence of
the superconducting order parameter. This is equivalent to the statement
that bosonic degrees of freedom are only present in the condensate for
this latter case.  This behavior is illustrated in Fig.~\ref{fig:2}.

\begin{figure}
\centerline{\includegraphics{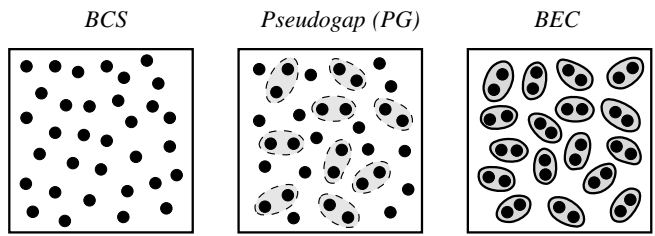}}
\caption{The character of the excitations in the BCS-BEC crossover both
  above and below $T_c$. The excitations are primarily fermionic
  Bogoliubov quasiparticles in the BCS limit and bosonic pairs (or
  ``Feshbach bosons") in the BEC limit. In the PG case the ``virtual
  molecules" consist primarily of ``Cooper'' pairs of fermionic atoms.}
\label{fig:3}
\end{figure}

Again, without doing any calculations we can make one more inference
about the nature of crossover physics at finite $T$. \textit{The
  excitations of the system must smoothly evolve from fermionic in the
  BCS regime to bosonic in the BEC regime}.  In the intermediate case,
the excitations are a mix of fermions and meta-stable pairs. Figure
\ref{fig:3} characterizes the excitations out of the condensate as well
as in the normal phase. This schematic figure will play an important
role in our thinking throughout this review.  In the BCS and BEC regimes
one is led to an excitation spectrum with a single component.  For the
PG case it is clear that the excitations now have two distinct (albeit,
strongly interacting) components.

\subsection{Introduction to high $T_c$ Superconductivity: Pseudogap
Effects}
\label{sec:1B}

This Review deals with the intersection of two fields and two important
problems: high temperature superconductors and ultracold fermionic atoms
in which, through Feshbach resonance effects, the attractive interaction
may be arbitrarily tuned by a magnetic field.  Our focus is on the
broken symmetry phase and how it evolves from the well known ground
state at $T=0$ to $T=T_c$. We begin with a brief overview
\cite{Timusk,LoramPhysicaC} of pseudogap effects in the hole doped high
temperature superconductors.  A study of concrete data in these systems
provides a rather natural way of building intuition about non-Fermi
liquid based superfluidity, and this should, in turn, be useful for the
cold atom community.

It has been argued by some \cite{Chen1,Micnas1,Ranninger,Strinati3,YY}
that a BCS-BEC crossover-induced pseudogap is the origin of the
mysterious normal state gap observed in high temperature
superconductors.  Although, this is a highly contentious subject, a
principal supporting rationale is that this above-$T_c$ excitation gap
seems to evolve smoothly into the fermionic gap within the
superconducting state.  This, in conjunction with a panoply of normal
state anomalies which appear to represent precursor superconductivity
effects, suggests that the pseudogap has more in common with the
superconducting phase than with the insulating magnetic phase associated
with the parent compound.

\begin{figure}
\includegraphics[bb=0 45 400 290, clip]{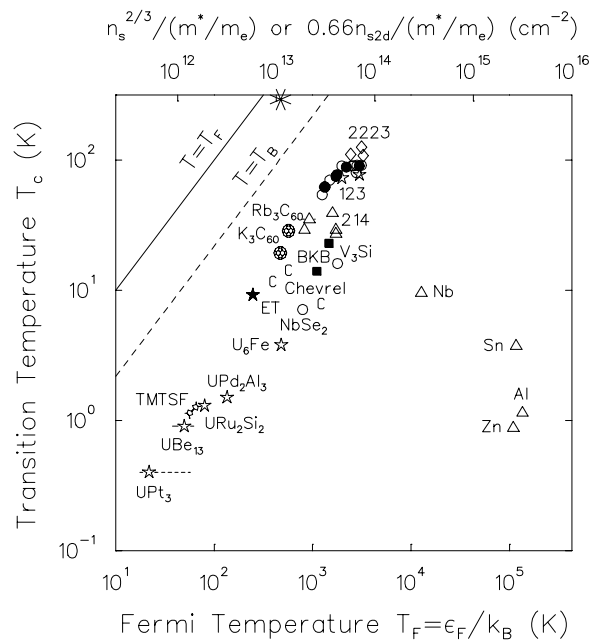}
\caption{Uemura plot \cite{Uemura} of the superconducting transition
  temperature as a function of Fermi temperature (lower axis) and of the
  low $T$ superfluid density (upper axis) for various superconductors.
  The plot indicates how ``exotic'' superconductors appear to be in a distinct
  category as compared with more conventional metal superconductors such
  as Nb, Sn, Al and Zn. Note the logarithmic scales.}
\label{fig:4}
\end{figure}

The arguments in favor of this (albeit, controversial) crossover
viewpoint rest on the following observations: (i) the transition
temperature is anomalously high and the coherence length $\xi$ for
superconductivity anomalously short, around $10 $\AA\ as compared more
typically with $1000$\AA.  (ii) These systems are quasi-two dimensional
so that one might expect pre-formed pairs or precursor superconductivity
effects to be enhanced.  (iii) The normal state gap has the same
$d$-wave symmetry as the superconducting order parameter
\cite{arpesstanford,arpesanl}, and (iv) to a good approximation its
onset temperature, called $T^*$, satisfies \cite{Fischer2,Oda} $ T^*
\approx 2 \Delta(0)/4.3 $ which satisfies the BCS scaling relation. (v)
There is widespread evidence for pseudogap effects both above
\cite{Timusk,LoramPhysicaC} and below \cite{Loram,JS1} $T_c$.  Finally,
it has also been argued that short coherence length superconductors may
quite generally exhibit a distinctive form of superconductivity
\cite{Uemura}.  Figure \ref{fig:4} shows a plot of data collected by
Uemura which seems to suggest that the traditional BCS superconductors
stand apart from other more exotic (and frequently short $\xi$) forms of
superconductivity. From this plot one can infer, that, except for high
$T_c$, there is nothing special about the high $T_c$ superconductors;
they are not alone in this distinctive class which includes the organics
and heavy fermion superconductors as well.  Thus, to understand them,
one might want to focus on this simplest aspect (short $\xi$) of high
$T_c$ superconductors, rather than invoke more exotic and less generic
features.

\begin{figure}
\includegraphics[width=2.3in,clip]{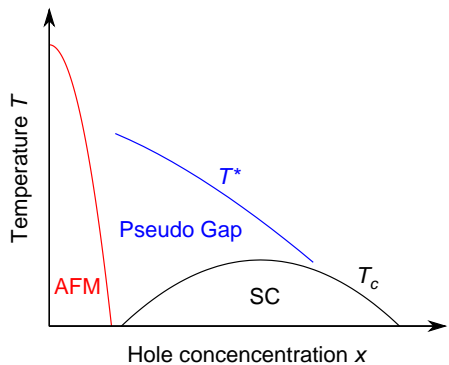}
\caption{Typical phase diagram of hole-doped high $T_c$
  superconductors.  There exists a pseudogap phase above $T_c$ in the
  underdoped regime. Here SC denotes superconductor, and $T^*$ is the
  temperature at which the pseudogap smoothly turns on.}
\label{fig:5}
\end{figure}

In Fig.~\ref{fig:5} we show a sketch of the phase diagram for the
copper oxide superconductors. Here $x$ represents the concentration of
holes which can be controlled by adding Sr substitutionally, say, to
La$_{1-x}$Sr$_{x}$CuO$_4$. At zero and small $x$ the system is an
antiferromagnetic (AFM) insulator. Precisely at half filling ($x=0$) we
understand this insulator to derive from Mott effects. These Mott
effects may or may not be the source of the other exotic phases
indicated in the diagram, as ``SC" for superconductivity and the
pseudogap phase.  Once AFM order disappears the system remains
insulating until a critical concentration (typically around a few
percent Sr) when an insulator-superconductor transition is encountered.
Here photoemission studies \cite{arpesstanford,arpesanl} suggest that
once this line is crossed, $\mu$ appears to be positive.  For $ x \leq
0.2$, the superconducting phase has a non-Fermi liquid (or pseudogapped)
normal state \cite{LoramPhysicaC}.  We note an important aspect of this
phase diagram at low $x$. As the pseudogap becomes stronger ($T^*$
increases), superconductivity as reflected in the magnitude of $T_c$
becomes weaker.  One way to think about this competition is through the
effects of the pseudogap on $T_c$. As this gap increases, the density of
fermionic states at $E_F$ decreases, so that there are fewer fermions
around to participate in the superconductivity.

\begin{figure}
  \includegraphics[width=2.8in]{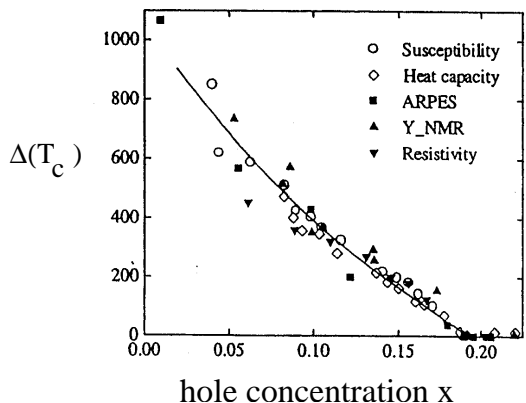}
\caption{Pseudogap magnitude at $T_c$ versus doping concentration in hole doped
  high $T_c$ superconductors measured with different experimental
  techniques (from \onlinecite{LoramPhysicaC}). Units on vertical axis
  are temperature in degrees K.}
\label{fig:6}
\end{figure}

Importantly, there is a clear excitation gap present at the onset of
superconductivity which is called $\Delta(T_c)$ and which vanishes at $x
\approx 0.2$.  The magnitude of the pseudogap at $T_c$ is shown in
Fig.~\ref{fig:6}, from \cite{LoramPhysicaC}.  Quite remarkably, as
indicated in the figure, a host of different probes seem to converge on
the size of this gap.

\begin{figure}
\centerline{\includegraphics[width=2.5in]{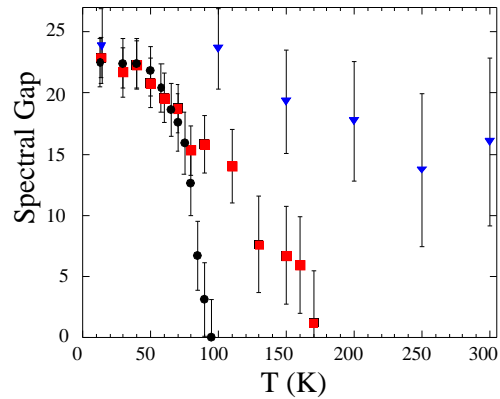}}
\caption{(Color) Temperature dependence of the excitation gap
  from ARPES measurements for optimally doped (filled black circles),
  underdoped (red squares) and highly underdoped (blue inverted
  triangles) single-crystal BSCCO samples (taken from
  \onlinecite{arpesanl1}). There exists a pseudogap phase above $T_c$ in
  the underdoped regime.}
\label{fig:7}
\end{figure}

Figure \ref{fig:7} indicates the temperature dependence of the
excitation gap for three different hole stoichiometries. These data were
taken \cite{arpesanl1} from angle resolved photoemission spectroscopy
(ARPES) measurement on single-crystal Bi$_2$Sr$_2$CaCu$_2$O$_{8+
  \delta}$ (BSCCO) samples.  For one sample shown as circles,
(corresponding roughly to ``optimal" doping) the gap vanishes roughly at
$T_c$ as might be expected for a BCS superconductor. At the other
extreme are the data indicated by inverted triangles in which an
excitation gap appears to be present up to room temperature, with very
little temperature dependence.  This is what is referred to as a highly
underdoped sample (small $x$), which from the phase diagram can be seen
to have a rather low $T_c$.  Moreover, $T_c$ is not evident in these
data on underdoped samples.  This is a very remarkable feature which
indicates that from the fermionic perspective there appears to be no
profound sensitivity to the onset of superconductivity.

\begin{figure}
\includegraphics[width=3.2in,clip]{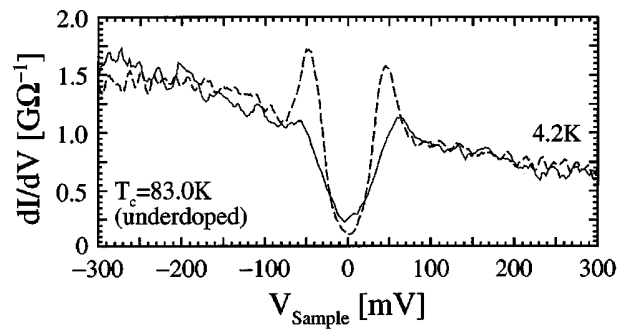}
\caption{STM measurements of the $dI/dV$ characteristics of an
  underdoped BSCCO sample inside (solid) and outside (dashed) a vortex
  core  at low $T$ from \onlinecite{Renner}. $dI/dV$ is proportional to
  the density of states.}
\label{fig:8}
\end{figure}

While the high $T_c$ community has focussed on pseudogap effects above
$T_c$, there is a good case to be made that these effects also persist
below.  Shown in Fig.~\ref{fig:8} are STM data \cite{Renner} taken
below $T_c$ within a vortex core (solid lines) and between vortices in
the bulk (dashed lines).  The quantity $dI/dV$ may be viewed as a
measure of the fermionic density of states at energy $ E$ given by the
voltage $V$.  This figure shows that there is a clear depletion of the
density of states around the Fermi energy ($V=0$) in the normal phase
within the core. Indeed the size of the inferred energy gap (or
pseudogap) corresponds to the energy separation between the two maxima
in $dI/dV$; this can be seen to be the same for both the normal and
superconducting regions (solid and dashed curves). This figure
emphasizes the fact that the existence of an energy gap has little or
nothing to do with the existence of phase coherent superconductivity. It
also suggests that pseudogap effects effectively persist below $T_c$;
the normal phase underlying superconductivity for $ T \leq T_c$ is not a
Fermi liquid.

\begin{figure}
\includegraphics[width=3.4in,clip]{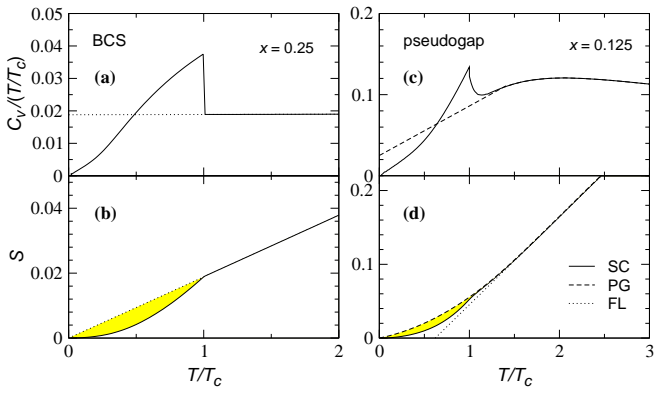}
\caption{Behavior of entropy $S$ and specific heat coefficient
  $\gamma\equiv C_v/T$ as a function of temperature for (a)-(b) BCS and
  (c)-(d) pseudogapped superconductors. Dotted and dashed lines are
  extrapolated normal states below $T_c$. The shaded areas were used to
  determine the condensation energy.}
\label{fig:9}
\end{figure}

\begin{figure*}[t]
\includegraphics[width=4.5in]{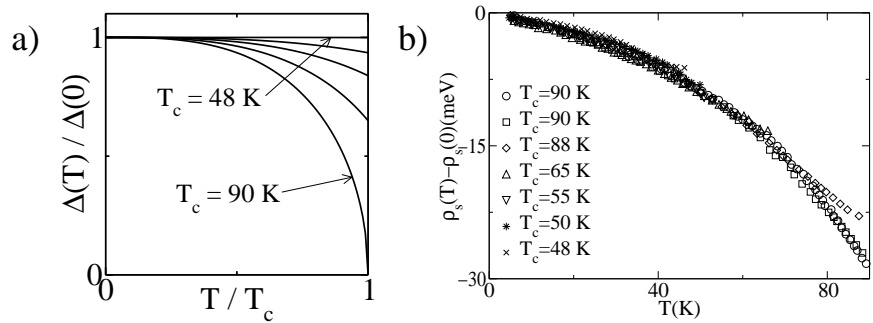}
\caption{Temperature dependence of (a) the fermionic excitation gap
  $\Delta$ and (b) superfluid density $\rho_s$ from experiment for
  various doping concentrations (from \cite{JS1}). Figure (a) shows a
  schematic plot of $\Delta$ as seen in the ARPES and STM tunneling
  data.  When $\Delta(T_c) \neq 0$, there is little correlation between
  $\Delta(T)$ and $\rho_s(T)$; this figure suggests that something other
  than fermionic quasiparticles (\textit{e.g.}, bosonic excitations)
  may be responsible for the disappearance of superconductivity with
  increasing $T$.  Figure (b) shows a quasi-universal behavior for the
  slope $d\rho_s/dT$ at different doping concentrations, despite the
  highly non-universal behavior for $\Delta(T)$.}
\label{fig:10}
\end{figure*}

Analysis of thermodynamical data \cite{Loram,LoramPhysicaC} has led to a
similar inference.  Figure \ref{fig:9} presents a schematic plot of the
entropy $S$ and specific heat for the case of a BCS superconductor, as
contrasted with a pseudogapped superconductor. Actual data are presented
in Fig.~\ref{fig:25}.  Figure \ref{fig:9} makes it clear that in a BCS
superconductor, the normal state which underlies the superconducting
phase, \textit{is} a Fermi liquid; the entropy at high temperatures
extrapolates into a physically meaningful entropy below $T_c$.  For the
PG case, the Fermi liquid-extrapolated entropy becomes negative. In this
way Loram and co-workers \cite{Loram} deduced that the normal phase
underlying the superconducting state is not a Fermi liquid.  Rather,
they claimed to obtain proper thermodynamics, it must be assumed that
this state contains a persistent pseudogap.  They argued for a
distinction between the excitation gap $\Delta$ and the superconducting
order parameter, within the superconducting phase.  To fit their data
they presumed a modified fermionic dispersion 
\begin{equation}
 \Ek = \sqrt{ (\ek -\mu)^2
  + \Delta^2 (T) } 
\label{eq:1b}
\end{equation}
where
\begin{equation}
\Delta^2(T) = \Delta_{sc}^2(T) + \Delta_{pg}^2
\label{eq:2a}
\end{equation}
Here $\Delta_{pg}$ is taken on phenomenological grounds to be
$T$-independent.
The authors argue that the pseudogap contribution may arise from physics
unrelated to the superconductivity.  While others
\cite{Laughlin,Nozieres2} have similarly postulated that $\Delta_{pg}$
may in fact derive from another (hidden) order parameter or underlying
band structure effect, in general the fermionic dispersion relation
$\Ek$ will take on a different character from that assumed above, which
is very specific to a superconducting origin for the pseudogap.

Finally, Fig.~\ref{fig:10} makes the claim for a persistent pseudogap
below $T_c$ in an even more suggestive way. Figure \ref{fig:10}(a)
represents a schematic plot of excitation gap data such as are shown in
Fig.~\ref{fig:7}.  Here the focus is on temperatures below $T_c$. Most
importantly, this figure indicates that the $T$ dependence in $\Delta$
varies dramatically as the stoichiometry changes. Thus, in the extreme
underdoped regime, where PG effects are most intense, there is very
little $T$ dependence in $\Delta$ below $T_c$.  By contrast at high $x$,
when PG effects are less important, the behavior of $\Delta$ follows
that of BCS theory.  What is most impressive however, is that these wide
variations in $\Delta(T)$ are \textit{not} reflected in the superfluid
density $\rho_s(T)$, which is proportional to the inverse square of the
London magnetic penetration depth and can be measured by, for example,
muon spin relaxation ($\mu$SR) experiments. Necessarily, $\rho_s(T)$
vanishes at $T_c$.  What is plotted \cite{JS1} in Fig.~\ref{fig:10}(b)
is $\rho_s(T) - \rho_s(0)$ versus $T$.  That these data all seem to sit
on a rather universal curve is a key point.  The envelope curve in
$\rho_s(T)- \rho_s(0)$ is associated with an ``optimal" sample where
$\Delta(T)$ essentially follows the BCS prediction.  Figure \ref{fig:10}
then indicates that, \textit{despite the highly non-universal behavior
  for $\Delta(T)$, the superfluid density does not make large excursions
  from its BCS- predicted form}. This is difficult to understand if the
fermionic degrees of freedom through $\Delta(T)$ are dominating at all
$x$. Rather this figure suggests that something other than fermionic
excitations is responsible for the disappearance of superconductivity,
particularly in the regime where $\Delta(T)$ is relatively constant in
$T$. At the very least pseudogap effects must persist below $T_c$.

Driving the superconductivity away is another important way to probe the
pseudogap. This occurs naturally with temperatures in excess of $T_c$,
but it also occurs when sufficient pair breaking is present through
impurities \cite{Tallon1997,Zhao,LoramPhysicaC} or applied magnetic
fields \cite{Boebinger}.  An important effect of temperature needs to be
stressed.  With increasing $T > T_c$, the $d$-wave shape of the
excitation gap is rapidly washed out \cite{arpesanl,arpesstanford}; the
nodes of the order parameter, in effect, begin to spread out, just above
$T_c$.  The inverse of this effect should also be emphasized: when
approached from above, \textit{$T_c$ is marked by the abrupt onset of
  long lived quasiparticles}.

One frequently, and possibly universally, sees a
superconductor-insulator (SI) transition when $T_c$ is driven to zero in
the presence of a pseudogap.  This suggests a simple scenario: that
\textit{the pseudogap may survive when superconductivity is suppressed}.
In this way the ground state is no longer a simple metal. This SI
transition is seen upon Zn doping \cite{Tallon1997,Zhao,LoramPhysicaC},
as well as in the presence of applied magnetic fields \cite{Boebinger}.
Moreover, the intrinsic change in stoichiometry illustrated in the phase
diagram of Fig.~\ref{fig:5} also leads to an SI transition. One can thus
deduce that the effects of pair-breaking on $T_c$ and $T^*$ are very
different, with the former being far more sensitive than the latter.

It is clear that pseudogapped fermions are in evidence in a multiplicity
of experiments.  Are there indications of bosonic degrees of freedom in
the normal state of high $T_c$ superconductors? The answer is
unequivocally yes: \textit{meta-stable bosons are observable as
  superconducting fluctuations}.  These bosons are intimately connected
to the pseudogap; in effect they are the other side of the same coin.
These bosonic effects are enhanced in the presence of the quasi-two
dimensional lattice structure of these materials.  Very detailed
analyses \cite{Larkinreview} of thermodynamic and transport properties
of the highest $T_c$ or ``optimal" samples reveal clearly these
pre-formed pairs. Moreover they are responsible \cite{Dorsey} for
divergences at $T_c$ in the dc conductivity $\sigma$ and in the
transverse thermoelectric \cite{Huse} response $\alpha_{xy}$.  These
transport coefficients are defined more generally in terms of the
electrical and heat currents by
\begin{equation}
 \vect{J^{elec}} = \sigma \vect{E} + \alpha \vect(-\nabla T)
\label{eq:tan1}
\end{equation}
\begin{equation}
 \vect{J^{heat}} = \tilde{\alpha} \vect{E} +\kappa \vect(-\nabla T)
\label{eq:tan2}
\end{equation}
Here $\sigma$ is the conductivity tensor, $\kappa$ the thermal
conductivity tensor, and $\alpha , \tilde{\alpha}$ are thermoelectric
tensors.  Other coefficients, while not divergent, exhibit precursor
effects, all of which are found to be in good agreement
\cite{Larkinreview} with fluctuation theory.  As pseudogap effects
become more pronounced with underdoping much of the fluctuation behavior
appears to set in at a higher temperature scale associated with $T^*$,
but often some fraction thereof \cite{Nernst,Tan}.  

\begin{figure}
\includegraphics[width=2.9in,clip]{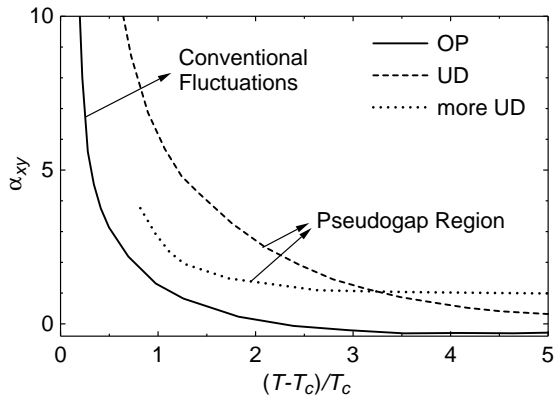}
\caption{Transverse thermoelectric response (which
  relates to the Nernst coefficient) plotted here in fluctuation regime
  above $T_c$.  Here OP and UD correspond to optimal and underdoping,
  respectively.  Data were taken from \onlinecite{Nernst}.}
\label{fig:11c}
\end{figure}

\begin{figure}
\includegraphics[width=2.9in,clip]{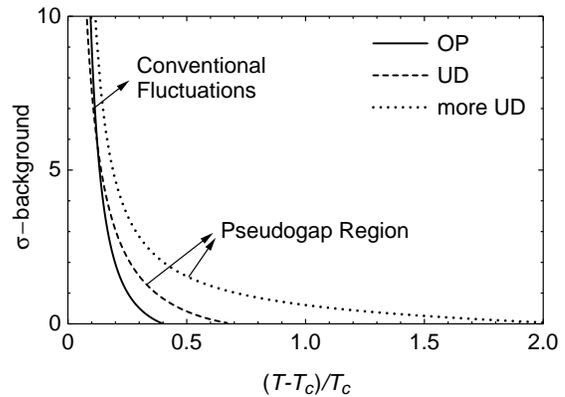}
\caption{Conductivity in fluctuation regime for optimally (OP) and underdoped
  (UD) cuprate superconductors. Data from \onlinecite{Watanabe}.}
\label{fig:11d}
\end{figure}

Figures \ref{fig:11c} and \ref{fig:11d} make the important point that
precursor effects in the transverse thermoelectric response
($\alpha_{xy}$) and conductivity $\sigma$ appear at higher temperatures
($\propto T^*$) as pseudogap effects become progressively more
important; the dotted lines which have the strongest pseudogap continue
to the highest temperatures on both Figures. Moreover, both transport
coefficients evolve smoothly from a regime where they are presumed to be
described by conventional fluctuations \cite{Larkinreview,Huse} as shown
by the solid lines in the figures into a regime where their behavior is
associated with a pseudogap.  A number of people have argued
\cite{Nernst,Corson1999} that normal state vortices are responsible for
the so called anomalous transport behavior of the pseudogap regime.
We will argue in this Review that these figures  
may alternatively be interpreted as suggesting
that bosonic degrees of freedom, not vortices, are responsible
for these anomalous transport effects.

\subsection{Introduction to High $T_c$ Superconductivity: 
Mott Physics and Possible Ordered States}
\label{sec:1C}

Most workers in the field of high $T_c$ superconductivity would agree
that we have made enormous progress in characterizing these materials
and in identifying key theoretical questions and constructs.
Experimental progress, in large part, comes from transport studies
\cite{Timusk,LoramPhysicaC} in addition to three powerful
spectroscopies: photoemission \cite{arpesanl,arpesstanford}, neutron
\cite{Aeppli1991,Keimer,Birgeneau,Mason,Aeppli1997,MookNature,Rossat-Mignod1991,Tranquada}
and Josephson interferometry \cite{Urbana,IBM,Maryland}.  These data
have provided us with important clues to address related theoretical
challenges.  Among the outstanding theoretical issues in the cuprates
are (i) understanding the attractive ``mechanism" that binds electrons
into Cooper pairs, (ii) understanding the evolution of the normal phase
from Fermi liquid (in the ``overdoped" regime) to marginal Fermi liquid
(at ``optimal" doping) to the pseudogap state, which is presumed to
occur as doping concentration $ x $ decreases, and (iii) understanding
the nature of that superconducting phase which evolves from each of
these three normal states.

The theoretical community has concentrated rather extensively on special
regions and $x$-dependences in the phase diagram which are presumed to
be controlled by ``Mott physics".  There are different viewpoints of
precisely what constitutes Mott physics in the metallic phases, and, for
example, whether or not this is necessarily associated with spin-charge
separation. For the most part, one presumes here that the insulating
phase of the parent compound introduces strong Coulomb correlations into
the doped metallic states which may or may not also be associated with
strong antiferromagnetic correlations.  There is a recent, rather
complete review \cite{LeeReview} of Mott physics within the scenario of
spin-charge separation, which clarifies the physics it entails.

Concrete experimental signatures of Mott physics are the observations
that the superfluid density $\rho_s(T=0, x) \rightarrow 0$ as $ x
\rightarrow 0$, as if it were reflecting an order parameter for a metal
insulator transition.  More precisely it is deduced that $\rho_s(0,x)
\propto x$, at low $x$.  Unusual effects associated with this linear-
in- $x$ dependence also show up in other experiments, such as the weight
of coherence features in photoemission data
\cite{arpesanl,arpesstanford}, as well as in thermodynamical signatures
\cite{LoramPhysicaC}.

While there is no single line of reasoning associated with these Mott
constraints, the low value of the superfluid density has been argued
\cite{Emery} to be responsible for soft phase fluctuations of the order
parameter, which may be an important contributor to the pseudogap.
However, recent concerns about this ``phase fluctuation scenario" for
the origin of the pseudogap have been raised \cite{Lee03}.  It is now
presumed by a number of groups that phase fluctuations alone may not be
adequate and an additional static or fluctuating order of one form or
another needs to be incorporated.  Related to a competing or co-existing
order are conjectures \cite{Laughlin} that the disappearance of
pseudogap effects around $ x \approx 0.2$ is an indication of a
``quantum critical point" associated with a hidden order parameter which
may be responsible for the pseudogap.  Others have associated small
\cite{Sachdev1999} $x$ or alternatively optimal \cite{Varma1999} $x$
with quantum critical points of a different origin. The nature of the
other competing or fluctuating order has been conjectured to be RVB
\cite{Vanilla}, ``d-density wave'' \cite{Laughlin}, stripes
\cite{diCastro,Zachar1997} or possibly antiferromagnetic spin
fluctuations \cite{Pines1997,Chubukov,SO5}. The latter is another
residue of the insulating phase.

What is known about the ``pairing mechanism''?  Some would argue that
this is an ill-defined question, and that superconductivity has to be
understood through condensation energy arguments based for example on
the data generated from the extrapolated normal state entropy
\cite{Loram,LoramPhysicaC} discussed above in the context of
Fig.~\ref{fig:9}.  These condensation arguments are often built around the
non-trivial assumption that there is a thermodynamically well
behaved but meta-stable normal phase which coexists with the
superconductivity.  Others would argue that Coulomb effects are
responsible for $d$-wave pairing, either directly \cite{Leggett2,Liu},
or indirectly via magnetic fluctuations \cite{Scalapino86,Pines1997}.
Moreover, the extent to which the magnetism is presumed to persist into
the metallic phase near optimal doping is controversial. Initially, NMR
measurements were interpreted as suggesting \cite{MMP} strong
antiferromagnetic fluctuations, while neutron measurements, which are
generally viewed as the more conclusive, do not provide compelling
evidence \cite{Si} for their presence in the normal phase. Nevertheless,
there are interesting neutron-measured magnetic signatures
\cite{Zha,Liu2,Kao} below $T_c$ associated with $d$-wave
superconductivity. There are also analogous STM effects which are
currently of interest \cite{DHLEE2}.

\subsection{Overview of the BCS-BEC Picture of High $T_c$
Superconductors in the Underdoped Regime}

We summarize here the physical picture associated with the more
controversial BCS-BEC crossover approach to high $T_c$ superconductors,
based on the standard ground state of Eq.~(\ref{eq:1a}).  One presumes
that the normal state (pseudo)gap arises from pre-formed pairs.  This
gap $\Delta(T)$ is thus present at the onset of superconductivity, and
reflects the fact that Cooper pairs form before they Bose condense, as a
result of sufficiently strong attraction. Below $T_c$, $\Delta$ evolves
smoothly into the excitation gap of the superconducting phase.

This behavior reflects the fact that $T_c$ corresponds to the
condensation of the $q =0$ pairs, while finite momentum pairs persist
into the order phase as excitations of the condensate. As the
temperature is progressively lowered below $T_c$, there are fewer and
fewer such bosonic excitations. Their number vanishes precisely at
$T=0$, at which point the system reaches the ground state of
Eq.~(\ref{eq:1a}). Here the superconducting order parameter and the
excitation gap become precisely equal.  This approach is based on a mean
field scheme and is thus not equivalent to treating phase fluctuations
of the order parameter.  In general, below $T_c$, the superconducting
order parameter $\Delta_{sc}$ and the pseudogap contributions add in
quadrature in the dispersion of Eq.~(\ref{eq:1b}) as
\begin{equation}
\Delta^2(T) = \Delta_{sc}^2(T) + \Delta_{pg}^2(T)
\label{eq:2cc}
\end{equation}
where 
$\Delta_{pg} \rightarrow 0$ as $ T
\rightarrow 0$; this behavior should be contrasted with
that in Eq.~(\ref{eq:2a}).

Transport and thermodynamics in the normal and superconducting phases
are determined by two types of quasiparticles: fermionic excitations
which exhibit a gap $\Delta(T)$, and noncondensed pairs or bosonic
excitations.  Contributions from the former are rather similar to their
(low $T$) counterparts in BCS theory, except that they are present well
above $T_c$ as well, with substantially larger $\Delta$ than in strict
BCS theory.  In this way, experiments which probe the opening of an
excitation gap (such as Pauli susceptibility in the spin channel, and
resistivity in the charge channel) will exhibit precursor effects, that
is, the gradual onset of superconductivity from around $T^*$ to below
$T_c$.

The contributions to transport and thermodynamics of the bosonic
excitations are rather similar to those deduced from the time dependent
Ginsburg-Landau theory of superconducting fluctuations, provided that
one properly extends this theory, in a quantum fashion, away from the
critical regime. It is these contributions which often exhibit
divergences at $T_c$, but unlike conventional superconductors, bosonic
contributions (pre-formed pairs) are present, up to high temperatures of
the order of $T^*/2$, providing the boson density is sufficiently high.
Again, these bosons do not represent fluctuations of the order
parameter, but they obey a generically similar dynamical equation of
motion, and, in the more general case which goes beyond
Eq.~(\ref{eq:1a}), they will be coupled to order parameter fluctuations.
Here, in an oversimplified sense there is a form of spin-charge
separation.  The bosons are spin-less (singlet) with charge $2e$, while
the fermions carry both spin and charge, but are non-Fermi-liquid like,
as a consequence of their energy gap.

In the years following the discovery of high $T_c$ superconductors,
attention focused on the so-called marginal Fermi liquid behavior of the
cuprates. This phase is associated with near-optimal doping.  The most
salient signature of this marginality is the dramatically linear in
temperature dependence of the resistivity from just above $T_c$ to very
high temperatures, well above room temperature. Today, we know that
pseudogap effects are present at optimal doping, so that the precise
boundary between the marginal and pseudogap phases is not clear.  As a
consequence, addressing this normal state resistivity in the BCS-BEC
crossover approach has not been done in detail, although it is unlikely
to yield a strictly linear $T$ dependence at a particular doping.
Nevertheless, the bosonic contributions lead to a \textit{decrease} in
resistivity as temperature is lowered, while the opening of a $d$-wave
gap associated with the fermionic terms leads to an \textit{increase},
which tends to compensate. These are somewhat analogous to
Aslamazov-Larkin and Maki-Thompson fluctuation contributions
\cite{Larkinreview}. The bosons generally dominate in the vicinity of
$T_c$, but at higher $T$, in the vicinity of $T^*$, the gapped fermions
will be the more important.

The vanishing of $T_c$ with sufficiently large $T^*$ ( extreme
underdoping) occurs in an interesting way in this crossover theory. This
effect is associated with the localization of the bosonic degrees of
freedom. Pauli principle effects in conjunction with the extended size
of $d$-wave pairs inhibit pair hopping, and thus destroy
superconductivity.  One can speculate that, just as in conventional Bose
systems, once the bosons are localized, they will exhibit an alternative
form of long range order.  Indeed, ``pair density wave" models have been
invoked \cite{Zhang3} to explain recent STM data \cite{Yazdani,Fu}.  In
contrast to the cold atom systems, for the $d$-wave case the BEC regime
is never accessed. This superconductor-insulator transition takes place
in the fermionic regime where $\mu \approx 0.8 E_F$.

In the context of BCS-BEC crossover physics, it is not essential to
establish the source of the attractive interaction.  For the most part
we skirt this issue in this Review by taking the pairing onset
temperature $T^*(x)$ as a phenomenological input parameter.  
At the simplest level one may argue that as the system approaches
the Mott insulating limit ($x \rightarrow 0$) fermions become
less mobile; this serves to increase the effectiveness of the
attractive interaction at low doping levels.  

The variable $x$ in the high $T_c$ problem should be viewed then,
as a counterpart to the magnetic field variable in the cold atom
problem which tunes the system through a Feshbach resonance;
both $x$ and field strength modulate the size of the attraction.
However, for the cuprates, in contrast to the cold atom systems, 
one never reaches the
BEC limit of the crossover theory.

It is reasonable to presume based on the evidence to date, that pairing
ultimately derives from Coulombic effects, not phonons, which are
associated with $l=0$ pairing.  While the widely used Hubbard model
ignores these effects, longer ranged screened Coulomb interactions have
been found \cite{Liu} to be attractive for electrons in a $d$-wave
channel.  In this context, it is useful to note that, similarly, in
${\rm He}^3$ short range repulsion destroys $s$-wave pairing, but leads
to attraction in a higher ($l=1$) channel \cite{AndersonBrinkman}.
There is, however, no indication of pseudogap phenomena in ${\rm He}^3$,
so that an Eliashberg extended form of BCS theory appears to be
adequate.  Eliashberg theory is a very different form of ``strong
coupling" theory from crossover physics, which treats in detail the
dynamics of the mediating boson.  Interestingly, there is an upper bound
to $T_c$ in both schemes.  For Eliashberg theory this arises from the
induced effective mass corrections \cite{Valls}, whereas in the
crossover problem this occurs because of the presence of a pseudogap at
$T_c$.

Other analogies have also been invoked in comparing the cuprates with
superfluid ${\rm He}^3$, based on the propensity towards magnetism. In
the case of the quantum liquid, however, there is a positive correlation
between short range magnetic ordering \cite{Valls} and superconductivity
while the correlation appears negative for the cuprates ( and some of
their heavy fermion ``cousins") where magnetism and superconductivity
appear to compete.

\subsubsection{Searching for the Definitive High $T_c$ Theory}

In no sense should one argue that the BCS-BEC crossover approach
precludes consideration of Mott physics. Nor should it be viewed as
superior. Indeed, many of the experiments we will address in this Review
have alternatively been addressed within the Mott scenario. Presumably
both components are important in any ultimate theory. We have chosen in
this Review to focus on one component only, and it is hoped that a
comparably detailed review of Mott physics will be forthcoming. In this
way one could compare the strengths and weaknesses of both schemes.  A
summary \cite{Carlson2} of some aspects of the Mott picture with
emphasis on one dimensional stripe states should be referred to as a
very useful starting point.  Also available now is a new Review
\cite{LeeReview} of Mott physics, based on spin charge separation.

Besides those discussed in this Review, there is a litany of other
experiments which are viewed as requiring explanation in order for a
high $T_c$ theory to be taken seriously.  Among these are the important
linear resistivity in the normal state (at optimal doping) and unusual
temperature and $x$ dependences in Hall data.  While there has been
little insight offered from the crossover scheme to explain the former,
aspects of the latter have been successfully addressed within pre-formed
pair scenarios \cite{Larkin}.  Using fermiology based schemes the major
features of photoemission\cite{Pepin} and of neutron scattering and NMR
data were studied theoretically and rather early on \cite{Si,Zha,Kao},
along with a focus on the now famous $41~meV$ feature \cite{Liu2}.
These and other related approaches to the magnetic data, make use of a
microscopically derived \cite{Si2} random phase approximation scheme
which incorporates the details of the Fermi surface shape and $d$-wave
gap.  Among remaining experiments in this litany is the origin of
incoherent transport along the $c$-axis \cite{Rojo}.

On the basis of the above cited literature which addressed magnetic and
$c$-axis \footnote{Here the $ab$-plane refers to the crystal plane in
  parallel with the layers of the cuprate compounds, and the $c$-axis
  refers to the axis normal to the layers.}  data we are led to conclude
that pseudogap effects and related crossover physics are not central to
their understanding.  The presence of a normal state gap will lead to
some degree of precursor behavior above $T_c$, but $d$-wave nesting
related effects such as reflected in neutron data \cite{Zha,Kao,Liu2}
will be greatly weakened above $T_c$ by the blurring out of the $d$-wave
gap nodes \cite{arpesstanford,arpesanl}.  It should be said, however,
that $c$-axis optical data do appear to reflect pseudogap effects
\cite{Ioffe} above $T_c$.  (The $ab$-plane counterparts are discussed in
Section \ref{sec:6F}).  In summary, we are far from the definitive high
$T_c$ theory and the field has much to gain by establishing a
collaboration with the cold atom community where BCS-BEC crossover as
well as Mott physics \cite{Demler2} can be further elucidated and
explored.

\subsection{Many Body Hamiltonian and Two Body Scattering Theory:
Mostly Cold Atoms}
\label{sec:hamiltonian}

\begin{figure}
\centerline{\includegraphics[width=2.2in]{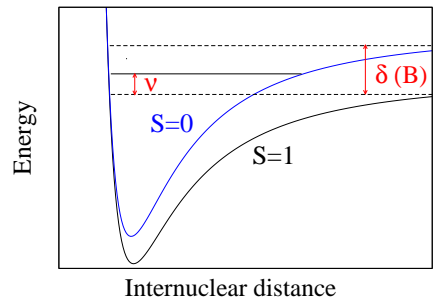}}
\caption{A schematic representation of the Feshbach resonance for $^6$Li. The
  effective interaction between the scattering of atoms in the triplet
  state is greatly enhanced, when the energy level of the singlet bound
  state approaches the continuum of the $S=1$ channel.}
\label{feshbach}
\end{figure}

The existence of Feshbach resonances in a gas of ultracold fermionic
atoms allows the interaction between atoms to be tuned
arbitrarily, via application of a magnetic field.  Figure \ref{feshbach}
presents a characteristic plot of the two body interaction potentials
which are responsible for this resonance.  These interaction potentials
represent an effective description of all Coulomb interactions between
the electrons and nuclei.  For definiteness, we consider the case of
$^6$Li here. In the region of magnetic fields near the ``wide" Feshbach
resonance, $\approx 837 G$, we may focus on the two lowest atomic levels
characterized by $|F, m_F \rangle$, where the two quantum
numbers correspond to the total spin and its $z$ component.
We specify the two states $|1\rangle \equiv |\frac{1}{2},
\frac{1}{2}\rangle$ and $|2\rangle \equiv |\frac{1}{2},-
\frac{1}{2}\rangle$.  In place of these quantum numbers 
one can alternatively characterize
the two states in terms of the projections of the electronic ($m_s$)
and nuclear ($m_i$) spins along the magnetic field.  In this way the
states 1 and 2 have $m_s = -\frac{1}{2}$ and $m_i =1$ and $0$,
respectively.

The curve in Fig.~\ref{feshbach} labelled $S=1$ represents the
interaction between these two states.  This corresponds to
the so-called ``open" channel and is associated with a triplet
(electronic) spin state. A Feshbach resonance requires that there be an
otherwise unspecified near-by state which we refer to as in
the ``closed"
channel.  The potential which gives rise to this singlet (electronic)
spin state is shown by the $S=0$ curve in Fig.~\ref{feshbach}.
A Feshbach resonance arises
when this bound state lies near the zero of energy corresponding to the
continuum of scattering states in the triplet or open channel.  The
electronic Zeeman coupling to the triplet state makes is possible to
shift the relative position of the two potentials by application of a
magnetic field. This shift is indicated by $\delta$ in
Fig.~\ref{feshbach}.  In this way, the resonance is magnetically tunable
and the scattering length of the system will vary continuously with
field, diverging at that magnetic field which corresponds to the
Feshbach resonance. We represent the deviation of this field from the
resonance condition by the ``detuning parameter" $\nu$.  When $\nu$ is
large and positive, the singlet state is substantially raised relative
to the triplet-- and it plays relatively little role.  By contrast when
the magnitude of $\nu$ is large, but with negative sign, then the
physics is dominated by the singlet or closed channel.

The most general form for the Hamiltonian contains two types of
particles: fermions which represent the open channel and ``molecular
bosons" which represent the two fermion bound state of the closed
channel.  We will also refer to the latter as ``Feshbach bosons".
These, in turn, lead to two types of interaction effects:
those associated with the direct attraction between fermions,
parameterized by $U$, and those associated with ``fermion-boson"
interactions, whose strength is governed by $g$.  The latter may be
viewed as deriving from hyperfine interactions which couple the (total)
nuclear and electronic spins.  In the high $T_c$ applications, one
generally, but not always, ignores the bosonic degrees of freedom.
Otherwise the Hamiltonians are the same, and can be written as
\begin{eqnarray}
\label{hamiltonian}
H&-&\mu N=\sum_{{\bf k},\sigma}(\epsilon_{\bf k}-\mu)\createa{{\bf
    k},\sigma}\destroya{{\bf k},\sigma}+\sum_{\bf q}(\epsilon_{\bf
    q}^{mb}+\nu-2 \mu)\createb{\bf q}\destroyb{\bf q}\nonumber\\ 
&+&\sum_{{\bf q},{\bf k},{\bf k'}}U({\bf k},{\bf k'})\createa{{\bf
    q}/2+{\bf k},\uparrow}\createa{{\bf q}/2-{\bf
    k},\downarrow}\destroya{{\bf q}/2-{\bf k'},\downarrow}\destroya{{\bf
    q}/2+{\bf k'},\uparrow}\nonumber\\ 
&+&\sum_{{\bf q},{\bf k}}\left(g({\bf k})\createb{{\bf q}}\destroya{{\bf
    q}/2-{\bf k},\downarrow}\destroya{{\bf q}/2+{\bf
    k},\uparrow}+h.c.\right)  
\label{eq:0c}
\end{eqnarray}
Here the fermion and boson kinetic energies are given by $\ek=k^2/(2
m)$, and $\epsilon_{\bf q}^{mb}=q^2/(2 M)$, and $\nu$ is an important
parameter which represents the magnetic field ``detuning". $M=2m$ is the
bare mass of the Feshbach bosons.  For the cold atom problem $\sigma =
\uparrow, \downarrow$ represents the two different hyperfine states
($|1\rangle$ and $|2\rangle$).  One may ignore \cite{Timmermans} the
non-degeneracy of these two states, since there is no known way for
state 2 to decay into state 1.  In the two channel problem the ground
state wavefunction is slightly modified
and given by
\begin{equation}
\bar{\Psi}_0 = \Psi_0 \otimes \Psi_0^B 
\end{equation}
where the normalized molecular boson contribution $\Psi_0^B$ is 
\begin{equation}
 \Psi_0^B=
 e^{-\lambda^2/2+\lambda b_0^{\dagger}}|0\rangle \,.
 \end{equation}
as discussed originally by 
\onlinecite{Ranninger2}.  Here $\lambda$ is a variational parameter. We
present the resulting variational conditions for the ground state in
Appendix \ref{App:Var}.

Whether both forms of interactions are needed in either the high $T_c$
or cold atom systems is still under debate. The bosons ($\createb{\bf
  k}$) of the cold atom problem \cite{Holland,Timmermans} will be
referred to as Feshbach bosons.  These represent a separate species, not
to be confused with the fermion pair ($\createa{\bf k} \createa{\bf
  -k}$) operators.  Thus we call this a ``two channel" model.  In this
review we will discuss the behavior of crossover physics both with and
without these Feshbach bosons (FB).  Previous studies of high $T_c$
superconductors have invoked a similar bosonic term
\cite{Ranninger,Micnas,Larkin} as well, although less is known about its
microscopic origin.  This fermion-boson coupling is not to be confused
with the coupling between fermions and a ``pairing-mechanism"-related
boson ($[b + b^\dagger]a^\dagger a$) such as phonons. The coupling
$b^\dagger a a$ and its conjugate represents a form of sink and source
for creating fermion pairs, inducing superconductivity in some ways, as
a by-product of Bose condensation. We will emphasize throughout that
crossover theory proceeds in a similar fashion with and without Feshbach
bosons.

\begin{figure}
\includegraphics[width=2.3in,clip]{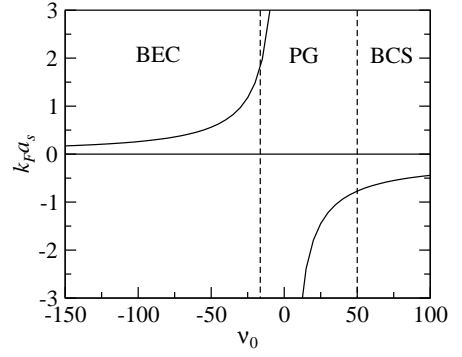}
\caption{Characteristic behavior of the scattering length
  $a_s$ in the three regimes. $a_s$ diverges when $\nu_0 = 2\mu$. Here,
  and in subsequent figures $\nu_0$ is always in units of $E_F$ and we
  estimate $1\ { \rm G} \approx 60 E_F$ for a typical $E_F = 2\, \mu$K,
  for either $^6$Li or $^{40}$K.}
\label{fig:9a}
\end{figure}

It is useful at this stage to introduce the $s$-wave scattering length,
$a_s$. This parameter is associated with the scattering amplitude
in the low energy limit, for a two body scattering process.
This scattering length can, in principle, be deduced from Figure
\ref{feshbach}.  It follows that $a_s$ is negative when there is no
bound state, it tends to $-\infty$ at the onset of the bound state and
to $+\infty$ just as the bound state stabilizes.  It remains positive
but decreases in value as the interaction becomes increasingly strong.
The magnitude of $a_s$ is arbitrarily small in both the extreme BEC and
BCS limits, but with opposite sign (See Fig.~\ref{fig:9a}).
\textit{The fundamental postulate of crossover theory is that even
  though the two-body scattering length changes abruptly at the unitary
  scattering condition ($|a_s| = \infty $), in the N-body problem the
  superconductivity varies smoothly through this point}.  Throughout
this Review, we presume an equivalence between the ``pseudogap'' and
unitary scattering regimes.

An important goal is to understand how to incorporate all the
constraints of the two body physics into the superfluid gap equation.
As a by-product, one avoids implicit divergences which would also appear
in the problem.  To begin we ignore Feshbach effects.  The way in which
the two body interaction $U$ enters to characterize the scattering (in
vacuum) is different from the way in which it enters to characterize the
N-body processes leading to superfluidity.  In each case, however, one
uses a $T$-matrix formulation to sum an appropriately selected but
infinite series of terms in $U$.  For the two-body problem in vacuum, we
introduce the scattering length, $a_s$,
\begin{equation}
\frac{m}{4 \pi a_s} \equiv \frac {1}{U_0} \,,
\label{eq:11g}
\end{equation}
which is related to $U$ via the Lippmann-Schwinger equation
\begin{equation}
\frac{m}{4 \pi a_s } \equiv \frac{1}{U} + \sum_{\bf k} \frac{1}{2 \ek} \,.
\label{eq:11c}
\end{equation}

In this way one may solve for the unknown $U$ in terms of $a_s$
or $U_0$. The two-body $T$-matrix equation 
 [Eq.~(\ref{eq:11c})]
can be rewritten as
\begin{equation}
U = \Gamma U_0, ~~~~\Gamma = \left(1+ \frac {U_0}{U_c}\right)^{-1} \,,
\label{eq:11h}
\end{equation}
where we define the quantity
$U_c$ as 
\begin{equation}
U_c^{-1} =- \sum_{\bf k} \frac{1}{2 \ek} \,.
\label{eq:Uc}
\end{equation}
Here $U_c$ is the critical value of the potential associated with the
binding of a two particle state in vacuum.  Specific evaluation of $U_c$
requires that there be a cut-off imposed on the above summation,
associated with the range of the potential.

Now let us turn to the Feshbach problem, where the superfluidity is
constrained by a more extended set of parameters $U$, $g$, $\nu$ and
also $\mu$. Provided we redefine the appropriate ``two body" scattering
length, Equation (\ref{eq:11c}) holds even in the presence of Feshbach
effects \cite{Griffin,Milstein}.  It has been shown that $U$ in the
above equations is replaced by $U_{eff}$ and $a_s$ by $a_s^*$
\begin{equation}
\frac {m}{4 \pi a_s^*} \equiv \frac {1}{U_{eff}} + \sum_{\bf k} \frac{1}
{2 \ek} \,.
\label{eq:11y}
\end{equation}
Here it is important to stress that many body effects, via the fermionic
chemical potential $\mu$ enter into $a_s^*$.

More precisely the effective interaction between two fermions is $Q$
dependent. It arises from a second order process involving emission and
absorption of a molecular boson. The net effect of the direct plus
indirect interactions is given by 
\begin{equation}
U_{eff}(Q) \equiv U + g^2 D_0 (Q) \;,
\label{eq:Ueff}
\end{equation}
where 
\begin{equation}
D_0(Q) \equiv 1/ [ i\Omega_n- \epsilon_{\bf q}^{mb}-\nu + 2 \mu
]
\label{eq:D0}
\end{equation}
is the non-interacting molecular boson propagator.  What appears in the
gap equation, however, is 
\begin{equation}
U_{eff}(Q=0) \equiv U_{eff}= U -
\frac{g^2}{\nu - 2\mu} \;.
\label{eq:Ueff0}
\end{equation}

Experimentally, the two body scattering length $a^*_s$ varies with
magnetic field $B$. Near resonance, it can be parameterized
\cite{Kokkelmans} in terms of three parameters of the two body problem
$U_0$, $g_0$ and the magnetic field which at resonance is given by
$B_0$. One fits the scattering length to $a_{bg}(1-\frac{w}{B-B_0})$ so
that the width $w$ is related to $g_0$ and the background scattering
length $a_{bg}$ is related to $U_0$.  Away from resonance, the crossover
picture requires that the scattering length vanish asymptotically as
\begin{equation}
a_s^* \approx 0 \,, ~~~~~\text{in~extreme~BEC~limit,}
\label{eq:11w}
\end{equation}
so that the system represents an ideal Bose gas.  Here it should be
noted that this BEC limit corresponds to $\nu \rightarrow \mu^+$.  The
counterpart of Eq.~(\ref{eq:11w}) also holds for the one channel
problem.  Note that the actual resonance of the many body problem will
be somewhat shifted from the atomic-fitted resonance parameter $B_0$.

It is convenient to define the parameter $\nu_0$ which is directly
related to the difference in the applied magnetic field $B$ and $B_0$
\begin{equation}
\nu_0 = (B-B_0) \Delta \mu^{0},
\label{eq:11x}
\end{equation}
where $\Delta \mu^{0}$ is the difference in the magnetic moment of the singlet 
and triplet paired  states.  
\textit{We will use the parameter $\nu_0$ as a measure of
magnetic field strength throughout this Review}.
The counterpart of Eq.~(\ref{eq:11g}) is then 
\begin{equation}
\frac{m}{4 \pi a^*_s} \equiv \frac {1}{U_0 - \frac{g_0^2}{\nu_0-2\mu}}
\equiv \frac{1}{U^*}\,,
\label{eq:11k}
\end{equation}
where $U_0=\frac{4 \pi a_{bg}}{m}$ and $g_0^2=U_0 \Delta \mu^0 w$.
For the purposes of this Review we will use the scattering length $a_s^*$,
although for simplicity we often drop the asterisk and write it
as $a_s$.

We can again invert the Lippmann-Schwinger or $T$-matrix scattering
equation \cite{Kokkelmans,Pethick} to arrive at the appropriate
parameters $g$ and $\nu$ which enter into the Hamiltonian and thus into
the gap equation.  Just as in Eq.~(\ref{eq:11h}), one finds a connection
between measurable parameters (with subscript $0$) and their
counterparts in the Hamiltonian (without the subscript):
\begin{equation}
 U=\Gamma U_0 , ~~~g = \Gamma g_0 , ~~~\nu - \nu_0 =  - \Gamma
 \frac{g_0^2}{U_c} \,.
\label{eq:200}
\end{equation}
To connect the various energy scales, which appear in the problem,
typically $1 \rm{G} \approx 60 E_F$ for both $^6$Li and $^{40}$K.
Here it is assumed that the Fermi energy is $E_F \approx 2\, \mu$K.

In Fig.~\ref{fig:9a} and using Eq.~(\ref{eq:11y}), we plot a
prototypical scattering length $k_Fa_s \equiv k_Fa^*_s$ as a function of
the magnetic field dependent parameter, $\nu_0$.  This figure indicates
the BEC, BCS and PG regimes. Here, as earlier, the PG regime is bounded
on the left by $\mu=0$ and on the right by $\Delta(T_c)\approx 0$.

A key finding associated with this plot is that \textit{the PG regime
  begins on the so-called ``BEC side of resonance", independent of
  whether Feshbach effects are included or not}.  That is, the fermionic
chemical potential reaches zero while the scattering length is positive.
This generic effect of the ground state self consistent equations arises
from the Pauli principle.  For an isolated system, two particle resonant
scattering occurs at $U=U_c$, where $U_c$ is defined in
Eq.~(\ref{eq:Uc}).  In the many body context, because of Pauli principle
repulsion between fermions, it requires an effectively stronger
attraction to bind particles into molecules.  Stated alternatively, the
onset of the fermionic regime ($\mu > 0$) occurs for more strongly
attractive interactions, than those required for two body resonant
scattering.  As a consequence of $\mu > 0$, \textit{there is very little
  condensate weight in the Feshbach boson channel near the unitary
  limit, for values of the coupling parameter $g_0$, appropriate to
  currently studied (and rather wide) Feshbach resonances}.

Finally, it is useful to compare the two interaction terms which we have
defined above, ($ U_{eff}$ and $U^*$), and to contrast their behavior in
the BEC and unitary limits.  The quantity $ U^* $ is proportional to the
scattering length; it reflects the two body physics and necessarily
diverges at unitarity, where $U_{eff} = U_c$.  By contrast, in the BEC
regime, $U^*$, or $a_s^*$ approaches zero, (under the usual presumption
that the interaction is a contact interaction).  This vanishing of $U^*$
should be compared with the observation that the quantity $U_{eff}$,
which reflects the many body physics, must diverge in this BEC limit.

\subsubsection{Differences Between One and Two Channel
Models: Physics of 
Feshbach bosons}

The main effects associated with including the explicit Feshbach
resonance are twofold. These Feshbach bosons provide a physical
mechanism for tuning the scattering length to be arbitrarily large, and
in the BEC limit, they lead to different physics from the one channel
problem.  As will be illustrated later via a comparison of Figures
\ref{fig:16} and \ref{fig:18}, in an artificial way one can capture most
of the salient physics of the unitary regime via a one channel approach
in which one drops all molecular boson related terms in the Hamiltonian,
but takes the interaction $U$ to be arbitrary.  In this way there is
very little difference between fermion-only and fermion-boson models.
There is an important proviso, though, that one is not dealing with
narrow resonances which behave rather differently near unitarity, as
will be seen in Fig.~\ref{fig:24x}.  In addition, as in the one
channel model a form of universality \cite{JasonHo} is found in the
ground state at the unitary limit, provided $g$ is sufficiently large.
This is discussed later in the context of Fig.~\ref{fig:23x}.

We emphasize that, within the BEC regime there are three important
effects associated with the Feshbach coupling $g$. As will become clear
later, (i) in the extreme BEC limit when $g$ is nonzero, there are no
occupied fermionic states. The number constraint can be satisfied
entirely in terms of Feshbach bosons. (ii) The absence of fermions will,
moreover, greatly weaken the inter-boson interactions which are presumed
to be mediated by the fermions.  In addition, as one decreases
$|U_{eff}|$ from very attractive to moderately attractive (i.e., increases
$a^*_s$ on the BEC side) the nature of the condensed pairs changes. Even
in the absence of Feshbach bosons (FB), the size of the pairs increases.
But in their presence the admixture of bosonic (i.e., molecular bosons)
and fermionic (i.e., Cooper pair) components in the condensate is
continuously varied from fully bosonic to fully fermionic. Finally (iii)
the role of the condensate enters in two very different ways into the
self consistent gap and number equations, depending on whether or not
there are FB.  The molecular boson condensate appears principally in the
number equation, while the Cooper pair condensate appears principally in
the gap equation.  To make this point clearer we can refer ahead to
Eq.~(\ref{number_equation}) which contains the FB condensate
contribution $n_b^0$. By contrast the Cooper condensate contribution
$\Delta_{sc}$ enters into the total excitation gap $\Delta$ as in
Eq.~(\ref{eq:2cc}) and $\Delta$ is, in turn, constrained via
Eq.~(\ref{eq:18}).  Points (i) and (iii), which may appear somewhat
technical, have essential physical implications, among the most
important of these is point (ii) above.

We stress, however, that \textit{for the PG and BCS regimes the
  differences with and without FB are, however, considerably less
  pronounced}.

\subsection{Current Summary of Cold Atom Experiments: Crossover in the
Presence of Feshbach Resonances}
\label{sec:exp_cold_summary}

There has been an exciting string of developments over the past few
years in studies of ultracold fermionic atoms, in particular, $^6{\rm
  Li}$ and $^{40}{\rm K}$, which have been trapped and cooled via
magnetic and optical means.  Typically these traps contain $\sim 10^5$
atoms at very low densities $\approx 10^{13}$ ${\rm cm}^{-3}$ when
cooled to very low $T$. Here the Fermi temperature $E_F$ in a trap can
be estimated to be of the order of a micro-Kelvin.  It was argued on the
basis of BCS theory alone \cite{Houbiers}, and rather early on (1997),
that the temperatures associated with the superfluid phases may be
attainable in these trapped gases.  This set off a search for the ``holy
grail'' of fermionic superfluidity.  That a Fermi degenerate state could
be reached at all is itself quite remarkable; this was was first
reported \cite{Jin} by Jin and deMarco in 1999.  By late 2002 reports of
unusual hydrodynamics in a degenerate Fermi gas indicated that strong
interactions were present \cite{ohara}.  This strongly interacting Fermi
gas (associated with the unitary scattering regime) has attracted
widespread attention independent of the search for superfluidity,
because it appears to be a prototype for analogous systems in nuclear
physics \cite{Baker,Heiselberg3} and in quark-gluon plasmas
\cite{Itakura,Heinz}.  Moreover, there has been a fairly extensive body
of analytic work on the ground state properties of this regime
\cite{Heiselberg2, Carlson}, which goes beyond the simple mean field
wave function ansatz.

\begin{figure}
\includegraphics[width=3.1in,clip]{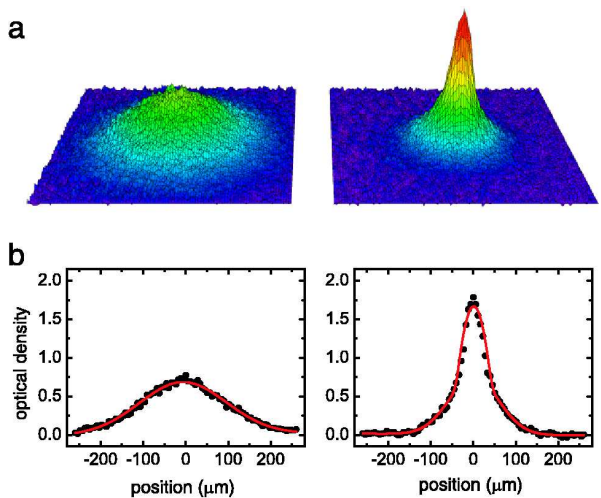}
\caption{(Color) Spatial density profiles of a
  molecular cloud of trapped $^{40}$K atoms in the BEC regime in the
  transverse directions after 20 ms of free expansion (from
  \onlinecite{Jin3}), showing thermal molecular cloud above $T_c$ (left)
  and a molecular condensate (right) below $T_c$. (a) shows the surface
  plots, and (b) shows the cross-sections through images (dots) with
  bimodal fits (lines).}
\label{fig:jin}
\end{figure}

As a consequence of attractive $s$-wave interactions between fermionic
atoms in different hyperfine states, it was anticipated that dimers
could also be made. Indeed, these molecules formed rather efficiently
\cite{Regal,Hulet3,Cubizolles} as reported in mid-2003 either via three
body recombination \cite{Jochim} or by sweeping the magnetic field
across a Feshbach resonance. Moreover, they are extremely long lived
\cite{Hulet3}. From this work it was relatively straightforward to
anticipate that a Bose condensate would also be achieved.  Credit goes
to theorists such as Holland \cite{Holland} and to Timmermans
\cite{Timmermans} and their co-workers for recognizing that the
superfluidity need not be only associated with condensation of long
lived bosons, but in fact could also derive, as in BCS, from fermion
pairs.  In this way, it was argued that a suitable tuning of the
attractive interaction via Feshbach resonance effects would lead to a
realization of a BCS-BEC crossover.

By late 2003 to early 2004, four groups
\cite{Jin3,Grimm,Ketterle2,Salomon3} had observed the ``condensation of
weakly bound molecules" (that is, on the $a_s > 0 $ side of resonance),
and shortly thereafter a number had also reported evidence for
superfluidity on the BCS side \cite{Jin4,Ketterle3,Thomas2,Grimm4}.  The
BEC side is the more straightforward since the presence of the
superfluid is reflected in a bi-modal distribution in the density
profile.  This is shown in Fig.~\ref{fig:jin} from \cite{Jin3}, and is
conceptually similar to the behavior for condensed Bose atoms
\cite{RMP}.  On the BEC side but near resonance, the estimated $T_c$ is
of the order of 500~nK, with condensate fractions varying from 20\% or
so, to nearly 100\%. The condensate lifetimes are relatively long in the
vicinity of resonance, and fall off rapidly as one goes deeper into the
BEC.  However, for $a_s < 0$ there is no clear expectation that the
density profile will provide a signature of the superfluid phase.

These claims that superfluidity may have been achieved on the BCS side
($a_s < 0$) of resonance were viewed as particularly exciting. The
atomic community, for the most part, felt the previous counterpart
observations on the BEC side were expected and not significantly
different from condensation in Bose atoms. The evidence for this new
form of ``fermionic superfluidity" rests on studies
\cite{Jin4,Ketterle3} that perform fast sweeps from negative $a_s$ to
positive $a_s$ across the resonance.  The field sweeps allow, in
principle, a pairwise projection of fermionic atoms (on the BCS side)
onto molecules (on the BEC side).  It is presumed that in this way one
measures the momentum distribution of fermionic atom pairs. The
existence of a condensate was thus inferred. Other experiments which
sweep across the Feshbach resonance adiabatically, measure the size of
the cloud after release \cite{Salomon3} or within a trap \cite{Grimm2}.
Evidence for superfluidity on the BCS side, which does not rely on the
sweep experiments, has also been deduced from collective excitations of
a fermionic gas \cite{Thomas2,Grimm3}. Pairing gap measurements with
radio frequency (RF) probes \cite{Grimm4} have similarly been
interpreted \cite{Torma2} as evidence for superfluidity, although it
appears more likely that these experiments establish only the existence
of fermionic pairs.  Quite recently, evidence for a phase transition has
been presented via thermodynamic measurements and accompanying theory
\cite{ThermoScience}.  The latter, like the theory \cite{Torma2} of RF
experiments \cite{Grimm4}, is based on the formalism presented in this
Review.

Precisely what goes on during the sweep is not entirely understood. It
should be stressed again that where the scattering length changes sign
is not particularly important to the N-body physics. \textit{Thus
  starting on the ``BCS side of resonance" and ending on the ``BEC side
  of resonance" involves a very continuous sweep in which the physics
  is not qualitatively changed}.
It has been speculated that for this sweep procedure to work a large
percentage of the Cooper pair partners must be closer than the
inter-atomic spacing.  Others have alternatively viewed the sweep as
having little to do with the large Cooper pairs of BCS
superconductivity.

\begin{figure}
\includegraphics[width=2.3in,clip]{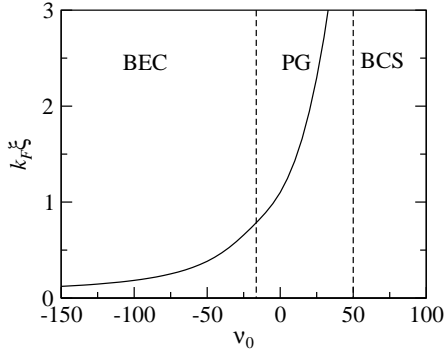}
\caption{Characteristic behavior of Cooper pair size $\xi$ in the condensate
  in the three regimes. The dotted vertical lines are not sharp
  transitions but indicate where $\mu = 0$ (left line) and where
  $\Delta(T_c) \approx 0$ (right line). The dimensionless $k_F\xi$ gives
  the ratio between pair size $\xi$ and the mean inter-particle spacing
  $1/k_F$. Pairs shrink and become tightly bound in the deep BEC regime.}
\label{fig:9b}
\end{figure}

It is useful to establish the various pair sizes here.  In the PG regime
the normal state consists of a significant number of small (compared to
BCS) pre-formed pairs. Below $T_c$, the condensate pair size is also
small.  Finally, there are excited pair states in the superfluid phase,
with characteristic size $\xi_{pg}$. The Cooper pair size $\xi$ as
plotted in Fig.~\ref{fig:9b} was deduced from the condensate
\cite{randeriareview} at $T=0$, but this can be shown to be rather
similar to $\xi_{pg}$, corresponding to the size of excited pair states
\cite{Chen2,ChenPhD} both above and below $T_c$.  It seems plausible
that during the sweep there is some rearrangement of excited states
(both fermionic and bosonic) and condensate.  In addition, all pairs
(excited and condensate) contract in size. When, at the end of a sweep,
they are sufficiently small, then they are more ``visible".  What
appears to be essential, however, is that the sweep time scales are
rapid enough so that the existence of a condensate at the initial
starting point is a necessary and sufficient condition for observing a
condensate at the end of a sweep.

\subsection{\textit{T}-Matrix-Based Approaches to BCS-BEC Crossover in the
Absence of Feshbach Effects}
\label{sec:1F}

In this subsection we introduce the concept of a $T$-matrix within the
pairing channel.  This is defined in Eq.~(\ref{eq:1c}) below and it
represents an effectively screened interaction in which the appropriate
polarizability ($\chi$) is associated with the particle-particle rather
than the particle-hole channel, as in the more conventional dielectric
susceptibility.  The $T$-matrix thus appears as an infinite series of
two particle ``ladder" diagrams.

Away from zero temperature, a variety of different many body approaches
have been invoked to address the physics of BCS-BEC crossover. For the
most part, these revolve around $T$-matrix schemes.  Here one solves self
consistently for the single fermion propagator ($G$) and the pair
propagator ($t$).  That one stops at this level without introducing
higher order Green's functions (involving three, and four particles,
etc) is believed to be adequate for addressing a leading order mean
field theory such as that represented by Eq.~(\ref{eq:1a}). One can see
that pair-pair (boson-boson) interactions are only treated in a mean
field averaging procedure; they arise exclusively from the fermions and
are presumed to be sufficiently weak so as not to lead to any incomplete
condensation in the ground state, as is compatible with Eq.~(\ref{eq:1a}).

One can view this approach as the first step beyond BCS in a hierarchy
of mean field theories.  In BCS, above $T_c$ one includes only the bare
fermion propagator $G_0$. Below $T_c$, pairs play a role but only through
the condensate.  At the next level one accounts for the interaction
between particles and noncondensed pairs in both the normal and
superconducting states. The pairs introduce a self energy $\Sigma$ into
the particles, which represents a correction to BCS theory.  The pairs
are treated at an effective mean field level.  By truncating the
equations of motion in this way, the effects of all higher order Green's
functions are subsumed into $t$ in an averaged way.

Below we demonstrate that at this $T$-matrix level there are a number of
distinct schemes which can be implemented to address BCS-BEC crossover
physics.  These same alternatives have also entered into a discussion
of pre-formed pairs as they appear in treatments of superconducting
fluctuations.  Above $T_c$, quite generally one writes for the
$T$-matrix
\begin{equation}
t(Q) = \frac{U}{1+U\chi(Q)}
\label{eq:1c}
\end{equation}
and theories differ only on what is the nature of the pair
susceptibility $\chi(Q)$, and the associated self energy of the
fermions.  Here and throughout we use $Q$ to denote a four-vector and
write $\sum_Q \equiv T \sum_{i \Omega_n} \sum_{\bf q}$, where $\Omega_n$
is a Matsubara frequency.  Below $T_c$ one can also consider a $T$-matrix
approach to describe the particles and pairs in the condensate. For the
most part we will defer extensions to the broken symmetry phase to
Section \ref{sec:2B}.

\subsubsection{Review of BCS Theory Using the $T$-matrix Approach}

In order to assess alternative schemes, it
is useful to review BCS theory within a $T$-matrix formalism.  In BCS
theory, pairs explicitly enter into the problem below $T_c$, but only
through the condensate.  These condensed pairs are associated with a
$T$-matrix given by
\begin{equation}
t_{sc}(Q) = - \Delta_{sc}^2 \delta(Q) / T
\label{eq:2c}
\end{equation}
with fermionic self energy 
\begin{equation}
\Sigma_{sc}(K) = \sum_Q t_{sc}(Q) G_0(Q-K)
\label{eq:2ca}
\end{equation}
so that $\Sigma_{sc}(K) = -\Delta_{sc}^2 G_0(-K)$.
Here, and throughout,
$G_0$ is the Green's function of the non-interacting system.
The number of fermions in a BCS superconductor is given by
\begin{equation}
n = 2 \sum_K G(K)
\label{eq:3c}
\end{equation}
and 
\begin{equation}
 G(K) = [ G_0^{-1}(K) - \Sigma_{sc}(K) ]^{-1}
\label{eq:12c}
\end{equation}
Doing the Matsubara summation in Eq.~(\ref{eq:3c}), one can then deduce
the usual BCS expression for the number of particles, which determines
the fermionic chemical potential
\begin{equation}
n= \sum_{\bf k} \left[ 1 - \frac{\epsilon_{\bf k}-\mu} {\Ek} + 2
  \frac{\epsilon_{\bf k}-\mu}{\Ek} f(\Ek)\right]
\label{eq:12ca}
\end{equation}
where
\begin{equation}
\Ek = \sqrt{ (\ek -\mu)^2 + \Delta_{sc}^2 (T) }
\label{eq:9c}
\end{equation}

We need, however, an additional equation to
determine $\Delta_{sc}(T)$. 
The BCS gap equation can be written as
\begin{equation}
 1+ U \chi_{BCS}(0) = 0,~~T \leq T_c
\label{eq:5c}
\end{equation}
where
\begin{equation}
\chi_{BCS}(Q) = \sum_K G(K)G_0(Q-K)
\label{eq:4c}
\end{equation}
This suggests that one consider the uncondensed or normal state pair
propagator to be of the form
\begin{equation}
t(Q) = \frac{U}{1 + U \chi_{BCS}(Q)}
\label{eq:6c}
\end{equation}
then in the superconducting state we have a BEC like condition on the
pair chemical potential $\mu_{pair}$ defined by
\begin{equation}
t^{-1}(Q=0) = \mu_{pair} \times const.
\label{eq:7c}
\end{equation}
where the overall constant is unimportant for the present purposes.
Thus we say that the pair chemical potential satisfies
\begin{equation}
\mu_{pair} = 0 , ~~~~ T \leq T_c\;.
\label{eq:7ca}
\end{equation}
That $\mu_{pair}$ vanishes at \textit{all} $T \leq T_c$ is a stronger
condition than the usual Thouless condition for $T_c$.  Moreover, it
should be stressed that \textit{BCS theory is associated with a
  particular asymmetric form for the pair susceptibility}.  These
uncondensed pairs play virtually no role in BCS superconductors but the
structure of this theory points to a particular choice for a pair
susceptibility.
We can then write the self consistent condition on $\Delta_{sc}$ which
follows from Eq.~(\ref{eq:5c}). 
Using Eqs.~(\ref{eq:2c}-
\ref{eq:4c}), we find the expected form for the BCS gap equation.
This represents a constraint on the order parameter $\Delta_{sc}$:
\begin{equation}
\Delta_{sc}(T) =-U \sum_{\bf k} \Delta_{sc}(T) \frac{1-2 f(\Ek)}{2 \Ek}\;.  
\label{eq:8c}
\end{equation}

The above discussion was presented in a somewhat different way in a
paper by Kadanoff and Martin \cite{Kadanoff}.

\subsubsection{Three Choices for the $T$-matrix of the Normal State:
Problems with the Nozi\'eres Schmitt-Rink Approach}

On general grounds we can say that there are three obvious choices for
$\chi(Q)$ which appears in the general definition of the $T$-matrix in
Eq.~(\ref{eq:1c}). Each of these has been studied rather extensively in
the literature.  All of these introduce corrections to BCS theory and
all were motivated by attempts to extend the crossover ground state to
finite $T$, or to understand widespread pseudogap effects in the high
$T_c$ superconductors.  In analogy with Gaussian fluctuations, one can
consider
\begin{equation}
\chi_0(Q) = \sum_K G_0(K)G_0(Q-K)
\label{eq:10c}
\end{equation}
with self energy
\begin{equation}
\Sigma_0(K) = \sum_Q t(Q) G_0(Q-K)
\label{eq:11d}
\end{equation}
which appears in $G$ in the analogue of Eq.~(\ref{eq:12c}). The number
equation is then deduced by using Eq.~(\ref{eq:3c}).

This scheme was adopted by Nozi\'eres and Schmitt-Rink (NSR), although
these authors \cite{NSR,randeriareview} approximated the number equation
\cite{Serene} by using a leading order series for $G$ in
Eq.~(\ref{eq:12c}) with
\begin{equation}
 G = G_0 + ~ G_0 \Sigma_0 G_0\;.
\label{eq:13c} 
\end{equation}
It is straightforward, however, to avoid this approximation in Dyson's
equation, and a number of groups \cite{Janko,Strinati2} have extended NSR
in this way.

Similarly one can consider 
\begin{equation}
\bar{\chi}(Q) = \sum_K G(K)G(Q-K)
\end{equation}
with self energy
\begin{equation}
\bar{\Sigma}(K) = \sum_Q t(Q) G(Q-K) \;.
\end{equation}
This latter scheme (sometimes known as fluctuation-exchange or FLEX)
has been also extensively discussed in the literature, by
among others, Haussmann \cite{Haussmann}, Tchernyshyov \cite{Tchern},
Micnas and colleagues \cite{Micnas95}
and Yamada and Yanatse \cite{YY}.

Finally, we can contemplate the asymmetric form \cite{Chen2,Berthod} for the
$T$-matrix, so that the coupled equations for $t(Q)$ and $G(K)$ are based
on
\begin{equation}
\chi(Q) = \sum_K G(K)G_0(Q-K)
\end{equation}
with self energy
\begin{equation}
\Sigma(K) = \sum_Q t(Q) G_0(Q-K)
\end{equation}

Each of these three schemes determines the superfluid transition
temperature via the Thouless condition. Thus $\mu_{pair} =0$ leads to a
slightly different expression for $T_c$, based on the differences in the
choice of $T$-matrix which appears in Eq.~(\ref{eq:7c}).

A central problem with the NSR scheme ($ \chi \approx G_0G_0$) is that
it incorporates self energy effects only through the number equation.
The absence of self energy effects in the gap equation is equivalent to
the statement that pseudogap effects only indirectly affect $T_c$: the
particles acquire a self energy from the pairs but these self energy
effects are not fed back into the propagator for the pairs.  On physical
grounds one anticipates that the pseudogap must have a direct effect on
$T_c$ as a consequence of the associated reduction in the density of
states. Other problems related to the thermodynamics were pointed out
\cite{Balseiro} when NSR was applied to a two dimensional system.
Finally, because only the number equation contains self energy effects,
it is not clear if one can arrive at a proper vanishing of the
superfluid density $\rho_s$ at $T_c$ within this approach.  This
requires a precise cancellation between the diamagnetic current
contributions (which depend on the number equation) and the paramagnetic
terms (which reflect the gap equation).

One might be inclined, then, to prefer \cite{Serene} the FLEX ($\chi
\approx GG$) scheme since it is $\phi$-derivable, in the sense of Baym
\cite{Baym}.  This means that it is possible to write down a closed form
expression for the thermodynamical potential, from which one derives
other physical quantities via functional derivatives of this potential.
Theoretical consistency issues in this approach have been rather
exhaustively discussed by Haussmann \cite{Haussmann} above $T_c$.  We
are not aware of a fully self consistent calculation of $\rho_s$ below
$T_c$, at the same level of completeness as Haussmann's normal state
analysis (or, for that matter, of the counterpart discussion to Section
\ref{sec:3A} and accompanying Appendix \ref{App:Ward}). However, in
principle, it should be possible to arrive at a proper expression for
the superfluid density, providing Ward identities are imposed. There is
some ambiguity \cite{Tremblay,Tchern,Micnas95} about whether pseudogap
effects are present in the FLEX approach; a consensus has not been
reached at this time.  Numerical work based on FLEX is more extensive
\cite{Hess,Tremblay} than for the other two alternative schemes.

\textit{It will be made clear in what follows that, if one's goal is to
  extend the usual crossover ground state of Eq.~(\ref{eq:1a}) to finite
  temperatures, then one must choose the asymmetric form for the pair
  susceptibility}.  Here $\chi \approx GG_0$.  This is different from
the approach of Nozi\'eres and Schmitt-Rink in which there is no evident
consistency between their finite $T$ treatment and the presumed ground
state.  Some have claimed \cite{randeriareview} that an NSR-based
approach, when extended below $T_c$, reproduces Eqs.~(\ref{eq:18}) and
(\ref{eq:19}). But elsewhere in the literature it has been shown that
these ground state self consistency conditions are derived from a lower
level theory \cite{Ranninger2}.  They arise from a $T=0$ limit of what
is called the saddle point approximation, or mean field approach. When
pairing fluctuations beyond the saddle point \cite{NSR} are included,
the number equation is changed.  Then the ground state is no longer that
of Eq.~(\ref{eq:1a}) \cite{Strinati2}.  In the cold atom literature, the
distinction between the NSR-extrapolated ground state and the ground
state of Eq.~(\ref{eq:1a}) has been pointed out in the context of atomic
density profiles \cite{Strinati5} within traps.

It should be evident from Eq.~(\ref{eq:4c}) and the surrounding
discussion that the asymmetric form is uniquely associated with the
BCS-like behavior implicit in Eq.~(\ref{eq:1a}).  The other two $T$-matrix
approaches lead to different ground states which should, however, be
very interesting in their own right. These will need to be characterized
in future.

Additional support for the asymmetric form, which will be advocated
here, was provided by Kadanoff and Martin. In their famous paper they
noted that several people had surmised that the form for $\chi(Q)$
involving $GG$ would be more accurate. However, as claimed in
\cite{Kadanoff}, ``This surmise is not correct.'' Aside from theoretical
counter-arguments which they present, the more symmetric combination of
Green's functions ``can also be rejected experimentally since they give
rise to a $T^2$ specific heat."  Similar arguments in support of the
asymmetric form were introduced \cite{Wilkins} into the literature in
the context of addressing specific heat jumps.

\subsection{Superconducting Fluctuations: a type of Pre-formed Pairs}
\label{sec:1G}

While there are no indications of bosonic degrees of freedom, (other
than in the condensate), within strict BCS theory, it has been possible
to access these bosons via probes of superconducting fluctuations.
Quasi-one dimensional, or quasi-2D superconductors in the presence of
significant disorder exhibit fluctuation effects \cite{Larkinreview} or
precursor pairing as seen in ``paraconductivity", fluctuation
diamagnetism, as well as other unusual behavior, often consisting of
divergent contributions to transport.  One frequently computes
\cite{Dorsey} these bosonic contributions to transport by use of a time
dependent Ginzburg-Landau (TDGL) equation of motion.  This is rather
similar to a Gross-Pitaevskii (GP) formalism except that the ``bosons"
here are highly damped by the fermions.

Alternatively $T$-matrix based approaches (involving all three choices of
$\chi(Q)$) have been extensively used to discuss conventional
superconducting fluctuations.  The advantage of these latter schemes is
that one can address both the anomalous bosonic and fermionic
contributions to transport through the famous Aslamazov-Larkin and
Maki-Thompson diagrams \cite{Larkinreview}.  We defer a discussion of
these issues until Appendix \ref{App:Ward}.

It is useful to demonstrate first how conventional superconducting
fluctuations behave at the lowest level of self consistency, called the
Hartree approximation. This scheme is closely associated\cite{JS} with a
$GG_0$ $T$-matrix, as is discussed in more detail in Appendix \ref{App:2}.
It is also closely associated with BCS theory, for one can show that, in
the spirit of Eq.~(\ref{eq:1a}), at this Hartree level the excitation
gap $\Delta(T_c)$ and $T_c$ lie on a specific BCS curve (specified by
$n$ and $U$).  What is different from strict BCS theory is that the
onset of superconductivity takes place in the presence of a finite
excitation gap (ie, pseudogap), just as shown in Fig.
\ref{fig:Delta_Deltasc}.  This, then, reflects the fact that there are
pre-formed pairs at $T_c$.  By contrast with high $T_c$ superconductors,
however, in conventional fluctuation effects, the temperature $T^*$ at
which pairs start to form is always extremely close to their
condensation temperature $T_c$. We thus say that there is a very narrow
critical region.

In Hartree approximated Ginzburg Landau theory \cite{Wilkins} the free
energy functional is given by
\begin{equation}
F[\Psi]=a_0(T-T^*)|\Psi|^2+b|\Psi| ^4 \approx a_0 (T-T^*)|\Psi|^2+2 b
\Delta ^2 |\Psi|^2 
\end{equation}
Here $\Delta^2$ plays the role of a pseudogap in the normal state. It is
responsible for a lowering of $T_c$ relative to the mean field value
$T^*$.  Collecting the quadratic terms in the above equation, it follows
that
\begin{equation}
T_c = T^* - \frac{2b}{a_0} \Delta^2(T_c)\;. 
\label{eq:36}
\end{equation}
Moreover, self consistency imposes a constraint on the magnitude of
$\Delta(T_c)$ via
\begin{equation}
\Delta ^2 (T) = \int D\Psi e^{-\beta F[\Psi]}\Psi ^2\Big/\int D\Psi e^{-\beta F[\Psi]}\;.
\label{eq:37}
\end{equation}

It should be noted that Eq.~(\ref{eq:36}) is consistent with the
statement that $\Delta(T_c)$ and $T_c$ lie on the BCS curve, since for
small separation between $T_c$ and $T^*$, this curve is, below $T_c$,
defined by
\begin{equation}
\Delta_{BCS}^2 (T) \approx \frac {a_0}{2b} (T^* - T)\;.
\label{eq:38}
\end{equation}
The primary effect of fluctuations at the Hartree level is that pairing
takes place in the presence of a finite excitation gap, reflecting the
presence of very short lived, but pre-formed pairs.  It should be clear
that $\Delta$ is not to be confused with the superconducting order
parameter $\Delta_{sc}$, as should be evident from Fig.
\ref{fig:Delta_Deltasc}.  We refer the reader to a more detailed
discussion which is presented in Appendix \ref{App:2}.

Figure \ref{fig:14c} illustrates how this pseudogap appears
experimentally, via a plot of the normalized density of states
\cite{Abeles} $\nu_1 \equiv N(E)$ in the normal state. This figure
derives from tunneling measurements on Al-based films. Note the
depression of $N(E_F)$ at low energies.  This is among the first
indications for a pseudogap reported in the literature.

\begin{figure}
\includegraphics[width=2.4in,clip]{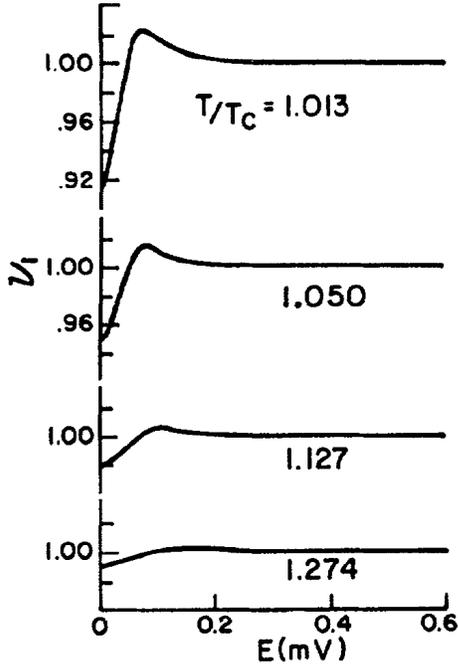}
\caption{Pseudogap in the density of states above $T_c$
  in conventional superconductors, from \onlinecite{Abeles}.
  $\nu_1\equiv N(E)$ is the normalized density of states.}
\label{fig:14c}
\end{figure}

Alternatively, fluctuations have also been discussed at the Gaussian
($G_0G_0$) level in which there is no need for self consistency, in
contrast to the above picture.  Here because the calculations are so
much more tractable there have been very detailed applications
\cite{Larkinreview} of essentially all transport and thermodynamic
measurements. This Gaussian analysis led to unexpected divergences in
the para-conductivity from the so-called Maki-Thompson term which then
provided a motivation to go beyond leading order theory.  Patton
\cite{Patton1971,Patton} showed how this divergence could be removed
within a more self consistent $GG_0$ scheme, equivalent to self
consistent Hartree theory. Others argued \cite{Schmid,Tucker} that
Hartree-Fock ($GG$) was more appropriate, although in this weak
coupling, narrow fluctuation regime, the differences between these
latter two schemes are only associated with factors of 2.  Nevertheless
with this factor of 2, $\Delta(T_c)$ and $T_c$ are no longer on the BCS
curve. It should be noted that a consensus was never fully reached
\cite{Wilkins} by the community on this point.

\subsubsection{Dynamics of Pre-formed Pairs Above $T_c$: Time Dependent
Ginzburg-Landau Theory}
\label{sec:TDGL}

We turn again to Fig.~\ref{fig:3} which provides a useful starting point
for addressing bosonic degrees of freedom and their implications for
experiment.  It is clear that transport in the unusual normal state,
associated with the pseudogap (PG) phase (as well as thermodynamics)
contains contributions from both fermionic and bosonic excitations. The
bosons are not infinitely long lived; their lifetime is governed by
their interactions with the fermions.
In some instances \cite{Larkinreview}, the bosonic contributions to
transport become dominant or even singular at $T_c$.  Under these
circumstances one can ignore the fermionic contributions except insofar
as they lead to a lifetime for the pairs.

To address the transport properties of these more well established
pre-formed pairs, one appeals to a standard way of characterizing the
dynamics of bosons -- a phenomenological scheme known as time dependent
Ginsburg- Landau (TDGL) theory.  This theory has some microscopic
foundations in diagrammatic $T$-matrix approaches in which one considers
only the Aslamazov-Larkin terms (which are introduced in Appendix
\ref{App:Ward}). At the Gaussian level both TDGL and the $T$-matrix
approaches are tractable.  At the Hartree level things rapidly become
more complicated, and it is far easier to approach the problem
\cite{Tan} by adopting a strictly phenomenological TDGL.

The generic Hartree-TDGL equation of motion for classical charged bosons
interacting with electromagnetic fields ($\phi , \bf {A}$) is given by
\begin{multline}
\label{eq:HartreeTDGL}
\gamma\InvariantEnergy\psi(\xt)=\frac{\InvariantMomentum^2}{2M^*}\psi(\xt)\\
-\mu_{pair}(T)\psi(\xt)+D(\xt),
\end{multline}
Here $\psi(\xt)$ is a classical field variable representing the bosons
which have vanishing chemical potential $\mu_{pair}$ at $T_c$. The
function $D(\xt)$ introduces a noise variable into the equation of
motion.
Generally $\gamma $ is complex.

We will explore the implications of this equation in our more
quantitative discussion of Section \ref{sec:6c}.

\section{Quantitative
Details of Crossover }
\label{sec:2}
\subsection{$T=0$, BEC Limit without Feshbach Bosons}
\label{sec:2A}

We begin by reviewing $T=0$ crossover theory in the BEC limit.  Our
starting point is the ground state wavefunction $\Psi_0$ of
Eq.~(\ref{eq:1a}), along with the self consistency conditions of
Eq.~(\ref{eq:2}).  For positive chemical potential $\mu$, the quantity
$\Delta_{sc}(0)$ also corresponds to the energy gap for fermionic
excitations. In the ground state the two energy scales $\Delta(0)$ and
$\Delta_{sc}(0)$ are degenerate, just as they are (at all temperatures)
in strict BCS theory.

It is convenient to rewrite these equations \footnote{To make the
  equations simpler, in much of what we present in this review we will
  consider a contact potential interaction. The momentum sums are
  assumed to run to infinity.  More generally, one can include a cutoff
  function $\phik$ for the momentum integrals.  The gap function thus
  becomes $\Delta_{\bf k} = \Delta\phik$. For the cuprate $d$-wave
  cuprate superconductors, $\phik = \cos k_x a - \cos k_y a$.  For our
  numerical work on atomic gases we use a Gaussian cutoff $\phik^2 =
  e^{-k^2/k_0^2}$ with $k_0$ taken to be as large as is numerically
  feasible.}
in terms of the inter-fermion scattering length $a_s$

\begin{eqnarray}
\frac{m}{4 \pi a_s }& =&  \mathop{\sum_{\bf k}}\left[ \frac{1}{2
  \ek } - \frac{1}{2E_{\bf k}} \right]\,, \label{eq:3d}\\ 
n & =& \sum _{\bf k} \left[ 1 -\frac{\ek - \mu}{\Ek} 
 \right] \,, ~~ T=0\,.
\label{eq:4d}
\end{eqnarray}
In the fermionic regime ($\mu > 0$) these equations are essentially
equivalent to those of BCS theory, although at weak coupling appropriate
to strict BCS, little attention is paid to the number equation since
$\mu = E_F$ is always satisfied.  The more interesting regime
corresponds to $ \mu \leq 0$ where these equations take on a new
interpretation. Deep inside the BEC regime it can be seen that
\begin{equation}
n =  \Delta^2_{sc}(0) \frac{  m^2} {4 \pi \sqrt{2m |\mu|}},
\label{eq:5}
\end{equation}
which, in conjunction with Eq.~(\ref{eq:3d}) (expanded in powers of
$\Delta^2_{sc}(0)/\mu^2$) :
\begin{equation}
\frac{m}{4 \pi a_s}=(2 m)^{3/2}
\frac{\sqrt{|\mu|}}{8 \pi} 
\left[1+\frac{1}{16} \frac{\Delta_{sc}^2(0)}{\mu^2}\right],
\label{eq:6}
\end{equation}
yields
\begin{equation}
\mu = - \frac{1}{2 m a_s^2} + \frac{a_s \pi n} {m}.
\label{eq:7}
\end{equation}

This last equation is equivalent \cite{Stringari,Strinati} to its
counterpart in Gross-Pitaevskii theory. This theory describes true
bosons, and is associated with the well known equation of state
\begin{equation}
n_{pairs} = \frac{m_B}{4 \pi a_B } \mu_B\,.
\label{eq:8}
\end{equation}
To see the equivalence we associate the number of bosons $n_{pairs} =
n/2$, the boson mass $m_B = 2 m$ and the bosonic scattering length $a_B
= 2 a_s$.  Here the bosonic chemical potential $\mu_B = 2\mu +
\epsilon_0$ and we define $\epsilon_0 = 1/(2 m a_s^2)$.  This factor of
2 in $a_B/a_s$ has received a fair amount of attention in the
literature.  Although it will be discussed later in the Review, it
should be noted now that this particular numerical value is particular
to the one channel description\cite{JS3}, and different from what is
observed experimentally \cite{Petrov}.

Despite these
similarities with GP theory, the fundamental parameters are the
fermionic $\Delta_{sc}(0)$ and chemical potential $\mu$.  It can be
shown that in this deep BEC regime the number of pairs is directly
proportional to the superconducting order parameter
\begin{equation}
n_{pairs} = \frac{n}{2} = Z_0 \Delta_{sc}^2(0)
\label{eq:9}
\end{equation}
where
\begin{equation}
Z_0 \approx \frac {m^2 a_s } { 8 \pi} \,.
\label{eq:10}
\end{equation}
One may note from Eq.~(\ref{eq:9}) that the ``gap equation" now
corresponds to a number equation (for bosons). Similarly the number
equation, or constraint on the fermionic chemical potential defines the
excitation gap for fermions, once the chemical potential is negative.
In this way the roles of the two constraints are inverted
\cite{randeriareview} relative to the BCS regime.

That a Gross-Pitaevskii approach captures the leading order physics at
$T=0$ can also be simply seen by rewriting the ground state
wavefunction, as pointed out by Randeria \cite{randeriareview}.  Define
$ v_{\bf k} / u_{\bf k} \equiv \eta_{\bf k}$
\begin{eqnarray}
\Psi_0 &=& const \times \Pi_k ( 1 + \eta_{\bf k } 
 c_k^{\dagger} c_{-k}^{\dagger})|0\rangle \nonumber\\
&=& const \times \exp \left(\sum_k \eta_{\bf k} 
c_k^{\dagger} c_{-k}^{\dagger}\right)|0\rangle \;.
\end{eqnarray}
Projecting onto a state with fixed particle number $N$ yields
\begin{equation}
\Psi_0= const \times \left(\sum_{\bf k} \eta_{\bf k} 
c_k^{\dagger} c_{-k}^{\dagger}\right)^{N/2}|0\rangle \;.
\label{eq:54a}
\end{equation}
This is a GP wavefunction of composite bosons with the important
proviso: that the characteristic size associated with the internal
wavefunction $\eta_{\bf k}$ is smaller than the inter-particle spacing.
It should be stressed that, except in the BEC asymptotic limit, (where
the ``bosons" are non-interacting) there are essential differences
between Eq.~(\ref{eq:54a}) and the Gross-Pitaevskii wavefunction.  Away
from this extreme limit, pair-pair interactions (arising indirectly
via the fermions and the Pauli principle), as well as the finite size of the
pairs, will differentiate this system from that of true interacting
bosons.  This same issue arises more concretely in the context of the
collective mode spectrum of ultracold gases, as discussed in Section
\ref{sec:updates}.

\subsection{Extending conventional
  Crossover Ground State to $T \neq 0$: BEC limit without Feshbach
  Bosons}
\label{sec:2B}

How do we extend \cite{Chen2} this picture to finite $T$?  In the BEC
limit, as shown in Fig.~\ref{fig:2}, fermion pairs are well established
or ``pre-formed" within the entire range of superconducting
temperatures.  The fundamental constraint associated with the BEC regime
is that: \textit{for all $T \leq T_c$, there should, thus, be no
  temperature dependence in fermionic energy scales. In this way
  Eqs.~(\ref{eq:3d}) and (\ref{eq:4d}) must be imposed at all
  temperatures $T$}.

\begin{eqnarray}
\frac{m}{4 \pi a_s } & = & \mathop{\sum_{\bf k}}\left[ \frac{1}{2
  \ek } - \frac{1}{2E_{\bf k}} \right]  \,, \label{eq:3a}\\  
n & =& \sum _{\bf k} \left[ 1 -\frac{\ek - \mu}{\Ek}
 \right] \,, ~~~~T \leq T_c\,.
\label{eq:4a}
\end{eqnarray}

It follows that the number of pairs at $T=0$ should be equal to the
number of pairs at $T=T_c$.  However, all pairs are condensed at $T=0$.
Clearly, the character of these pairs changes so that at $T_c$, all
pairs are noncondensed.  To implement these physical constraints (and
to anticipate the results of a more microscopic theory) we write
\begin{eqnarray}
n_{pairs} &=&\frac{n}{2}= Z_0 \Delta^2 
\label{eq:11} \\
n_{pairs} &=& n_{pairs}^{condensed}(T) + n_{pairs} ^ {noncondensed}(T) 
\label{eq:12}
\end{eqnarray}
so that we may decompose the excitation gap into two contributions
\begin{equation}
\Delta^2  = \Delta_{sc}^2 (T) + \Delta_{pg}^2 (T) 
\label{eq:13}
\end{equation}
where $\Delta_{sc}(T)$ corresponds to condensed and $\Delta_{pg}(T)$ to
the noncondensed gap component. Each of these are proportional to the
respective number of condensed and noncondensed pairs with
proportionality constant $Z_0$.  At $T_c$,
\begin{equation}
n_{pairs}^{noncondensed} = \frac{n}{2}= \sum_{\bf q} b (\Omega_q,T_c) \,,
\label{eq:52}
\end{equation}
where $b(x)$ is the usual Bose-Einstein function and $\Omega_q$
is the unknown dispersion of the noncondensed pairs.
Thus
\begin{equation}
\Delta_{pg}^2 (T_c) = Z_0^{-1} \sum b(\Omega_q, T_c) = \frac{n}{2}
Z_0^{-1} \,.
\label{eq:14}
\end{equation}
We may deduce directly from Eq.~(\ref{eq:14}) that $\Delta_{pg}^2 =
-\sum_Q t(Q)$, if we presume that below $T_c$, the noncondensed pairs
have propagator
\begin{equation}
t(Q) = \frac{Z_0^{-1}}{\Omega- \Omega_q} \,.
\label{eq:15}
\end{equation}

It is important to stress that the dispersion of the pairs $\Omega_q$
cannot be put in by hand.  It is not known a priori.  Rather, it has be
to \textit{derived} according to the constraints imposed by
Eqs.~(\ref{eq:3a}) and (\ref{eq:4a}).  We can only arrive at an
evaluation of $\Omega_q$ after establishing the nature of the
appropriate $T$-matrix theory. In what follows we derive this dispersion
using the two channel model.  We state the one channel results here for
completeness. In the absence of FB, $\Omega_q \equiv B_0 q^2$, where
$B_0 \equiv 1/(2 M_0^*)$.  As a consequence of a ``bosonic" self energy,
which arises from fermion-boson (as distinguished from boson-boson)
interactions, the bosonic dispersion is found to be quadratic.  This
quadratic dependence persists for the two channel case as well.

\subsection{Extending Conventional Crossover Ground State to
$T \neq 0$: $T$-matrix Scheme in the Presence of Feshbach Bosons }
\label{sec:2C}

To arrive at the pair dispersion for the noncondensed pairs,
$\Omega_q$, we need to formulate a generalized $T$-matrix based scheme
which is consistent with the ground state conditions, and with the $T$
dependence of strict BCS theory.  It is useful from a pedagogical point
to now include the effects of Feshbach bosons \cite{JS2}. Our intuition
concerning how true bosons condense is much better than our intuition
concerning the condensation of fermion pairs, except in the very limited
BCS regime.  We may assume that $\Delta$ and $\mu$ evolve with
temperature in such a way as to be compatible \textit{with both the
  temperature dependences of BCS and with the above discussion for the
  BEC limit}. We thus take
\begin{eqnarray}
\Delta(T)& =&-U_{eff} \sum_{\bf k} \Delta(T)
 \frac{1-2 f(\Ek)}{2 \Ek}\,, \label{eq:18}\\ 
n & =& \sum _{\bf k} \left[ 1 -\frac{\ek - \mu}{\Ek} 
+2\frac{\ek - \mu}{\Ek}f(\Ek)  \right] \,,
\label{eq:19}
\end{eqnarray}
where $\Ek = \sqrt{ (\ek -\mu)^2 + \Delta^2 (T) }$.
Note that in the presence of Feshbach bosons, we use $n$ to denote the
density of fermionic atoms which is to be distinguished from the
constituents of the Feshbach bosons.  The total density of particles
will be denoted by $n^{tot}$.  \textit{Equations (\ref{eq:18}) and
  (\ref{eq:19}) will play a central role in this review. They have been
  frequently invoked in the literature, albeit under the presumption
  that there is no distinction between $\Delta(T)$ and
  $\Delta_{sc}(T)$}.

Alternatively one can rewrite Eq.~(\ref{eq:18}) as
\begin{equation}
\frac{m}{4 \pi a^*_s } =  \mathop{\sum_{\bf k}}\left[ \frac{1}{2
  \ek } - \frac{1- 2 f(\Ek)}{2E_{\bf k}} \right] \,.
\label{eq:18a}
\end{equation}

Clearly Eqs.~(\ref{eq:18}) and (\ref{eq:19}) are consistent with
Eqs.~(\ref{eq:3a}) and (\ref{eq:4a}) since Fermi functions are
effectively zero in the BEC limit.  Our task is to find a $T$-matrix
formalism which is compatible with Eqs.~(\ref{eq:18}) and (\ref{eq:19}),
and to do this we focus first on the \textit{noncondensed} molecular
bosons.  Their propagator may be written as
\begin{equation}
D(Q) \equiv  \frac{1}{i\Omega_n - \epsilon_q^{mb} - \nu + 2 \mu  -
  \Sigma_B(Q)} \,. 
\end{equation}
We presume that the self energy of these molecules can be written
in the form
\begin{equation}
 \Sigma_B (Q) \equiv -g^2 \chi(Q) / [ 1 + U \chi(Q) ] \,,
\label{eq:sigmaB}
\end{equation}
where $\chi(Q)$ is as yet unspecified. This RPA-like self energy arises
from interactions between the molecular bosons and the fermion pairs,
and this particular form is required for self-consistency.

\textit{noncondensed bosons in equilibrium with a condensate must
necessarily have zero chemical potential.}
\begin{equation}
\mu_{boson}(T) =0,\qquad T \leq T_c.
\label{eta_gg0}
\end{equation}
This is equivalent to the Hugenholtz-Pines condition that
\begin{equation}
D^{-1}(0) =  0, ~~ T \leq T_c.
\label{eta_gg01}
\end{equation}
Using Eqs.~(\ref{eq:sigmaB}) and (\ref{eta_gg01}) it can be seen that
\begin{equation}
 U_{eff}^{-1}(0)+\chi(0) = 0, \qquad T \le T_c  \;.
\label{eq:3}
\end{equation}
This equation can be made compatible with our fundamental constraint in
Eq.~(\ref{eq:18}) provided we take
\begin{equation}
\chi(Q) =  \mathop{\sum_K} G(K)G_0(Q-K) \,,
\label{eq:chi}
\end{equation}
where $G(K)$ 
\begin{equation}
G(K) \equiv [ G_0^{-1}(K) - \Sigma(K)]^{-1}
\label{eq:62}
\end{equation}
includes a self energy given by the BCS-form 
\begin{equation}
\Sigma(K) = - G_0(-K) \Delta^2 \,,
\label{eq:sigma}
\end{equation}
which now involves the quantity $\Delta$ to be distinguished from the
order parameter $\Delta_{sc}$.  Note that with this form for
$\Sigma(K)$, the number of fermions is indeed given by Eq.~(\ref{eq:19}).

More generally, we may write the constraint on the total number of
particles as follows.  The number of noncondensed molecular bosons is
given directly by 
\begin{equation}
n_b(T) = -\sum_Q D(Q)  \,.
\label{eq:62a}
\end{equation}
For $ T \le T_c$, the number of
fermions is given via Eq.~(\ref{eq:19}).
The \textit{total} number ($n^{tot}$) of particles is then
\begin{equation}
n + 2 n_b + 2 n^0_b = n^{tot} \;,
\label{number_equation}
\end{equation}
where $n^0_b=\phi _m^2$ is the number of molecular bosons in the
condensate, and $\phi_m$ is the order parameter for the molecular
condensate.

Thus far, we have shown that the condition that noncondensed molecular
bosons have zero chemical potential, can be made consistent with
Eqs.~(\ref{eq:18}) and (\ref{eq:19}) provided we constrain $\chi(Q)$ and
$\Sigma(K)$ as above.  We now want to examine the counterpart condition
on the fermions and their condensate contribution.  The analysis leading
up to this point should make it clear \textit{that we have both
  condensed and noncondensed fermion pairs, just as we have both
  condensed and noncondensed molecular bosons}. Moreover, these fermion
pairs and Feshbach bosons are strongly admixed, except in the very
extreme BCS and BEC limits.  Because of noncondensed pairs, we will see
that the excitation gap is distinct from the superconducting order
parameter, although in the literature this distinction has not been
widely recognized.

\begin{figure}
\includegraphics[width=2.5in]{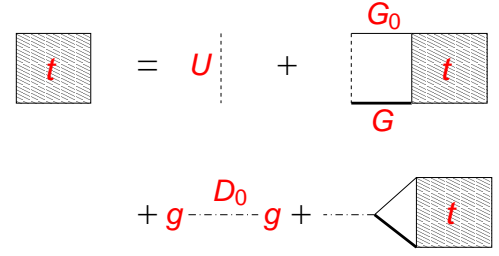}
\caption{Diagrammatic scheme for present $T$-matrix theory. Here $U$ is
  the open-channel interaction, $g$ is the coupling between open and
  closed channels. $G_0$, $G$ and $D_0$ are bare fermion, full fermion,
  and bare Feshbach boson Green's functions, respectively. $U$ and
  $g_0^2 D_0$ yield an effective pairing interaction, $U_{eff}$.}
\label{T-matrix-scheme}
\end{figure}

Just as the noncondensed molecular bosons have zero chemical potential
below $T_c$ we have the same constraint on the noncondensed fermion
pairs which are in chemical equilibrium with the noncondensed bosons
\begin{equation}
\mu_{pair} = 0 , \qquad T \le T_c \;. 
\label{eq:21}
\end{equation}
The $T$-matrix scheme in the presence of the Feshbach bosons is shown in
Fig.~\ref{T-matrix-scheme}.
Quite generally, the $T$-matrix consists of two contributions: from the
condensed ($sc$) and noncondensed or ``pseudogap"-associated ($pg$)
pairs,
\begin{eqnarray}
t &=& t_{pg} + t_{sc} \,, \label{t-matrix}\\
t_{pg}(Q)&=& \frac{U_{eff}(Q)}{1+U_{eff}(Q) \chi(Q)}, \qquad Q \neq 0 \,,
\label{t-matrix_pg}\\
t_{sc}(Q)&=& -\frac{\tilde{\Delta}_{sc}^2}{T} \delta(Q) \,,
\label{t-matrix_sc}
\end{eqnarray}
where we write
$\tilde{\Delta}_{sc}=\Delta _{sc}-g \phi _m$, with $\Delta
_{sc}=-U \sum _{\bf k}\langle a_{-{\bf k}\downarrow}a_{{\bf
    k}\uparrow}\rangle$ and $\phi _m=\langle b_{{\bf
    q}=0} \rangle$.
Here, the order parameter is a linear combination of
both paired fermions and condensed molecules.    
Similarly, we have two contributions for the fermion self energy
\begin{equation}
\Sigma(K) = \Sigma_{sc}(K) + \Sigma_{pg}(K) = \sum_Q t(Q) G_0 (Q-K) \,,
\label{eq:sigma2}
\end{equation}
where, as in BCS theory, 
\begin{equation}
\Sigma_{sc}(K) = \sum_Q t_{sc}(Q) G_0(Q-K) \,.
\label{eq:70}
\end{equation}

Without loss of generality, we choose order parameters
$\tilde{\Delta}_{sc}$ and $\phi_m$ to be real and positive with $g<0$.
Importantly, these two components are connected \cite{Kokkelmans} by the
relation $\phi _m=g \Delta _{sc}/[(\nu -2\mu) U]$.

The vanishing of the pair chemical potential implies
that
\begin{equation}
t_{pg}^{-1} (0) = U_{eff}^{-1}(0) + \chi(0) = 0, \qquad T \le T_c \;.
\label{eq:3e}
\end{equation}
This same equation was derived from consideration of the bosonic
chemical potential. Most importantly, we argued above that this was
consistent with Eq.~(\ref{eq:18}) provided the fermion self energy
assumes the BCS form. We now verify the assumption in
Eq.~(\ref{eq:sigma}).  A vanishing chemical potential means that
$t_{pg}(Q)$ is strongly peaked around $Q=0$. Thus, we may approximate
\cite{Maly1} Eq.~(\ref{eq:sigma2}) to yield
\begin{equation}
\Sigma (K)\approx -G_0 (-K) \Delta^2 \,,
\label{eq:sigma3}
\end{equation}
where 
\begin{equation}
\Delta^2 (T) \equiv \tilde{\Delta}_{sc}^2(T)  + \Delta_{pg}^2(T) \,,
\label{eq:sum}
\end{equation}
and we define the pseudogap $\Delta_{pg}$
\begin{equation}
\Delta_{pg}^2 \equiv -\sum_{Q\ne 0} t_{pg}(Q). 
\label{eq:delta_pg}
\end{equation}
This bears a close resemblance to Eq.~(\ref{eq:62a}), if we write the
number of noncondensed fermion pairs and molecular bosons as
\begin{equation}
n_p(T) = Z \Delta_{pg}^2 (T) \,.
\label{eq:62b}
\end{equation}

Note that in the normal state (where $\mu_{pair}$ is nonzero) one cannot
make the approximation of Eq.~(\ref{eq:sigma3}).  Referring back to our
discussion of Hartree-approximated TDGL, a strong analogy between
Eq.~(\ref{eq:delta_pg}) and (\ref{eq:37}) should be observed. There is a
similar analogy between Eq.~(\ref{eq:18}) and (\ref{eq:38}); more
details are provided in Appendix \ref{App:2}.  We thus have that
Eqs.~(\ref{eq:3d}) and (\ref{eq:sigma3}) with Eq.~(\ref{eq:62}) are
alternative ways of writing Eqs.~(\ref{eq:18}) and (\ref{eq:19}).  Along
with Eq.~(\ref{eq:delta_pg}) we now have a closed set of equations for
addressing the ordered phase.  Moreover the propagator for noncondensed
pairs can now be quantified, using the self consistently determined pair
susceptibility.  At moderately strong coupling and at small four-vector
$Q$, we may expand to obtain
\begin{equation}
t_{pg}(Q) = \frac { Z^{-1}}{\Omega - \Omega_q +\mu_{pair} + i \Gamma_Q},
\label{eq:expandt}
\end{equation}
Consequently, one can rewrite Eq.~(\ref{eq:delta_pg})
as 
\begin{equation}
\Delta_{pg}^2 (T) = Z^{-1}\sum b(\Omega_q, T) \,.
\label{eq:81}
\end{equation}

Here we introduce a simple notation for the small $Q$ expansions: 
\begin{eqnarray}
\chi(Q)-\chi(0) &=& Z_0 (i\Omega_n  -B_0 q^2), \nonumber \\
U^{-1}_{eff}(Q) - U^{-1}_{eff}(0) &=& Z_g (i\Omega_n  -B_g q^2) \nonumber
\end{eqnarray}
with
\begin{eqnarray}
Z_0 &=& \frac{1}{2\Delta^2} \left[n -2 \sumk f(\ek-\mu)\right] , \nonumber \\
Z_g &=& \frac{g^2}{[(2\mu-\nu)U + g^2]^2}\;. \nonumber
\end{eqnarray}
We have that
$Z=Z_0+Z_g$, $Z_b = (Z_0 + Z_g)/ Z_0$, 
and the $q^2$ coefficient $B \equiv 1/(2 M^*)$ in $\Omega_q \equiv
B q^2$
is such that
\begin{equation}
B=\frac{B_0Z_0+\frac{1}{2 M}
     Z_g}{Z} \;.\nonumber
\end{equation}

The strong hybridization between the molecular bosons and fermion
pairs should be stressed. Indeed, there is just one branch
$\Omega_q$ for bosonic-like excitations, so that at small $Q$ 
\begin{equation}
D(Q) = \frac { Z_b^{-1}}{\Omega - \Omega_q +\mu_{boson} + i \Gamma_Q},
\label{eq:expandD}
\end{equation}
where below $T_c$, $\mu_{pair} = \mu_{boson} =0$. Here $\Gamma_Q$
contains all the imaginary part of the inverse $T$-matrix and it
vanishes rapidly as $Q \rightarrow 0$. As the coupling is decreased from
the BEC limit towards BCS, the character of the pairs changes from
pre-dominantly molecular-boson like to (noncondensed) Cooper pairs,
composed only of the fermionic atoms.

\subsection{Nature of the Pair Dispersion:
  Size and Lifetime of noncondensed Pairs Below $T_c$}
\label{sec:2D}

We may rewrite the pair susceptibility \cite{Chen2,Chen1} of
Eq.~(\ref{eq:chi}) (after performing the Matsubara sum and analytically
continuing to the real axis) in a relatively simple form as
\begin{eqnarray}
 \chi (Q) &=& \sum_{\bf k} \Big[ \frac{1-f(E_{\bf k})-f(\xi_{\bf k-q})}
 {E_{\bf k}+\xi_{\bf k-q}-\Omega -i 0^+}u_{\bf k}^2 \nonumber \\
  &&{}-\frac{f(E_{\bf k})-f(\xi_{\bf k-q})}{E_{\bf k}-\xi_{\bf k-q}+
  \Omega +i 0^+}v_{\bf k}^2 \Big] \,,
\label{chi_expr}
\end{eqnarray}
where $u_{\bf k}^2$ and $v_{\bf k}^2$ are given by their usual BCS
expressions in terms of $\Delta$ and $\xi_{\bf k} \equiv \epsilon_{\bf
  k}-\mu$.  In the long wavelength, low frequency limit, the inverse of
$t_{pg}$ can be written as
\begin{equation}
 a_1\Omega^2 + Z_0(\Omega - \frac{q^2}{2 M^*}
+ \mu _{pair}  +i \Gamma_Q).
\label{Omega_q:exp}
\end{equation}

We are interested in the moderate and strong coupling cases, where we
can drop the $a_1 \Omega^2$ term in Eq.~(\ref{Omega_q:exp}), and hence
we have Eq.~(\ref{eq:expandt})
with 
  \begin{equation}
  \Omega_{\mathbf{q}} \equiv \frac{q^2} {2 M^*}=\frac{q^2 \xi_{pg}^2}{Z_0} \,.
  \end{equation}
  This establishes a quadratic dispersion and defines the effective pair
  mass, $M^*$.  Analytical expressions for this mass are possible via a
  small ${\bf q}$ expansion of $\chi$, in Eq.~(\ref{chi_expr}).
  It is important to note that the pair mass reflects the effective
  \textit{size} $\xi_{pg}$ of noncondensed pairs.  This serves to
  emphasize the fact that the $q^2$ dispersion derives from the
  compositeness of the ``bosons", in the sense of their finite spatial
  extent. A description of the system away from the BEC limit must
  accommodate the fact that the pairs have an underlying fermionic
  character.  This pair mass has a different origin from the mass
  renormalization associated with real interacting bosons.  There one
  finds a mass shift which comes from a Hartree approximate treatment of
  their $q$-dependent interactions.  Finally, we note that $\xi_{pg}$ is
  comparable to the size $\xi$ of pairs in the condensate. In the weak
  coupling BCS limit, a small $Q$ expansion of $\chi_0$ shows that,
  because the leading order term in $\Omega$ is purely imaginary, the
  $\Omega^2$ contribution cannot be neglected. The $T$-matrix, then,
  does not have the propagating $q^2$ dispersion, discussed above.
  
  It follows from Eq.~(\ref{chi_expr}) that the pair lifetime
\begin{eqnarray}
\Gamma_Q &=& \frac{\pi}{Z_0} \sumk  
 \left[1-f(\Ek)-f(\xi_{\bf k-q})\right] \uk^2 \delta (\Ek+\xi_{\bf
 k-q}-\Omega)  \nonumber\\
 &&{} +\left[f(\Ek)-f(\xi_{\bf k-q})\right] \vk^2  \delta(\Ek-\xi_{\bf
 k-q}+\Omega) \,.
\label{eq:83} 
\end{eqnarray}
%
Here $\Gamma_Q $ reflects the rate of decay of noncondensed bosons into
a bare and dressed fermion. Note that the excitation gap in $E_{\bf k}$
significantly restricts the contribution from the $\delta$ functions.
Thus, \textit{the decay rate of pair excitations is greatly suppressed,
  due to the excitation gap (for fermions) in the superconducting
  phase}.  Even in the normal phase pairs live longer than one might
have anticipated from ``Pauli blocking" arguments which are based on
evaluating $\Gamma_Q$ in a Fermi liquid state.
However, the same equations are not strictly valid above $T_c$, because
Eq.~(\ref{eq:sigma3}) no longer holds. To compute the pair lifetime in
the normal state requires a more extensive calculation involving the
full $T$-matrix self consistent equations and their numerical
solution \cite{Janko,Maly1,Maly2}.

\subsection{$T_c$ Calculations: Analytics and Numerics}
\label{sec:2E}

Calculations of $T_c$ can be performed using Eqs.~(\ref{eq:18}) and
(\ref{eq:19}) along with Eq.~(\ref{eq:delta_pg}).  The $T$-matrix is
written in the expanded form of Eq.~(\ref{eq:expandt}), which is based
on the pair dispersion as derived in the previous section.  For the most
part the calculations proceed numerically.

\begin{figure}
\includegraphics[width=2.3in,clip]{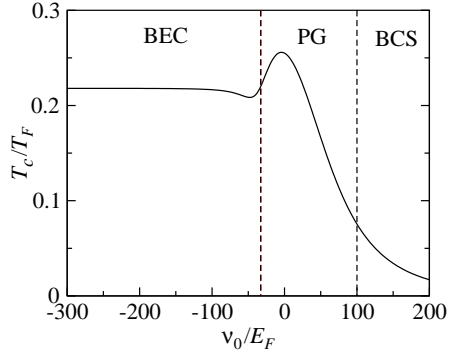}
\caption{Typical behavior of $T_c$ in the homogeneous case. For the
  purposes of illustration, this calculation includes Feshbach effects.
  $T_c$ follows BCS for large magnetic detuning parameter $\nu_0$ and
  approaches the BEC asymptote $0.218$ for large negative $\nu_0$. It
  reaches a maximum at unitarity, and has a minimum when the fermionic
  chemical potential $\mu$ changes sign. }
\label{fig:Tc}
\end{figure}

A typical curve is plotted in Fig.~\ref{fig:Tc}, where the
three regimes BCS, BEC and PG are indicated.
For positive, but decreasing the magnetic detuning $\nu_0$, $T_c$
follows the BCS curve until the ``pseudogap" or $\Delta(T_c)$ becomes
appreciable.  After its maximum (near the unitary limit, at slightly
negative $\nu_0$), $T_c$ decreases, (as does $\mu$), to reach a minimum
at $\mu \approx 0$.  This decrease in $T_c$ reflects the decreasing
number of low energy fermions due to the opening of a pseudogap.  Beyond
this point, towards negative $\nu_0$, the system is effectively bosonic,
and the superconductivity is no longer hampered by pseudogap effects.
In the presence of FB, the condensate consists of two contributions,
although the weight of the fermion pair component rapidly disappears.
Similarly $T_c$ rises, although slowly, towards the ideal BEC asymptote,
following the inverse effective boson mass.  The corresponding curve
based on the NSR approach \cite{NSR} has only one extremum, but
nevertheless the overall magnitudes are not so different
\cite{Griffin,Milstein}.

Analytic results are obtainable in the near-BEC limit only.  The general
expression for the effective mass
$1/M_0^*$ in this limit is given by
\begin{equation}
\frac{1}{M_0^*}=\frac{1}{Z_0 \Delta^2} \sum_{\bf k}
\left[ \frac{1}{ m} \vk^2
- \frac{4 \Ek k^2}{3m^2 \Delta^2} \vk^4\right] \,,
\label{eq:B}
\end{equation}
where here FB effects have been dropped for simplicity.  In what
follows, we expand Eq.~(\ref{eq:B}) in powers of $na_s^3$ and obtain after
some algebra
\begin{equation}
M_0^* \approx 2m \left( 1 + \frac{\pi a_s^3 n}{2}\right) \,.
 \label{eq:mass_change}
 \end{equation}
 We now invoke an important constraint, derived earlier in
 Eq.~(\ref{eq:52}) which corresponds to the fact that at $T_c$ all
 fermions are constituents of uncondensed pairs
 \begin{equation}
 \frac{n}{2}=\sum_{\bf q} b(\Omega_{\bf q}, T_c) \,.
 \label{eq:totaln}
 \end{equation}
 From the
 above equation it follows that $(M^* T_c)^{3/2} \propto n = const.$
 which, in conjunction with Eq.~(\ref{eq:mass_change}) implies
 \begin{equation}
 \frac{ T_c - T_c^0}{T_c^0} = - \frac{\pi a_s^3 n}{2}.
 \end{equation}
 Here $T_c^0$ is the transition temperature of the ideal Bose gas with
 $M_0=2m$. This downward shift of $T_c$ follows the effective mass
 renormalization, much as expected in a Hartree treatment of GP theory
 at $T_c$. Here, however, in contrast to GP theory for a homogeneous
 system with a contact potential \cite{RMP}, there is a non-vanishing
 renormalization of the effective mass.
 This is a key point which underlines the importance in this approach of
 the fermionic degrees of freedom, even at very strong coupling.

\section{Self Consistency Tests}
\label{sec:3}

All $T$-matrix approaches to the BCS-BEC crossover problem must be
subject to self consistency tests.  Among these, one should demonstrate
that \textit{normal state} self energy effects within a $T$-matrix
scheme are not associated with superfluidity or superconductivity
\cite{Kosztin1}.  While this seems at first sight straightforward, all
theories should be put to this test. Thus for a charged system, a
meaningful result for the superfluid density $\rho_s$ requires that
there be an exact cancellation between diamagnetic and paramagnetic
current contributions at $T_c$.  In this way self energy effects in the
number equation and gap equation must be treated in a consistent
fashion.  In the following we address some of these self consistency
issues in the context of the asymmetric ($\chi \approx GG_0$) BCS-BEC
crossover scheme.  Other crossover theories with different forms for the
pair susceptibility should also be subjected to these tests.  We begin
with a summary of calculations \cite{Kosztin2,JS} of the Meissner
effect, followed by a treatment of the collective mode spectrum.  Our
discussion in this section is presented for the one channel problem, in
the absence of Feshbach effects.

\subsection{Important check: behavior of $\rho_s$}
\label{sec:3A}

The superfluid density may be expressed in terms of the local (static)
electromagnetic response kernel $K(0)$
\begin{equation}
  \label{eq:ns2}
  n_s = \frac{m}{e^2} K(0) = n - \frac{m}{3\,e^2} P_{\alpha\alpha}(0) \;,
\end{equation}
with the current-current correlation function given by 
\begin{eqnarray}
  \label{eq:ns3}
  P_{\alpha\beta}(Q) &=& -2\,e^2 \sum_K
  \lambda_{\alpha}(K,K+Q)\,G(K+Q)\nonumber\\ &&
  \times\,\Lambda^{EM}_{\beta}(K+Q,K)\,G(K) \;.
\end{eqnarray}
Here the bare vertex ${\bf \lambda}(K,K+Q) = \frac{1}{m}
({\bf k}+{\bf q}/2)$ 
and we consider $Q=({\bf q},0)$, with  
${\bf q}\rightarrow 0$. 
We write
${\bf \Lambda}^{EM} = {\bf \lambda} + \delta {\bf \Lambda}_{pg}
+\delta {\bf \Lambda}_{sc}$, where the pseudogap contribution
$\delta {\bf \Lambda}_{pg}$ to the vertex correction 
will be shown in Appendix \ref{App:Ward} to satisfy
a Ward identity below $T_c$
\begin{equation}
  \delta {\bf \Lambda}^{pg}(K,K)=\frac{\partial\Sigma_{pg}(K)}{\partial
  {\bf k}}\;. 
\label{eq:Ward}
\end{equation}
By contrast for the superconducting contributions, one has
\begin{equation}
 \delta {\bf \Lambda}^{sc}(K,K) =
 -\frac{\partial\Sigma_{sc}(K)}{\partial {\bf k}} \,.
\label{eq:79}
\end{equation}
This important difference in sign is responsible for the fact that the
Meissner effect is associated with superconductivity, and not with a
normal state self energy.

The particle density $n$, 
after partial integration can be rewritten as $n=-(2/3)\sum_K {\bf
  k}\cdot\partial_{\bf k}{G(K)}$. Then, as a result of Dyson's equation,
one arrives at the following general expression which relates to the
diamagnetic contribution
\begin{equation}
  \label{eq:ns11}
  n = - \frac{2}{3} \sum_K G^2(K) \left[\frac{k^2}{m} +
    {\bf k}\cdot\frac{\partial\Sigma_{pg}(K)}{\partial {\bf k}} +
    {\bf k}\cdot\frac{\partial\Sigma_{sc}(K)}{\partial {\bf k}}\right] \;.
\end{equation}
Now, inserting Eqs.~(\ref{eq:ns11}) and (\ref{eq:ns3}) into
Eq.~(\ref{eq:ns2}) one can see that the pseudogap contribution to $n_s$
drops out by virtue of Eq.~(\ref{eq:Ward}); we find
\begin{equation}
  \label{eq:n_s-0}
  n_s = \frac{2}{3} \sum_K
  G^2(K)\,{\bf k}\cdot\left(\delta{\bf \Lambda}_{sc} -
    \frac{\partial\Sigma_{sc}}{\partial{\bf k}}\right) \;.
\end{equation}
We emphasize that the cancellation of this pseudogap contribution to the
Meissner effect is the central physics of this analysis, and it depends
on treating self energy effects in a Ward-identity- consistent fashion.

We also have that
\begin{equation}
  \label{eq:dL_sc}
  \delta{\bf \Lambda}_{sc}(K+Q,K) = \Delta_{sc}^2
  G_0(-K-Q) G_0(-K) {\bf \lambda}(K+Q,K)\;.
\end{equation}
Inserting  
Eqs.~(\ref{eq:70}), (\ref{eq:sigma3}), and (\ref{eq:dL_sc}) into
Eq.~(\ref{eq:n_s-0}), after calculating the Matsubara sum, one arrives at
\begin{equation}
  \label{eq:n_s}
  n_s = \frac{4}{3} \sum_{\bf k}
  \frac{\Delta_{sc}^2}{\Ek^2}
    \ek \,
  \left[\frac{1-2\,f(E_{\bf{k}})}{2\,E_{\bf{k}}} + f'(E_{\bf{k}})
  \right] \;.
\end{equation}
This expression can be simply rewritten in terms of the BCS result for
the superfluid density
\begin{equation}
\left( \frac{n_s}{m} \right)  = \frac{\Delta_{sc}^2}{\Delta^2} 
\left ( \frac{n_s}{m} \right)^{BCS}  \:.
\label{Lambda_BCS_Eq}
\end{equation}
Here $(n_s/m)^{BCS}$ is just $(n_s/m) $ with the overall prefactor
$\Delta_{sc}^2$ replaced with $\Delta^2$.

Finally we can rewrite Eq.~(\ref{Lambda_BCS_Eq}) using Eq.~(\ref{eq:sum})
as
\begin{equation}
\left( \frac{n_s}{m} \right)  = \left[1-\frac{\Delta_{pg}^2}{\Delta^2}\right]
\left ( \frac{n_s}{m} \right)^{BCS}  \:.
\label{Lambda_BCS_Eq2}
\end{equation}
In this form it is evident that (via $\Delta_{pg}^2$)
\textit{pair excitations out of
the condensate are responsible
for a suppression of the superfluid density relative to
that obtained from fermionic excitations only \cite{JS1}.}

\subsection{Collective Modes and Gauge Invariance}
\label{sec:3B}

The presence of pseudogap self energy effects greatly complicates the
computation of collective modes \cite{Kosztin2}. This is particularly
apparent at nonzero temperature.  Once dressed Green's functions $G$
enter into the calculational schemes, the collective mode
polarizabilities and the electromagnetic (EM) response tensor must
necessarily include vertex corrections dictated by the form of the
self-energy $\Sigma$, which depends on the \textit{T}-matrix which, in
turn depends on the form of the pair susceptibility $\chi$.  These
necessary vertex corrections are associated with gauge invariance in the
same way, as was seen for $\rho_s$, and discussed in Appendix
\ref{App:Ward}.  Collective modes are important in their own right,
particularly in neutral superfluids, where they can be directly detected
as signatures of long range order.  They also must be invoked to arrive
at a gauge invariant formulation of electrodynamics. It is relatively
straightforward to introduce these collective mode effects into the
electromagnetic response in a completely general fashion that is
required by gauge invariance.  The difficulty is in the implementation.

In the presence of a weak externally applied EM field, with four-vector
potential $A^{\mu} = (\phi, {\bf A})$, the four-current density $J^{\mu}
= (\rho, {\bf J})$ is given by
\begin{equation}
\label{eq:em1}
    J^{\mu}(Q) = K^{\mu\nu}(Q) A_{\nu}(Q)\;,
\end{equation}
where $K^{\mu\nu}(Q)$ is the EM response kernel.

The incorporation of gauge invariance into a general microscopic theory
may be implemented in several ways.  Here we do so via a generalized
matrix Kubo formula  \cite{Kulik} in which the perturbation of the
condensate is included as additional contributions $\Delta_1+i\Delta_2$
to the applied external field.  These contributions are self
consistently obtained (by using the gap equation) and then eliminated
from the final expression for $K^{\mu\nu}$.  We now implement this
procedure.  Let $\eta_{1,2}$ denote the change in the expectation value
of the pairing field $\hat{\eta}_{1,2}$ corresponding to $\Delta_{1,2}$.
For the case of an $s$-wave pairing interaction $U<0$, the
self-consistency condition $\Delta_{1,2}=U\eta_{1,2}/2$ leads to the
following equations:

\begin{subequations}
\label{eq:em9}
 \begin{eqnarray}
  \label{eq:em9a}
 J^{\mu} &=& K^{\mu\nu}A_{\nu} = K_0^{\mu\nu}A_{\nu} + R^{\mu
   1}\Delta_1 + R^{\mu 2}\Delta_2\,,\\
  \label{eq:em9b}
  \eta_1 &=& -\frac{2\Delta_1}{|U|} = R^{1\nu}A_{\nu} + Q_{11}\Delta_1 +
   Q_{12}\Delta_2\;, \\
    \label{eq:em9c}
  \eta_2 &=& -\frac{2\Delta_2}{|U|} = R^{2\nu}A_{\nu} + Q_{21}\Delta_1 +
     Q_{22}\Delta_2\;,
\end{eqnarray}
\end{subequations}
 where
 \begin{equation}
 \label{eq:em2}
  K_0^{\mu\nu}(\omega,{\bf q}) = P^{\mu\nu}(\omega,{\bf q}) +
  \frac{ne^2}{m}
  g^{\mu\nu}(1-g^{\mu 0})
 \end{equation}
  is the usual Kubo expression for the electromagnetic response.  We define
  the current-current correlation function $P^{\mu\nu}(\tau,{\bf q}) =
  -i\theta(\tau)\langle[j^{\mu}(\tau,{\bf q}),j^{\nu}(0,-{\bf q})]\rangle$.
   In the above equation, $g^{\mu\nu}$ is a (diagonal) metric tensor with
   elements $(1,-1,-1,-1)$.
   We define
\begin{equation}
R^{\mu i}(\tau,{\bf q})=-i\theta(\tau)\langle[j^{\mu}(\tau,{\bf q}),
  \hat{\eta}_i(0,-{\bf q})]\rangle
  \end{equation}
  with $\mu=0,\ldots,3$, and $i,j =1,2$; and
  \begin{equation}
  Q_{ij}(\tau,{\bf q})=-i\theta(\tau)\langle[\hat{\eta}_i(\tau,{\bf q}),
      \hat{\eta}_j(0,-{\bf q})]\rangle
      \end{equation}
      Finally, it is convenient to define 
\begin{equation}
\tilde{Q}_{ii} = 2/U+Q_{ii}\;.
\end{equation}

In order to demonstrate gauge invariance and reduce the number of
component polarizabilities, we first rewrite $K^{\mu\nu}$ in a way which
incorporates the effects of the amplitude contributions via a
renormalization of the relevant generalized polarizabilities,
\begin{equation}
  \label{eq:em14b}
    {K'}_0^{\mu\nu} = K_0^{\mu\nu}-\frac{R^{\mu
          1}R^{1\nu}}{\tilde{Q}_{11}}\;,
\end{equation}

It can be shown, after some analysis, that the gauge invariant
form for the response tensor is given by
\begin{equation}
  \label{eq:em15}
    K^{\mu\nu} = {K'}_0^{\mu\nu} -
    \frac{\left({K'}_0^{\mu\nu'}q_{\nu'}\right)
    \left(q_{\nu''}{K'}_0^{\nu''\nu}\right)}{q_{\mu'}{K'}_0^{\mu'\nu'}q_{
    \nu'}}\;.
\end{equation}

The above equation satisfies two important requirements: it is
manifestly gauge invariant and, moreover, it has been reduced to a form
that depends principally on the four-current-current correlation
functions.  (The word ``principally'' appears because in the absence of
particle-hole symmetry, there are effects associated with the order
parameter amplitude contributions that add to the complexity of the
calculations).

The EM response kernel of a superconductor contains a pole structure
that is related to the underlying Goldstone boson of the system.  Unlike
the phase mode component of the collective mode spectrum, this
Anderson-Bogoliubov (AB) mode is independent of Coulomb effects.  The
dispersion of this amplitude renormalized AB mode is given by
\begin{equation}
  \label{eq:em19}
    q_{\mu}{K'}_0^{\mu\nu}q_{\nu} = 0\;.
\end{equation}
For an isotropic system ${K'}_0^{\alpha\beta} = {K'}_0^{11}
\delta_{\alpha\beta}$, and Eq.~(\ref{eq:em19}) can be rewritten as
\begin{equation}
      \label{eq:em20}
        \omega^2 {K'}_0^{00} + {\bf q}^2 {K'}_0^{11} - 2\omega q_{\alpha}
          {K'}_0^{0\alpha} = 0\;,
\end{equation}
with $\alpha=1,2,3$, and in the last term on the left hand side of
Eq.~(\ref{eq:em20}) a summation over repeated Greek indices is assumed.

It might seem surprising that from an analysis which incorporates a
complicated matrix linearized response approach, the dispersion of the
AB mode ultimately involves only the amplitude renormalized four-current
correlation functions, namely the density-density, current-current and
density-current correlation functions. The simplicity of this result is,
nevertheless, a consequence of gauge invariance.

At zero temperature ${K'}_0^{0\alpha}$ vanishes, and the sound-like AB
mode has the usual linear dispersion $\omega=\omega_{\bf q}=c|{\bf q}|$
with the ``sound velocity'' given by
\begin{equation}
  \label{eq:em22}
    c^2 = {K'}_0^{11}/{K'}_0^{00} \;.
\end{equation}
We may, thus, interpret the AB mode as a special type of collective mode
which is associated with $A_{\nu} = 0 $. This mode corresponds to free
oscillations of $\Delta_{1,2}$ with a dispersion $\omega= c q$ given by
the solution to the equation
\begin{equation}
\label{eq:em17}
\text{det}|Q_{ij}| = \tilde{Q}_{11}\tilde{Q}_{22}-Q_{12}Q_{21} = 0\;.
\end{equation}
The other collective modes of this system are derived by including the
coupling to electromagnetic fields.  Within self-consistent linear
response theory the field $\delta\phi$ must be treated on an equal
footing with $\Delta_{1,2}$ and formally can be incorporated into the
linear response of the system by adding an extra term $K_0^{\mu
  0}\delta\phi$ to the right hand side of Eqs.~(\ref{eq:em9a}),
(\ref{eq:em9b}), and (\ref{eq:em9c}).  Note that, quite generally, the
effect of the ``external field'' $\delta\phi$ amounts to replacing the
scalar potential $A^0=\phi$ by $\bar{A^0} = \bar{\phi} =
\phi+\delta\phi$. In this way one arrives at the following set of three
linear, homogeneous equations for the unknowns $\delta\phi$, $\Delta_1$,
and $\Delta_2$
\begin{subequations}
\label{eq:em23}
\begin{eqnarray}
0 &=& R^{10}\delta\phi + \tilde{Q}_{11}\Delta_1 + Q_{12}\Delta_2\;,
\label{eq:em23a}
 \\ 0 &=& R^{20}\delta\phi + Q_{21}\Delta_1 +
   \tilde{Q}_{22}\Delta_2\;,
\label{eq:em23b} \\ 
\delta\rho =
 \frac{\delta\phi}{V} &=& K_0^{00}\delta\phi + R^{01}\Delta_1 +
        R^{02}\Delta_2 \;,
\label{eq:em23c}
\end{eqnarray}
\end{subequations}
where $V$ is an effective particle-hole interaction which may derive
from the pairing channel or, in a charged superconductor, from the
Coulomb interaction.  
The dispersion of the collective modes of the system is given by the
condition that the above equations have a nontrivial solution
\begin{equation}
  \label{eq:em24}
     \left|
         \begin{array}{ccc}
        Q_{11}+1/U & Q_{12} & R^{10} \\
      Q_{21} & Q_{22}+1/U & R^{20} \\
           R^{01} & R^{02} & K_0^{00} - 1/V
        \end{array}
                  \right| = 0\;.
\end{equation}
In the BCS limit where there is particle-hole symmetry
$Q_{12}=Q_{21}=R^{10}=R^{01}=0$ and, the amplitude mode decouples from
the phase and density modes; the latter two are, however, in general
coupled.
 
The above formalism (or its equivalent RPA variations
\cite{Cote,Randeria2}) has been applied to address collective modes in
the crossover scenario. The most extensive studies have been at $T=0$
based on the ground state of Eq.~(\ref{eq:1a}).  There one finds a
smooth change in the character of the Anderson-Bogoliubov (AB) mode. At
weak coupling one obtains the usual BCS value $ c = v_F/\sqrt{3}$.  By
contrast at strong coupling, the collective mode spectrum reflects
\cite{Randeria2} an effective boson-boson interaction deriving from the
Pauli statistics of the constituent fermions.  This is most clearly seen
in jellium models \cite{Griffin2,Kosztin2} where the AB sound velocity
is equivalent to that predicted for a 3D interacting Bose gas $c = [ 4
\pi n a_B / m_B^2 ]^{1/2}$. Here, as earlier, the inter-boson scattering
length is twice that of the inter-fermionic counterpart, at strong
coupling.\footnote{In the neutral case, for the full collective modes of
  Eq.~(\ref{eq:em24}), there are numerical differences (of order unity)
  in the prefactors of the mode frequency, so that the collective modes
  of the crossover theory are not strictly the same as in GP theory.  In
  these calculations $V$ in Eq.~(\ref{eq:em24}) should be associated
  with the pairing interaction, which enters
in the context of fermionic density-density
correlation effects within the particle-hole channel.}

\begin{figure}
\includegraphics[width=3.0in,clip]{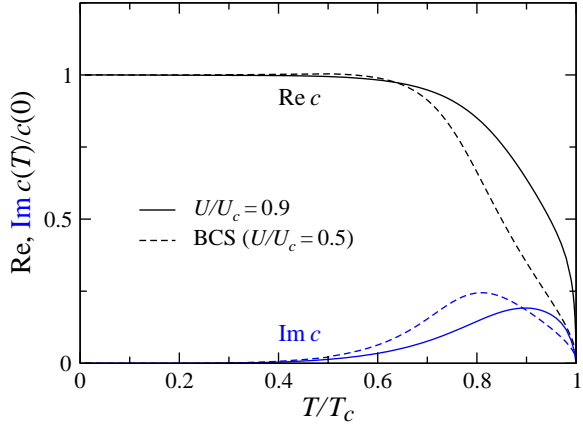}
\caption{Temperature dependence of the real and imaginary
parts of the AB collective mode velocity for moderate (solid lines)
and weak (dashed lines) coupling. The velocity vanishes at $T_c$ for
both cases.}
\label{fig:16z}
\end{figure}

The effects of finite temperature on the AB mode have been studied in
\cite{Kosztin2}, by making the approximation that the temperature
dependence of the order parameter amplitude contribution is negligible.
Then the calculation of this dispersion reduces to calculations of the
electromagnetic response, as discussed, for example in $\rho_s$. See
also Appendix \ref{App:Ward}.  At finite $T$, the AB mode becomes damped
and the real and imaginary parts of the sound velocity have to be
calculated numerically.  Here one finds, as expected, that the complex $
c \rightarrow 0 $ as $ T \rightarrow T_c$.  The results are plotted in
Fig.~\ref{fig:16z} for both moderate and weak coupling.  Because the
real and imaginary parts are comparable there, the mode is highly damped
near $T_c$.

\subsection{Investigating the Applicability of a Nambu 
Matrix Green's Function Formulation}

Diagrammatic schemes appropriate to BCS superconductors are based on a
Nambu matrix Green's function approach.  The off-diagonal or anomalous
Green's functions in this matrix are given by
\begin{equation}
F(K) \equiv \Delta_{sc} G(K)G_0(-K)
\label{eq:Nambu}
\end{equation}
as well as its Hermitian conjugate.  This Nambu formalism was designed
to allow study of perturbations to the BCS state, due to, for example,
external fields or impurities.  For these perturbations the central
assumption is that they act on both the ``normal" (with BCS Green's
function $G(K)$ from Eq.~(\ref{eq:12c})) and anomalous channels in a
symmetrical way.

Understanding as we now do that BCS theory is a very special case of
superconductivity, this raises the caution that once one goes beyond
BCS, some care should be taken to justify this Nambu approach.  At the
very least the distinction between the order parameter $\Delta_{sc}$ and
the excitation gap $\Delta$ raises ambiguity about applying a
diagrammatic theory based on Nambu-Gor'kov Green's functions.

More importantly, in treating the effects of the pseudogap self energy
$\Sigma_{pg}$ as we do here, it should be clear that this self energy
does \textit{not} play a symmetric role in the anomalous and normal
channels. It is viewed here as an entirely normal state effect.  However
this theory accommodates the equivalent \cite{Kosztin2} of the
Nambu-Gor'kov ``F"- function in general response functions such as in
the Maki-Thompson diagrams of Appendix \ref{App:Ward} through the
asymmetric combination $GG_0$ which always arises in pairs (for example,
as the $FF$ combinations of BCS theory).
In this regard the $GG_0$ formalism appears to differ from all other
$T$-matrix schemes which are designed to go below $T_c$, in that the
Nambu scheme is not assumed at the start. Nevertheless, many features of
this formalism seem to naturally arise in large part because of
Eq.~(\ref{eq:Nambu}), which demonstrates an intimate connection between
$GG_0$ and the conventional diagrams of BCS theory.

\section{Other Theoretical Approaches to the Crossover Problem}
\label{sec:4}
\subsection{$T \neq 0$ Theories}
\label{sec:4a}

We have noted in Section \ref{sec:1F} that there are three main
theoretical approaches to the crossover problem based on $T$-matrix
theories. Their differences are associated with different forms for the
pair susceptibility $\chi$. The resulting calculations of $T_c$ show
similar variations.  When two full Green's functions are present in
$\chi$ (as in the FLEX approach), $T_c$ varies monotonically
\cite{Haussmann} with increasing attractive coupling, approaching the
ideal gas Bose-Einstein asymptote from below.  When two bare Green's
functions are present in $\chi$, as in the work of Nozi\'eres and
Schmitt-Rink, and of Randeria and co-workers, then $T_c$ overshoots
\cite{NSR,randeriareview,deMelo} the BEC asymptote and ultimately approaches it
from above.  Finally when there is one bare and one dressed Green's
function, $T_c$ first overshoots and then decreases to a minimum around
$\mu =0$, eventually approaching the asymptote from below \cite{Maly2}.
This last appears to be a combination of the other two approaches.
Overall the magnitudes are relatively similar and the quantitative
differences are small.  There is, finally, another body of work
\cite{Berthod} based on the $GG_0$ scheme which should be noted here.
The approach by the Swiss group seeks to make contact with the phase
fluctuation scenario for high $T_c$ superconductors and addresses STM
and ARPES data.

Bigger differences appear when these $T$-matrix theories are extended
below $T_c$.  Detailed studies are most extensive for the NSR approach.
Below $T_c$ one presumes that the $T$-matrix (or $\chi$) contains only
bare Green's functions, but these functions are now taken to correspond
to their Nambu matrix form, with the order parameter $\Delta_{sc}$
appearing in the dispersion relation for excited
fermions \cite{randeriareview,Strinati2}.  A motivation for generalizing
the below- $T_c$ $T$-matrix in this way is that one wants to connect to
the collective mode spectrum of the superconductor, so that the
dispersion relation for pair excitations is $\Omega_q \approx cq$. In
this way the system would be more directly analogous to a true Bose
system.

Self energy effects are also incorporated below $T_c$ but only in the
number equation, either in the approximate manner of NSR
\cite{randeriareview,Griffin2} or through use of the full Dyson
resummation \cite{Strinati2} of the diagonal Nambu-matrix component
$G(\Sigma_0)$. For the latter scheme, Strinati's group has addressed
pseudogap effects in some detail with emphasis on the experimentally
observed fermionic spectral function.  Some concern can be raised that
the fermionic excitations in the gap equation do not incorporate this
pseudogap, although these pairs are presumed to emerge out of a normal
state which has a pseudogap.  Indeed, this issue goes back to the
original formulation of NSR, which includes self energy effects in the
number equation, and not in the gap equation.

Tchernyshyov \cite{Tchern} presented one of the first discussions below
$T_c$ for the FLEX-based $T$-matrix scheme. He also addressed pseudogap
effects and found a suppression of the fermion density of states at low
energy which allows for long-lived pair excitations inside this gap.  At
low momenta and frequencies, their dispersion is that of a
Bogoliubov-sound-like mode with a nonzero mass.
Some of these ideas have been recently extended \cite{Ranninger03}
to apply to the boson-fermion model.

The work of Micnas and collaborators \cite{Micnas95} which is also based
on the FLEX scheme predates the work of Tchernyshyov, although their
initial focus was above $T_c$ in two dimensional Hubbard models. Here
the fermionic spectral functions were studied in detail, using a
numerically generated solution of the $T$-matrix equations.  A nice review
of their large body of work on ``real space pairing" in the cuprates was
presented based on a variety of different model Hamiltonians
\cite{Micnas1}.

An extensive body of work on the FLEX scheme has been contributed by Yamada
and Yanatse \cite{YY} both above and below $T_c$. They point out
important distinctions between their approach and that of NSR.  The
effects of the broken symmetry are treated in a generalized Nambu
formalism, much as assumed 
by Pieri \textit{et al.} (\cite{Strinati2}) and by Griffin and Ohashi
(\cite{Griffin2}), but here the calculations involve self energy effects
in both the number and gap equations.  Their work has emphasized the
effects of the pseudogap on magnetic properties, but they have discussed
a wide variety of experiments in high $T_c$ and other exotic
superconductors.

The nature of the ground states which result from these two alternatives
($\chi \approx GG$ and $\chi \approx G_0G_0$ ) has yet to be clearly
established.  It should be noted that, despite the extensive body of
work based on the NSR or $G_0G_0$ scheme, no detailed picture has been
presented in the literature of the nature of the ground state which
results when their calculations at $T_c$ are extrapolated
\cite{Strinati2} down to $T=0$.  In large part, the differences between
other work in the literature and that summarized in Sections
\ref{sec:2C} are due to the whether (as in NSR-based papers) or not (as
here) the superconducting order parameter $\Delta_{sc}$ alone
characterizes the fermionic dispersion below $T_c$.

One might well ask the question: because the underlying Hamiltonian
[Eq.~(\ref{eq:0c})] is associated with inter-fermionic, not
inter-bosonic interactions, will this be reflected in the near-BEC limit
of the crossover problem?  The precise BEC limit is, of course, a
non-interacting, or ideal Bose gas, but away from this limit, fermionic
degrees of freedom would seem to be relevant in ways that may not be
accounted for by the analogue treatment of the weakly interacting Bose
gas.  The most extensive study (albeit, above $T_c$) of this issue is
due to Pieri and Strinati \cite{Strinati3}. Their work, importantly,
points out the inadequacies of $T$-matrix schemes, particularly at
strong coupling.  While the ground state in their calculations is
unknown, it is necessarily different from the conventional crossover
state of Eq.~(\ref{eq:1a}).  There is much intuition to be gained by
studying this simplest of all ground states, as outlined in this Review,
but it will clearly be of great value in future to consider states, in
which, for example, there is less than full condensation.

Finally, we note that studies of non-$s$ wave orbital pairing states in
the context of the BCS-BEC crossover problem have rapidly proliferated
in the literature. This is, in large part because of the observed
$d$-wave pairing in the cuprates. Indeed, Leggett's initial work
\cite{Leggett}, addressed $p$-wave pairing in $^3 He $.  Randeria and
co-workers \cite{randeriareview} also considered non-$s$ wave pairing.
Studies \cite{Andrenacci,Hertog} of this crossover for a $d$-wave
lattice case are plentiful at $T=0$. Even earlier work was done
\cite{Chen1} on extensions of this ground state to include nonzero
temperature.  Moreover, a $d$-wave continuum model has also been
considered \cite{Gyorffy}.  It is expected that in the near future cold
atom experiments will begin to address $ l \neq 0$ pairing via
appropriate choices of the Feshbach resonances \cite{Ho2}.

\subsection{Boson-Fermion Models for High $T_c$ Superconductors}
\label{sec:4b}
The notion that it might be appropriate to include bosonic degrees
of freedom in high $T_c$ superconductors goes back to T.D. Lee and R.
Friedberg who introduced \cite{TDLee1,TDLee2} the ``boson-fermion" model
within three years of the discovery of cuprate superconductivity.  The
work of Lee and colleagues is based on the Hamiltonian of
Eq.~(\ref{eq:0c}) without the direct fermion-fermion interaction $U$. To
our knowledge this is the first study which argues for something akin to
``real space pairing" in these materials. Interestingly, this work was
motivated only by the observed short coherence length.  Pseudogap
effects, which make this case much stronger, had not yet been properly
characterized.  The analogies with He$^4$ and a more BEC-like
picture of superconductivity were discussed, as well as the superfluid
density, and energy gap structure, albeit for $s$-wave pairing.

Subsequent work on this model Hamiltonian in the context of the high
$T_c$ copper oxides was pursued by Ranninger and Micnas and their
collaborators. Indeed, Ranninger and Robaszkiewicz \cite{RR} were the
first to write down the boson-fermion model, although initially, in an
entirely different context, the bosons were associated with phonons
rather than electron pairs, and they were presumed to be localized.  In
this context they addressed mean field theoretic solutions below $T_c$
as well as more recently \cite{Ranninger03} an approach based on
renormalization group techniques. The physical picture which emerges
leads to two gap structures below $T_c$, one associated with a pseudogap
and another with the superconductivity. In this way it appears to be
very different from approaches based on Eq.~(\ref{eq:1a}).

\begin{figure}
\includegraphics[width=3.4in,clip]{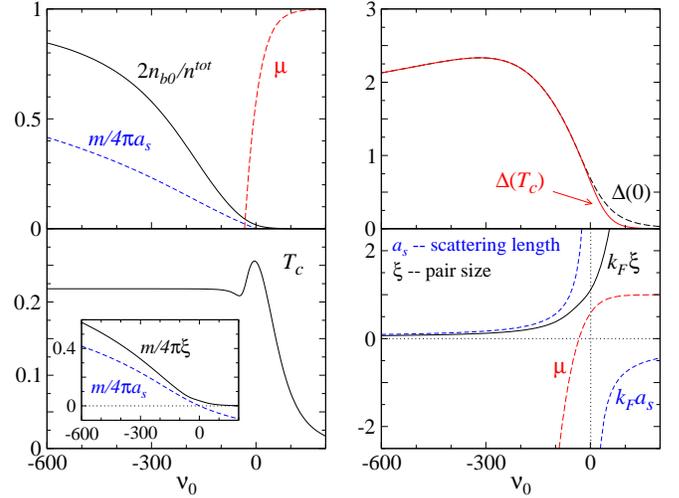}
\caption{(Color) Characteristics of the ground state and $T_c$ as a
  function of the magnetic detuning $\nu_0$ in a homogeneous case. Shown
  are the chemical potential $\mu$, fraction of molecular condensate
  $2n_{b0}/n^{tot}$, inverse scattering length $m/4\pi a_s$, gap
  $\Delta(0)$, inverse pair size $m/4\pi \xi$, dimensionless relative
  pair size $k_F\xi$ at $T=0$ as well as $T_c$ and $\Delta(T_c)$. The
  pair size is roughly the same as inter-particle spacing, $k_F\xi = 1$,
  at unitarity. As $\nu_0$ decreases, the gap $\Delta$ increases and
  reaches a maximum, and then decreases slowly and approaches its BEC
  asymptote (not shown) from above.  Here we take $g_0= -40
  E_F/k_F^{3/2}$ and $U_0=-3 E_F/k_F^3$, and we use the convention for units
  $E_F=k_F=\hbar=k_B=2m=1$. A large exponential momentum cutoff,
  $k_c/k_F=80$ is assumed in the calculations.}
\label{fig:16}
\end{figure}

\section{Physical Implications: Ultracold Atom Superfluidity}
\label{sec:5}
\subsection{Homogeneous case}
\label{sec:5A}

In this section we summarize the key characteristics of fermionic
superfluidity in ultracold gases, in the homogeneous situation, without
introducing the trap potential.  Our results are based on numerical
solution of the coupled Eqs.~(\ref{eq:18}), (\ref{eq:19}) and
(\ref{eq:delta_pg}).  The upper left panel of Fig.~\ref{fig:16} plots
the fermionic chemical potential, the Feshbach boson condensate ratio
and the inverse scattering length as a function of $\nu_0$. Here we have
chosen what we believe is the physically appropriate value for the
Feshbach coupling $g_0 = -40 E_F/k_F^3$ for $^{40}$K, as determined from
the experimentally measured scattering length data and resonance
width.\footnote{We note that the precise values of $g_0$ and $U_0$ are
  not critical for either $^6$Li or $^{40}$K as long as they give
  roughly the right (large) resonance width $\Delta B$ and background
  scatter length $a_{bg}$. In fact, these parameters usually change
  during experiment since the total particle number $N$ (and hence $E_F$
  and $k_F$) decreases in the process of evaporative cooling. In order
  to compare with experiment, one needs to compare at the correct value
  of $k_F a_s$.} This parameter is roughly an order of magnitude larger
for $^{6}$Li.  Here $\nu_0$, the gaps and $\mu$ are all in units of
$E_F$, and the plots, unless indicated otherwise, are at zero
temperature.

The upper right panel plots the excitation gap at $T_c$ as well as the
gap at $T=0$.  The lower left hand panel indicates $T_c$ along with the
inverse pair size $\xi^{-1}$ in the condensate. Finally the lower right
panel plots $\xi$ itself, along with $a_s$.

One can glean from the figure that the Feshbach boson fraction decreases,
becoming negligible when the chemical potential passes through zero.
This latter point marks the onset of the PG regime, and in this regime
the condensate consists almost entirely of fermionic pairs. The upper
limit of the PG regime, that is, the boundary line with the BCS phase,
is reached once the pseudogap, $\Delta(T_c)$ is essentially zero.  This
happens when $\mu$ is close to its saturation value at $E_F$. At this
point the pair size rapidly increases.

\begin{figure}
\includegraphics[width=3.4in,clip]{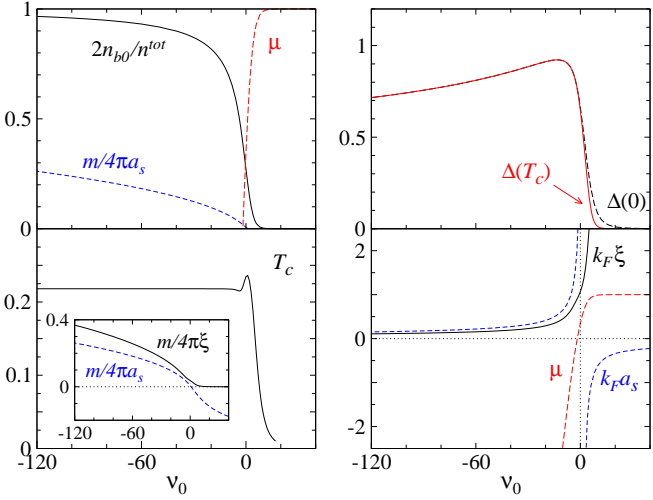}
\caption {(Color) Characteristics of the ground state and $T_c$, 
  as in Fig.~\ref{fig:16} except for a smaller $g_0$. In comparison with
  Fig.~\ref{fig:16}, the Feshbach resonance becomes much narrower. Here
  $g_0= -10 E_F/k_F^{3/2}$ and $U_0=-3 E_F/k_F^3$.}
\label{fig:17}
\end{figure}

\begin{figure}
\includegraphics[width=3.4in,clip]{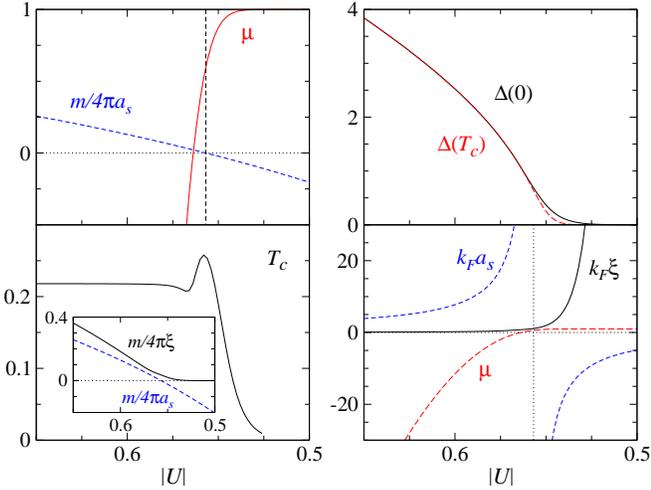}
\caption{(Color) Characteristics of the ground state and $T_c$ in the
  one-channel model, as in Fig.~\ref{fig:16} but as a function of
  the attraction strength $U$.  Here $g_0= 0$. Different from the
  two-channel shown in Figs.~\ref{fig:16}-\ref{fig:17} case, the gap
  $\Delta$ increases indefinitely with $|U|$.  The units for $U$ are
  $E_F/k_F^3$. Note that $|U|$ increases to the left.}
\label{fig:18}
\end{figure}

These results can be compared with those derived from a smaller value of
the Feshbach coupling constant $g_0 = -10 E_F/k_F^{3/2}$ shown in
Fig.~\ref{fig:17}.  Now the resonance is effectively narrower. Other
qualitative features remain the same as in the previous figure.

One can plot the analogous figures in the absence of Feshbach effects.
Here the horizontal axis is the inter-fermionic interaction strength
$|U|$, as is shown in Fig.~\ref{fig:18}.  Three essential differences
can be observed. Note, first the obvious absence of the Feshbach boson
or molecular condensate $n_{b0}$.  There is, of course, a condensate
associated with pairs of fermions ($\Delta_{sc}$) and these pairs will
become bound into ``fermionic molecules" for sufficiently strong
attraction.  Secondly, note that the excitation gap $\Delta(0)$ is
monotonically increasing as $|U|$ increases towards the BEC limit. By
contrast, from the upper right panels of Fig.~\ref{fig:17} one can see
that when Feshbach effects are present $\Delta(0)\approx \Delta(T_c) $
decreases towards zero in the extreme BEC limit.  It can also be seen
that the shape of the scattering length curve $vs$ $U$ is different from
the plots in the previous two figures. Here $a_s$ more rapidly increases
in magnitude on either side of the unitary limit.  Nevertheless it is
important to stress that, \textit{except for the nature of the Bose
  condensate in the BEC limit, the physics of the Feshbach
  resonance-induced superfluidity is not so qualitatively different from
  that associated with a direct inter-fermionic attraction}.

\begin{figure}
\includegraphics[width=3.4in,clip]{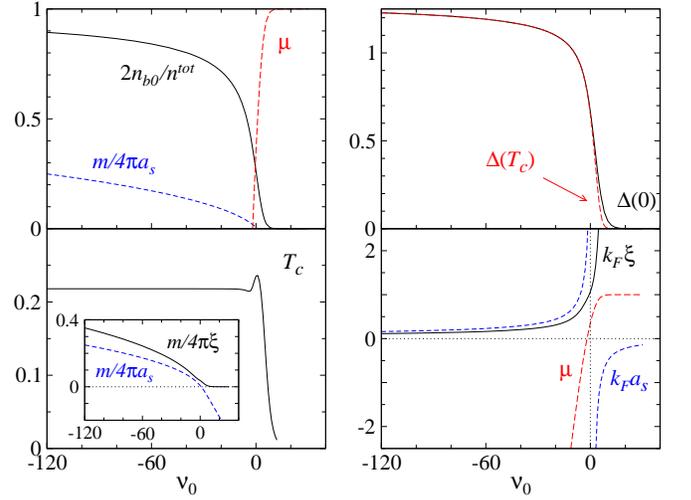}
\caption{(Color) Characteristics of the ground state and $T_c$, as in 
  Fig.~\ref{fig:17}, except now $U_0= 0$. As a consequence of the vanishing
  of $U_0$, the gap $\Delta$ is monotonic and approaches its BEC
  asymptote from below as $\nu_0$ decreases. Here $g_0 = -10
  E_F/K_F^{3/2}$. }
\label{fig:18z}
\end{figure}

\begin{figure} 
\includegraphics[width=3.3in,clip]{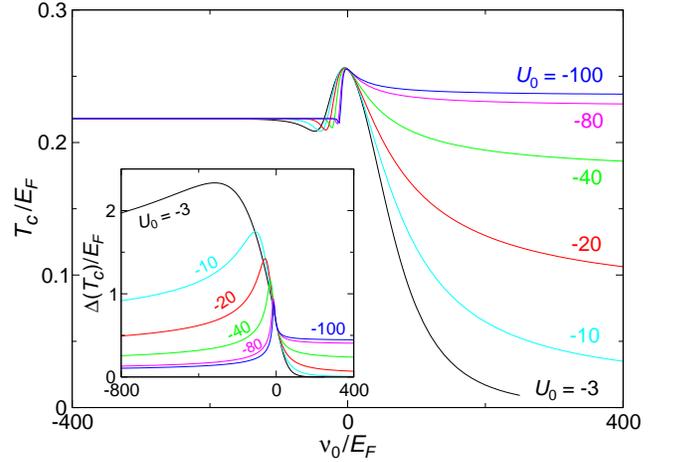}
\caption{(Color) Behavior of $T_c$ (main figure) and $\Delta(T_c)$
  (inset) at fixed $g_0=-40 E_F/k_F^{3/2}$ with variable $U_0$. As $U_0
  \rightarrow -\infty$, unitarity is approached from the BCS side at
  large positive $\nu_0$. At the same time, $\Delta \propto |g|$ also
  decreases.}
\label{fig:19a}
\end{figure}

Figure \ref{fig:18z} presents the analogous plots for the case when the
direct attraction between fermions $U$ vanishes.  Here the coupling
constant $g_0$ is the same as in Fig.~\ref{fig:17}, and pseudogap
effects are present here just as in all other cases. The most notable
difference in the physics is that the formation of the Cooper condensate
is driven exclusively by coupling to the molecular bosons. A contrast
between the two figures is in the behavior of $\Delta(T_c)$ and
$\Delta(0)$. For the $U=0$ case and in the BEC limit, $\Delta(0)$
approaches the asymptotic value ($|g|\sqrt{n_b^0}$) from below.
When the system is the fermionic regime ($\mu > 0$), aside from
quantitative details, there is very little to distinguish the effects of
$U$ from those of $g$, except when these parameters are taken to
very extreme limits.

Finally, in Fig.~\ref{fig:19a} we plot $T_c$ vs $\nu_0$ in the presence
of Feshbach effects with variable $U_0$, from weak to strong background
coupling.  The lower inset shows the behavior of the excitation gap at
$T_c$. With very strong direct fermion attraction $U$, we see that $T_c$
has a very different dependence on $\nu_0$. In this limit there is a
molecular BEC to PG crossover (which may be inaccessible in actual
experiments, since $U$ is not sufficiently high).  Nevertheless, it is
useful for completeness to illustrate the entire range of theoretical
behavior.

\begin{figure}
\includegraphics[width=3.2in,clip]{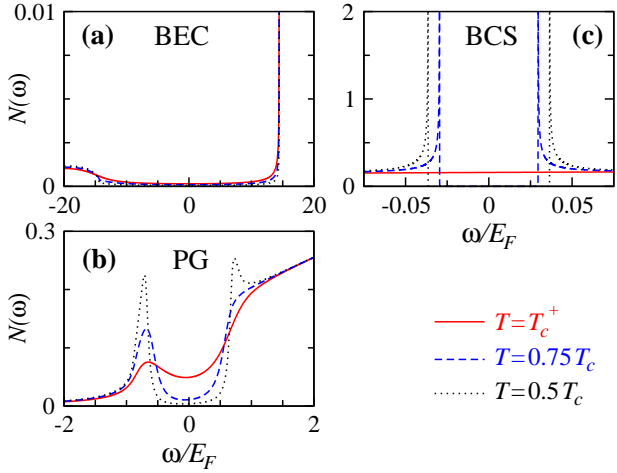}
\caption{Typical density of states (DOS) $N(\omega)$ $vs$ energy
  $\omega$ in the three regimes at indicated temperatures. There is no
  depletion of the DOS at $T_c$ in the BCS case.  By contrast, in the BEC
  case, there is a large gap in DOS at $T_c$, roughly determined by
  $|\mu|$. In the intermediate, i.e., pseudogap, regime, there is a
  strong depletion of the DOS already at $T_c$. Once $T$ decreases below
  $T_c$, the DOS decreases rapidly, signalling the superfluid phase
  transition.}
\label{fig:dos}
\end{figure}

Figures \ref{fig:dos}(a)-\ref{fig:dos}(c) show the fermionic density of
states for the BEC, PG and BCS limits.  These figures are important in
establishing a precise visual picture of a ``pseudogap".  The
temperatures shown are just above $T_c$, and for $ T = 0.75 T_c$ and $ T
= 0.5 T_c$.  The methodology for arriving at these plots will be
discussed in the following section, in the context of high $T_c$
superconductors.  Only in the BCS case is there a clear signature of
$T_c$ in the density of states, but the gap is so small and $T_c$ is so
low, that this is unlikely to be experimentally detectable. Since the
fermionic gap is well established in the BEC case, very little
temperature dependence is seen as the system goes from the normal to the
superfluid states.  Only the PG case, where $T_c$ is maximal, indicates
the presence of superfluidity, not so much at $T_c$ but once superfluid
order is well established at $ T = 0.5 T_c$, through the presence of
sharper coherence features, much as seen in the cuprates.

\begin{figure}
\includegraphics[width=3.in,clip]{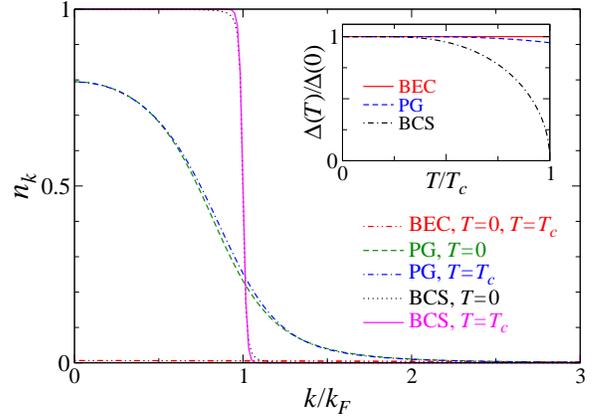}
\caption{Typical behavior of the fermionic momentum distribution
  function $n_k$ at $T=0$ and $T_c$ in the three regimes for
  a homogeneous system. The inset
  indicates the corresponding normalized gaps as a function of $T/T_c$. In
  all three regimes, $n_k$ barely changes between $T=T_c$ and $T= 0$.}
\label{fig:nk}
\end{figure}

The inset of Fig.~\ref{fig:nk} plots the temperature dependence of
$\Delta(T)$.  It should be stressed that $T_c$ is only apparent in
$\Delta(T)$ in the BCS case.  To underline this point, in the main body
of Fig.~\ref{fig:nk} we plot the fermionic momentum distribution
function $n_k$, which is the summand in the number
equation, Eq.~(\ref{eq:19}), at $T=0$ and $T=T_c$.  The fact that there
is very little change from $ T = 0$ to $ T= T_c$ makes the important
point that this momentum distribution function in a homogeneous system
is not a good indicator of phase coherent pairing.  For the PG case,
this, in turn, derives from the fact that $\Delta(T)$ is nearly
constant.  For the BEC limit the excitation gap, which is dominated by
$\mu$, similarly, does not vary through $T_c$. In the BCS regime,
$\Delta(T)$ is sufficiently small so as to be barely perceptible on the
scale of the figure.

\subsubsection{Unitary Limit and Universality}
\label{sec:5c}

There has been considerable emphasis on the behavior of fermionic gases
near the unitary limit where $a_s = \pm \infty$.  Attention
\cite{JasonHo,Stringari} has concentrated on this limit in the absence
of Feshbach bosons. We now want to explore the more general two-channel
problem here.
At unitarity we require that 
\begin{equation}
\frac{4\pi a_s^*}{m} =  U_0 + \frac{g_0^2}{2 \mu -\nu_0}
\label{eq:U1}
\end{equation}
diverges. This will be consistent, provided
\begin{equation}
\nu_0 = 2 \mu \,.
\label{eq:U4}
\end{equation}
It should be clear that the parameter $U_0$ is of no particular
relevance in the unitary limit.  This is to be contrasted with the one
channel problem, where Feshbach bosons are absent and unitarity is
achieved by tuning the direct inter-fermion interaction, $U$.

In the unitary limit, the $T=0$ gap and number equations 
become
\begin{equation}
1 +U_c \sumk \frac{1} {2\Ek} =0, 
\label{eq:U5}
\end{equation}
and 
\begin{equation}
n^{tot} = 2n_b^0 + n = \frac{2\Delta_0^2}{g^2_0} + \sumk \left( 1-
\frac{\ek -\mu}{\Ek}\right) \,,
\label{eq:U7}
\end{equation}
where we have used the fact that at unitarity
\begin{equation}
n_b^0 = \frac{g^2 \Delta_0^2} {[(2\mu-\nu)U+g^2]^2} =
\frac{\Delta_0^2}{g^2_0} \,,
\label{eq:U6}
\end{equation}
and we define 
$\Delta_0 \equiv \tilde{\Delta}_{sc} (T=0)$.
Here $U_c$ is the critical coupling, given in Eq.~(\ref{eq:Uc}).
From these equations it follows that in this limit the values of
$\mu$ and $\Delta_0$ in the ground state are entirely
determined by the Feshbach coupling $g_0$.

\begin{figure}
\includegraphics[width=3.3in,clip]{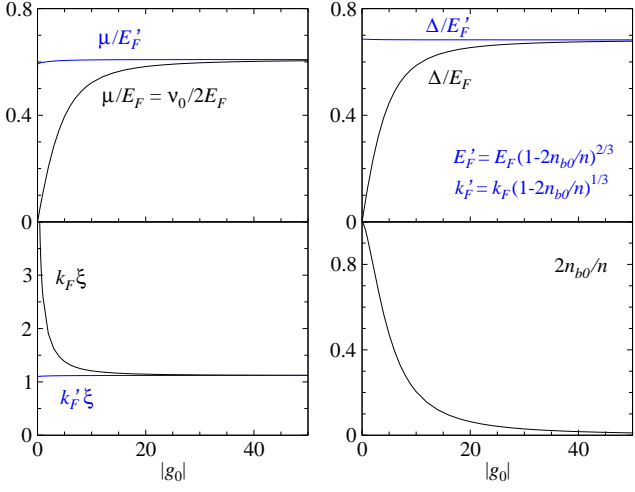}
\caption{(Color) Characteristics of the unitary limit at $T=0$ as a
  function of $|g_0|$ in the two-channel problem. Here $n_{b0}$ is the
  density of condensed Feshbach bosons. $E_F^\prime = k_F'^2/2m$ defines
  an effective Fermi energy of the fermionic component with a density $n
  = n^{tot} - 2n_{b0}$. After rescaling, $\mu/E_F^\prime$,
  $\Delta/E_F^\prime$, and $k_F'\xi$ are essentially a constant in
  $g_0$. This indicates that at unitarity, the system can be treated
  effectively as a two-fluid model at $T=0$. The units for $g_0$ are
  $E_F/k_F^{3/2}$.}
\label{fig:23x}
\end{figure}

Figure \ref{fig:23x} presents a plot of zero temperature properties
showing their dependence on $g_0$.  The two endpoints of the curves are
of interest.  For arbitrarily small but finite $g_0$ the unitary limit
corresponds roughly to $\mu \approx 0$.  Moreover, here the condensate
consists exclusively of Feshbach bosons. At large $g_0$, (which appear
to be appropriate to current experiments on $^6$Li and $^{40}$K), the
behavior at unitarity is closer to that of the one channel problem
\cite{JasonHo}.  Here one finds the so-called universal behavior with
parameters $\mu/E_F = 0.59$, and $\Delta/E_F = 0.69$ associated with the
simple crossover ground state and a contact potential
interaction\footnote{In our plots we find small numerical corrections
  associated with our non-infinite cutoff.}.
%

To what extent is the unitary limit of the two channel problem
universal? It clearly is not in an absolute sense.  However, there is a
kind of universality implicit in these $T=0$ calculations, as can be
seen from Fig.~\ref{fig:23x} if we take into account the fact that
there are a reduced number of fermions, as a result of the Bose
condensate.  We, thus, divide the two zero temperature energy scales by
an effective Fermi energy defined by $E_F^\prime= E_F (n/n^{tot})^{2/3}$
with $ n= n^{tot} -2n_b^0$.  Figure \ref{fig:23x} shows that as $g_0$
decreases, the effective or rescaled Fermi energy $E_F^\prime$ decreases
while the Bose condensate fraction increases.
The rescaled quantities $\mu/E_F^\prime$ and $\Delta/E_F^\prime$ are
universal, that is, they are essentially independent of $g_0$, as seen
in the Figure. A similar scaling can be done for the Cooper pair size
$\xi$. With a reduced effective Fermi momentum $k_F^\prime =
k_F(n/n^{tot})^{1/3}$, the dimensionless product $k_F^\prime \xi$ is
also independent of $g_0$. We can arrive at some understanding of this
modified universality, by referring back to Sec.~\ref{sec:5A}.  In
summary, the breakdown of universality seen in Fig.~\ref{fig:23x} is
to be associated with a reduced number of fermions, which, in turn
derives from the Bose condensate $n_b^0$.  Once $n_b^0$ is negligible,
the distinction between one and two channel models disappears.

\begin{figure}
\includegraphics[width=3.3in,clip]{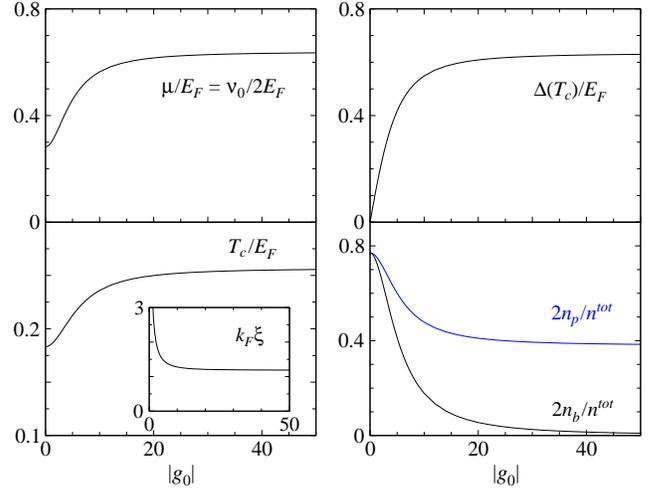}
\caption{(Color) Characteristics of the unitary limit at $T=T_c$ as a
  function of $|g_0|$ in the two channel problem. In contrast to the
  previous figure for $T=0$, there is no rescaling which would lead to
  universal behavior of $\mu$, $\Delta$ or $\xi$ with respect to $g_0$.
  More than $ 20\%$ of the system is composed of fermionic
  quasiparticles when $|g_0| \rightarrow 0$. Here $n_b$ and $n_p$ denote
  the density of noncondensed Feshbach bosons and of the sum of FB and
  fermion pairs, respectively. The units for $g_0$ are $E_F/k_F^{3/2}$.}
\label{fig:23z}
\end{figure}

Finite temperature properties, such as $T_c$ and $\Delta(T_c)$ can also
be calculated in the unitary limit, as a function of $g_0$. These are
plotted in Fig.~\ref{fig:23z}.  Just as for the zero temperature
energy scales, these parameters [along with $\mu(T_c)$] increase
monotonically with $g_0$.  However, at finite temperatures there are
noncondensed Feshbach bosons as well as noncondensed fermion pairs,
which interact strongly with each other.  These noncondensed
states are represented in the lower right inset. Plotted there is the
fraction of noncondensed bosons [$n_b(T_c)$] and sum of both
noncondensed fermion pairs and noncondensed bosons [$n_p(T_c)$] at
$T_c$.  These quantities correspond to the expressions shown in
Eqs.~(\ref{eq:62a}) and (\ref{eq:62b}), respectively.

These noncondensed particles are responsible for the fact that at $T_c$
a simple rescaling will not yield the universality found at $T=0$.
Moreover, these noncondensed bosons and fermion pairs influence
calculations of $T_c$, itself, via pseudogap effects which enter as
$\Delta(T_c)$. These pseudogap effects, which are generally ignored in
the literature, compete with superconductivity, so that $T_c$ is bounded
with its a maximum given by $\approx 0.25T_F$, just as seen in
Fig.~\ref{fig:18}, for the one channel problem.  Despite the fact that
there is no universality, even in the rescaled sense, at $T_c$ one can
associate the large $g_0$ limits with the behavior of the one channel
system which has been rather extensively studied in the unitary limit
\cite{JasonHo}.  This large $g_0$ limit seems to be reasonably
appropriate for the currently studied Feshbach resonances in $^6$Li and
$^{40}$K.  In this sense, these particular atomic resonances are well
treated by a one channel problem, near the unitary limit.

In summary, comparison between Fig.~\ref{fig:23x} and Figure
\ref{fig:23z} reveals that (i) at $T=0$, all Feshbach bosons and pairs
are condensed and the two (fermionic and bosonic) condensates are
relatively independent of each other. (ii) By contrast, at $T=T_c$,
fermionic quasiparticles, noncondensed fermion pairs and Feshbach bosons
all coexist and interact strongly with each other. This destroys the
simple scaling found at $T=0$.

\begin{figure}
\includegraphics[width=3.3in,clip]{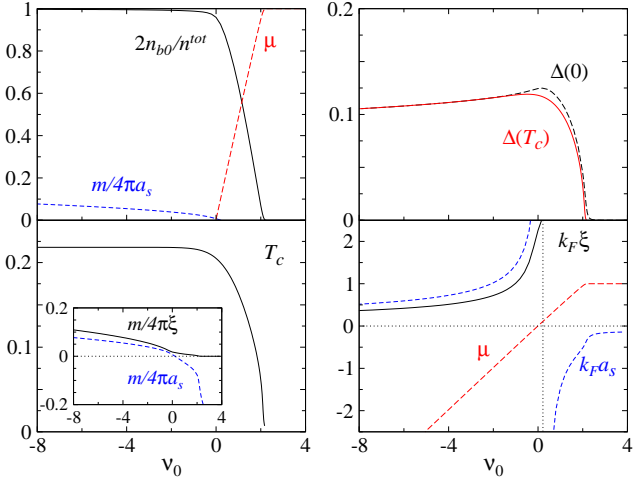}
\caption{(Color) Characteristics of the small $g_0$ limit as a function
  of $\nu_0$ in the two-channel model. Unless indicated otherwise the
  plots correspond to $T=0$. The molecular condensate fraction is very
  high and there is no maximum at unitarity. The gap $\Delta$ is very
  small ($\propto g_0$). Here we take $g_0 = -1 E_F/k_F^{3/2}$ and $U_0 =
  -3 E_F/k_F^3$. Note the very different scales for the abscissa when
  comparing with Figs.~\ref{fig:16} and \ref{fig:17}.}
\label{fig:24x}
\end{figure}

\subsubsection{Weak Feshbach Resonance Regime: $|g_0| = 0^+$}
\label{sec:sub1}

The effects of very small yet finite $g_0$ lead to a nearly bi-modal
description of crossover phenomena.  Figure \ref{fig:24x} presents the
behavior for very narrow Feshbach resonances. It shows that there is a
rather abrupt change as a function of magnetic
field (via $\nu_0$) from a regime in which the
condensate is composed only of Feshbach bosons, to one in which it
contains only fermion pairs.  Depending on the size of $g_0$, this
transition point (taken for concreteness to be where $2n_b^0/n^{tot} \approx
0.5$, for example) varies somewhat, but it seems to occur close to the
unitary limit and, importantly, for negative scattering lengths
\textit{i.e.}, on the BCS side. This should be contrasted with the large
$g_0$ case). This small $g_0$ regime of parameter space was explored by
Ohashi and Griffin \cite{Griffin}.  Indeed the behavior shown in the
figure is similar to the physical picture which Falco and Stoof
\cite{Stoof} presented.

This figure represents a natural evolution in the behavior already
indicated by Figs.~\ref{fig:16} and \ref{fig:17}.  Here, however, in
contrast to these previous figures the behavior of $T_c$ is essentially
monotonic, and the weak maximum and minimum found earlier has
disappeared. For this regime, the small pseudogap which can be seen in
the upper right panel arises predominantly from the contribution of
Feshbach bosons via Eqs.~(\ref{t-matrix_pg}) and (\ref{eq:delta_pg}).
At the same time it should be noted that the transition from mostly
molecular bosons to mostly fermion pairs in the condensate is
continuous.  While it may look sharp particularly for a plot with a
more extended range of $\nu_0$ than shown here, this closeup view gives
a more precise picture of a smooth crossover transition.

\begin{figure*}[t]
\centerline{\includegraphics[angle=0,width=6.0in,clip]{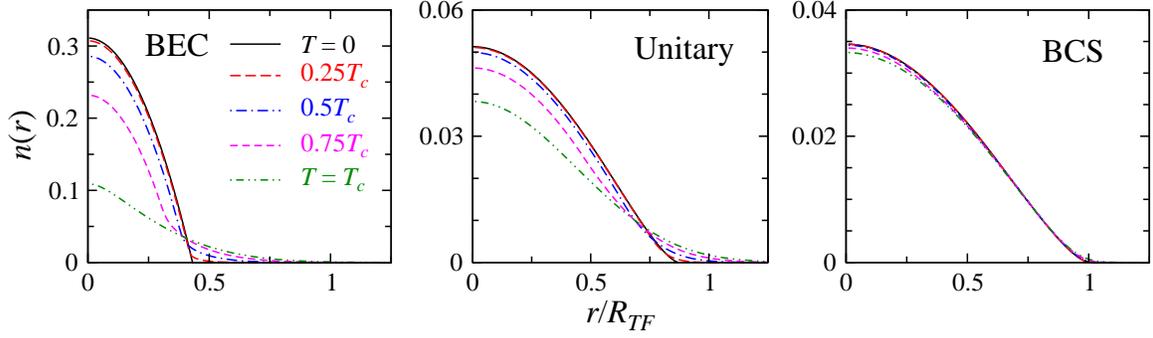}}  
\caption{(Color) Representative spatial profiles of fermions in a
  harmonic trap potential at various temperatures from $T_c$ to $T=0$ in
  the three indicated regimes. As the system evolves from the BCS to the
  deep BEC regime, the fermion cloud shrinks, and a bimodal distribution
  of the density develops below $T_c$. Note the different scales for
  the vertical axes. }
\label{fig:trap}
\end{figure*}

We end by noting that in this small $g_0$ regime, the system is far from
the one channel limit obtained when $g_0$ is identically 0, as shown in
Fig.~\ref{fig:18}.  It should be clear from the above discussion that
\textit{Feshbach bosons play a more important role in the regime of
  unitary scattering the smaller is $g_0$, provided that it does not
  strictly vanish}.

\subsection{Superfluidity in Traps}
\label{sec:5B}

Most experiments on fermionic superfluids are done in the presence of a
trap potential $V(r)$. For the most part, this is most readily included
at the level of the local density approximation (LDA).  In this section,
we summarize the self consistent equations for spherical traps
$V(r)=\frac{1}{2}m\omega^2 r^2$.  $T_c$ is defined as the highest
temperature at which the self consistent equations are satisfied
precisely at the center.  At a temperature $T$ lower than $T_c$ the
superfluid region extends to a finite radius $R_{sc}$. The particles
outside this radius are in a normal state.  Also important is the radius
$R_{\Delta}$ beyond which the excitation gap is effectively zero, to a
chosen level of numerical accuracy.

The self consistent equations (with simplified units $\hbar=1$, fermion mass
$m=\frac{1}{2}$, and Fermi energy $E_F=\hbar \omega (3
N)^{\frac{1}{3}}=1$, where $\omega$ is the trap frequency) are given in
terms of the Feshbach coupling constant $g$ and inter-fermion attractive
interaction $U$ by a gap equation
%
\begin{equation}
1+\left[U+\frac{g^2}{2 \mu-2 V(r)-\nu}\right] \sum_{\bf k} \frac{1-2
  f(\Ek)}{2 \Ek}=0\,\,. 
\label{gap_eq_trap}
\end{equation}
which characterizes
the excitation gap $\Delta(r)$ at positions $r$ such that
$\mu_{pair}(r) = 0$. Moreover, the pseudogap contribution to
$\Delta^2(T) = \tilde{\Delta}_{sc}^2(T) + \Delta_{pg}^2(T)$ is given by
\begin{equation}
\Delta_{pg}^2=\frac{1}{Z} \sum_{\bf q}\, b(\Omega_q -\mu_{pair})\,\,.
\label{eq:1}
\end{equation}
Here $b$ represents the usual
Bose function. The density of particles at radius $r$:
\begin{eqnarray}
n(r)&=& 2 n_b^0 + \frac{2}{Z_b} \sum_{\bf q} b(\Omega_q
-\mu_{boson})\nonumber\\ 
&&{}+ 2 \sum_{\bf k}\left [\vk^2 (1-f(\Ek))+\uk^2 f(\Ek)\right] \,\,.
\label{number_eq_trap:above}
\end{eqnarray}

The important quantity $\mu_{pair} =\mu_{boson}$ is identically zero in
the superfluid region $r < R_{sc}$, and must be solved for self
consistently at larger radii.  Here $M $ is the bare boson mass equal to
$2m$, and the various residues, $Z$ and $Z_b$ and pair dispersion
$\Omega_q$ are described in Section \ref{sec:2C}.

The transition temperature $T_c$ is numerically determined as follows:
(i) A chemical potential is assigned to the center of the trap $\mu(0)$.
The chemical potential at radius $r$ will then be $\mu(r)=\mu(0)-V(r)$.
(ii) The gap equation [Eq.~(\ref{gap_eq_trap})] and pseudogap equation
[Eq.~(\ref{eq:1})] are solved at the center (setting
$\Delta_{pg}=\Delta$) to find $T_c$ and $\Delta(0,T_c)$ (iii) Next the
radius $R_{\Delta}$ is determined.  (It can be defined as the point
where the gap has a sufficiently small value, such as
$\Delta(0,T_c)/1000$).  (iv) Following this, the pseudogap equation (Eq.
(\ref{eq:1})) is solved for $\Delta(r,T_c)$ up to the radius
$R_{\Delta}$. Then $n(r)$ is determined using
Eq.~(\ref{number_eq_trap:above}).  (v) Finally, $n(r)$ is integrated
over all space ($\int d^3{\mathbf r}n(r)$) and compared with the given
total number of fermions $N$.  The procedure is repeated until the
prescribed accuracy has been reached.  To extend this algorithm below
$T_c$ a similar approach is used which now includes an additional step
in which $R_{sc}$ is computed.

Figure \ref{fig:trap} shows a plot of the density profiles obtained from
solving the self consistent equations. Here the Thomas-Fermi radius,
called $R_{TF}$, is calculated for the non-interacting case. In the BCS
regime there is virtually no temperature dependence in the density
profile as expected from the small values for $\Delta$.  In the unitary
and BEC regimes as the system approaches $T_c$, $n(r)$ can be reasonably
well fit to a Gaussian. A bi-modality in the distribution is evident in
the BEC regime, thus providing a clear signature of the superfluid
phase.  However, in the PG or unitary it is more difficult to establish
superfluidity, since the profiles do not show a clear break at the
position where the condensate first appears \cite{JS5}.

These observations differ somewhat from other theoretical predictions in
the literature \cite{JasonHo,Chiofalo}. The differences between the
present theory and this other work can be associated with the inclusion
of noncondensed pairs or pseudogap effects.  These noncondensed pairs
smooth out the otherwise abrupt transition between the condensate and
the fermionic excitations. In contrast to earlier work \cite{Strinati5}
these studies, which are based on the standard crossover ground state
\cite{Leggett}, yield only monotonic density distributions.  Future
experiments will be needed to look for the secondary maxima predicted by
Strinati and co-workers \cite{Strinati4} in the BEC and unitary regimes.

\begin{figure}
 \includegraphics[width=2.8in,clip]{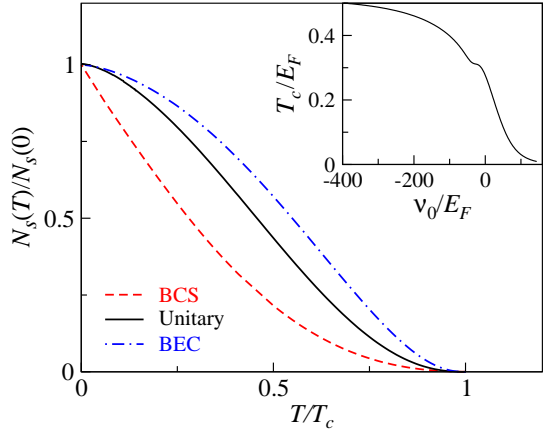}
 \caption{Behavior of superfluid density $N_s$ as a function of $T$ in
   the three different regimes. Note that the negative curvatures at low
   $T$ for the unitary and BEC cases are different from the linear
   behavior in the BCS case. In the BEC limit, $N-N_s \propto T^{3/2}$.
   In the inset we show $T_c$ as a function of magnetic detuning $\nu_0$
   in a trap.  We have taken $U_0 = -0.9E_F/k_F^3$ and $g_0 = - 35
   E_F/k_F^{3/2}$, based on experimental data for $^{40}$K.}
 \label{fig:traps2}
 \end{figure}

Finally, Fig.~\ref{fig:traps2} shows the behavior of $T_c$ (inset) in
a trap, together with the superfluid density (main figure).  The latter
is computed following the discussion in Sec.~\ref{sec:3}.  From the inset
one can read off that $T_c/E_F \approx 0.3$ at resonance, which is very
close to \cite{Strinati4}, or slightly smaller than 
other estimates \cite{Griffin3} in the literature.

The structure seen in the plot of $T_c$ for the trapped gas reflects the
minimum and maximum found for the homogeneous solution
(Fig.~\ref{fig:16}); these are related, respectively, to where $\mu
\approx 0$ and where unitary scattering occurs. The $T_c$ curve follows
generally the decrease in the density at the center of the trap from BEC
to BCS.  The temperature dependences of the (integrated over $r$)
superfluid density reflect the nature of excitations of the condensate.
The BEC limit has only bosonic, while the BCS has only fermionic
excitations.  The unitary case is somewhere in between.  The fermionic
contributions are apparent in the BCS regime, where $N_s$ decreases
almost linearly at low $T$. For an $s$-wave superfluid this reflects the
contributions from low $\Delta(r) \leq T$ regions, near the edge of the
trap.  These are present to a lesser degree in the unitary case.  The
bosonic contribution to $N_s$ and to thermodynamics (in both the BEC and
unitary regimes) is governed by the important constraint that
$\mu_{pair}(r)$ vanishes at positions $r$ where there is a co-existent
superfluid.  A similar constraint is imposed at the level of
Hartree-Fock-Bogoliubov theory for true bosons \cite{Griffin4}.  These
gapless pair excitations are responsible for a $T^{3/2}$ temperature
dependence in $N_s$.  In this way the low $T$ power law dependences are
similar to their analogues in the homogeneous case of Section
\ref{sec:6B}, much as found for true Bose gases \cite{Carr} in a trap.

As for the BEC regime there is an interesting observation to make about
the contribution to the density profile associated with the condensate.
Experimentalists generally assume a Gaussian form throughout the trap
for the noncondensed pairs. In this way they estimate the condensate
fraction. Here we find \cite{JS5} that this Gaussian form is not correct
in the superfluid region of the trap, and that this Gaussian
approximation will underestimate (by $\approx$ a factor of 1.5) the
condensate fraction in the BEC regime.

\subsection{Comparison Between Theory and Experiment: Cold Atoms}
\label{sec:updates}

As time progresses it will be important to assess the assumed ground
state wavefunction \cite{Leggett} and its two channel variant
\cite{Milstein,Griffin,JS3} in more detail. A first generation set of
measurements and theories have indicated an initial level of success in
addressing four experiments: RF pairing gap spectroscopy and collective
mode measurements, as well as more recently, the particle density
profiles \cite{JS5} and thermodynamics \cite{ThermoScience} of Fermi
gases in the unitary regime.  The first of these experiments
\cite{Grimm4} seem to provide some support for the fact that the
Bogoliubov excitations with dispersion, $E_{\bf k} \equiv \sqrt{ (\ek
  -\mu)^2 + \Delta^2 (T) }$ have a ``gap parameter" or pseudogap,
$\Delta(T)$ which is in general different from the superconducting order
parameter, as a consequence of noncondensed pairs at and above $T_c$.
T\"orm\"a and colleagues \cite{Torma1} have used the formalism outlined in
Section \ref{sec:2B}, to address RF spectroscopy.  Subsequently
\cite{Torma2} they have combined this analysis with an approximate LDA
treatment of the trap to make contact with RF experiments \cite{Grimm4}.
Important to their work is the fact that the edges of a trap have a free
atom character, where there is a vanishing $\Delta(r)$.

Additional support for the wavefunction comes from the collective mode
behavior near $T=0$.  The mode frequencies can be connected to an
equation of state \cite{Stringari2}. Numerical
studies \cite{Tosi,Heiselberg} of this equation for the standard
crossover ground state demonstrate a good fit to the measured breathing mode
frequencies \cite{Thomas2,Grimm3}.  Importantly, in the near-BEC limit
the frequencies decrease with increasing magnetic field; this is
opposite to the behavior predicted for true bosons \cite{Stringari2}, and
appears to reinforce the notion that for the magnetic fields
studied thus far, the ``BEC" regime should be associated with
composite rather than real bosons.

At $T=0$, calculations of the collective mode frequencies are based on
an equation of state.  Moreover, the zero temperature equation of state
associated with the mean field ground state can be computed analytically
in the near-BEC limit
\begin{equation}
\mu=-\frac{1}{2ma_s^2}+\frac{\pi a_sn}{m}-\frac{3(\pi
  na_s^3)^2}{4ma_s^2}+O(n^3), \nonumber
\label{Leggett}
\end{equation}
where the first two terms on the right hand side were previously discussed
following Eq.~(\ref{eq:7}). Importantly, the third term is responsible
for the observed agreement between theory and collective mode experiments,
in the near-BEC regime.

A serious shortcoming of the wavefunction ansatz has been emphasized in
the literature. This is associated with the ratio of the inter-boson to
the fermionic scattering length near the unitary regime, but on the BEC
side of the Feshbach resonance.  Experimentally the numbers are
considerably smaller \cite{Grimm2,Ketterle3} ($0.6$) than the factor of
2 predicted for the one channel model discussed below Eq.~(\ref{eq:7}).
They are consistent, however, with more exact few body calculations
\cite{Petrov}.  These few body calculations apply to a regime of
magnetic fields where $k_F a_s$ is smaller than about $0.3$ -$0.4$. This
seems to be close to the boundary for the ``quasi-BEC" regime which can
be accessed experimentally \cite{Grimm2} in $^6$Li.

It is interesting to repeat the $T=0$ many body calculations now with
the inclusion of FB, since these two channel effects begin to be
important at $k_F a_s$ smaller than around $0.2$ -$0.25$.  Here, one
sees \cite{JS3}, that the factor of 2 found in the one channel
calculations is no longer applicable, and the values of the ratio (of
bosonic to fermionic scattering lengths) are considerably smaller, and,
thus, tending in the direction which is more in line with experiment.
This reflects the decrease in the number of fermionic states which
mediate the interaction between molecular bosons.

In this way the mean field theory presented here looks promising. 
Nevertheless, uncertainties remain.  Current theoretical work
\cite{Torma2} on the RF experiments is limited to the intermediate
coupling phase, while experiments are also available for the near-BEC
and near-BCS regimes, and should be addressed theoretically. In
addition, the predictions of alternative approaches \cite{Strinati5}
which posit a different ground state will need to be compared with these
RF data.  The collective mode calculations \cite{Tosi,Heiselberg} which
have been performed have not yet considered the effects of Feshbach
bosons, which may change the behavior in the near-BEC regime.

It is, finally, quite possible that incomplete $T=0$ condensation will
become evident in future experiments. If so, an alternative wavefunction
will have to be contemplated. Indeed, it has been argued \cite{Holland2}
that the failure of the one channel model to match the calculations of
Petrov and co-workers \cite{Petrov} is evidence that there are serious
weaknesses.  At an experimental level, new pairing gap spectroscopies
appear to be emerging at a fairly rapid pace \cite{Jin5,BaymBruun} which
will further test these and subsequent theories.

\begin{figure}
\includegraphics[width=2.8in,clip]{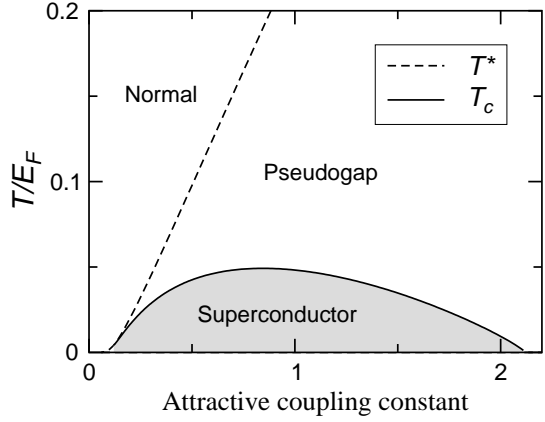}
\caption{Phase diagram for a quasi-two-dimensional (2D) $d$-wave
  superconductor on a lattice. Here the horizontal axis corresponds to
  the strength of attractive interaction $-U/4t$, where $t$ is the
  in-plane hopping matrix element.}
\label{fig:23}
\end{figure}

\section{Physical Implications: High $T_c$ Superconductivity}
\label{sec:6}

We begin with an important caveat that many of the experiments outlined
in this section can alternatively be addressed from a Mott physics point
of view \cite{LeeReview}.  Here we summarize the interpretation of these
experiments within BCS-BEC crossover theory with the hope that this
discussion will be of value to the cold atom community, and that
ultimately, it will help to establish precisely how relevant is the
BCS-BEC crossover scenario for understanding high temperature
superconductivity.

\subsection{Phase Diagram and Superconductor-Insulator Transition:
Boson Localization Effects}
\label{sec:6A}

The high $T_c$ superconductors are different from the ultracold
fermionic superfluids in one key respect; they are $d$-wave
superconductors and their electronic dispersion is associated with a
quasi-two dimensional tight binding lattice. In many ways this is not a
profound difference from the perspective of BCS-BEC crossover.  Figure
\ref{fig:23} shows a plot of the two important temperatures $T_c$ and
$T^*$ as a function of increasing attractive coupling.  On the left is
BCS and the right is PG. The BEC regime is not apparent. This is because
$T_c$ disappears before it can be accessed.  At the point where $T_c$
vanishes, $\mu/E_F \approx 0.8$.

\begin{figure}
\includegraphics[width=2.8in,clip]{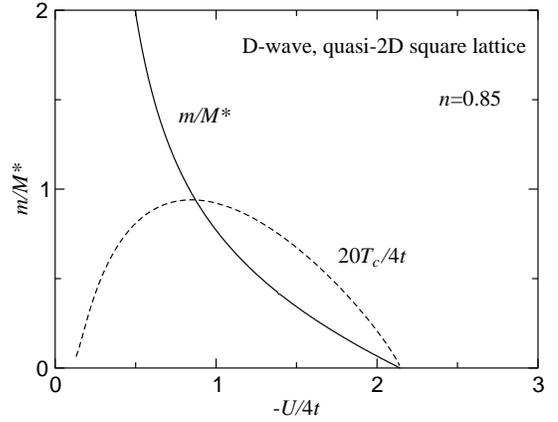}
\caption{Inverse pair mass (solid line) and $T_c$ (dashed line)  as a
  function of coupling strength $-U/4t$ for a $d$-wave superconductor on
  a quasi-2D square lattice, where $t$ is the in-plane hopping matrix
  element.  $T_c$ vanishes when the pair mass diverges, before the
  bosonic regime is reached.  The electron density is taken to be $n =
  0.85$ per unit cell, corresponding to optimal hole doping in the
  cuprate superconductors.}
\label{fig:30}
\end{figure}

A competition between increasing $T^*$ and $T_c$ is also apparent in
Fig.~\ref{fig:23}. This is a consequence of pseudogap effects.  There are
fewer low energy fermions around to pair, as $T^*$ increases.  It is
interesting to compare Fig.~\ref{fig:23} with the experimental phase
diagram plotted as a function of $x$ in Fig.~\ref{fig:5}. If one inverts
the horizontal axis (and ignores the AFM region, which is not addressed
here) the two are very similar. To make an association between the
coupling $U$ and the variable $x$, it is reasonable to simply fit
$T^*(x)$. We do not dwell on this last step here, since we wish to
emphasize crossover effects which are not complicated by ``Mott
physics''.

\begin{figure*}[t]
\includegraphics[width=4.5in,clip]{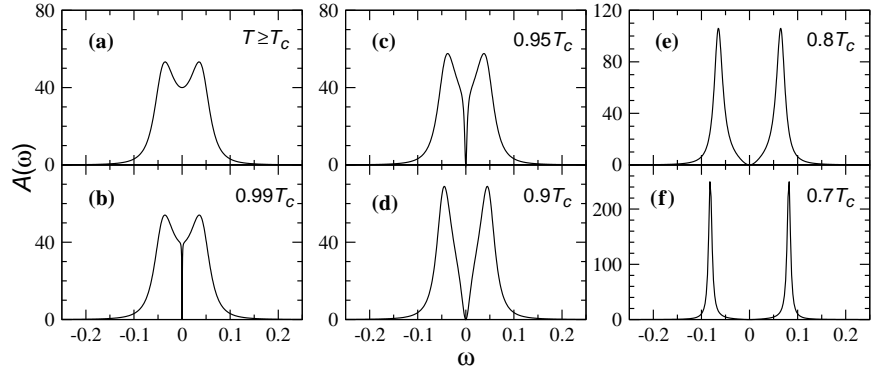}
\caption{Evolution of spectral functions of a pseudogapped
  superconductor with temperature. As soon as $T$ drops below $T_c$, the
  spectral peaks split up rapidly, signalling the onset of phase
  coherence.}
\label{fig:24}
\end{figure*}

Because of quasi-two dimensionality, the energy scales of the vertical
axis in Fig.~\ref{fig:23} are considerably smaller than their three
dimensional analogues.  Thus, pseudogap effects are intensified, just as
conventional fluctuation effects are more apparent in low dimensional systems.
This may be one of the reasons why the cuprates are among the first
materials to clearly reveal pseudogap physics.  Moreover, the present
calculations show that in a strictly $2D$ material, $T_c$ is driven to
zero, by bosonic or fluctuation effects.  This is a direct reflection of
the fact that there is no Bose condensation in $2D$.

Figure \ref{fig:30} presents a plot \cite{Chen1} of boson localization
effects associated with $d$-wave pairing.  A divergence in the effective
mass $m/M^*$ occurs when the coupling strength, or equivalently $T^*$
becomes sufficiently large.  Once the bosons localize the system can no
longer support superconductivity and $T_c$ vanishes.  It is generally
believed that a system of bosons with short range repulsion is either a
superfluid or a Mott insulator in its ground state \cite{MPFisher}.
Moreover, the insulating phase is presumed to contain some type of long
range order.  Interestingly, there are recent reports of ordered phases
\cite{Fu,Yazdani} in the very underdoped cuprates.  These appear on the
insulating side and possibly penetrate across the
superconductor-insulator boundary; they have been interpreted by some as
Cooper pair density waves \cite{Zhang3,Tesanovic1,Tesanovic2}.  The
observation of boson localization appears to be a natural consequence of
BCS-BEC theory; it should be stressed that it is associated with the
$d$-wave lattice case. Because $d$-wave pairs are more spatially
extended (than their $s$-wave counterparts) they experience stronger
Pauli repulsion \cite{Chen1} and this inhibits their hopping.

\subsection{Superconducting Coherence Effects}
\label{sec:6E}

The presence of pseudogap effects raises an interesting set of issues
surrounding the signatures of the transition which the high $T_c$
community has wrestled with, much as the cold atom community is doing
today.  For a charged superconductor there is no difficulty in measuring
the superfluid density, through the electrodynamic response. Thus one
knows with certainty where $T_c$ is.  Nevertheless, people have been
concerned about precisely how the onset of phase coherence is reflected
in thermodynamics, such as $C_V$ or in the fermionic spectral function.
One understands how phase coherence shows up in BCS theory, since the
ordered state is always accompanied by the appearance of an excitation
gap. When a gap is already well developed at $T_c$, how do these
signatures emerge?

\begin{figure*}
\centerline{\includegraphics{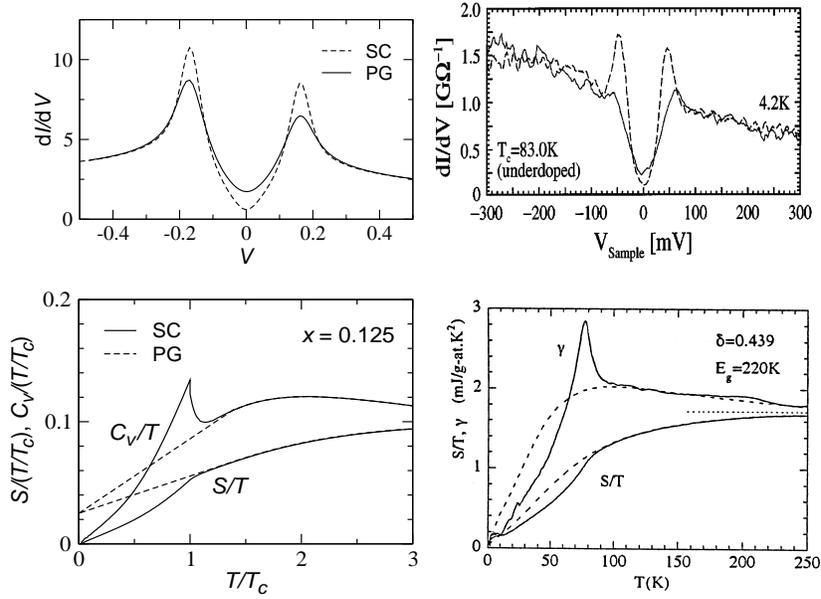}}
\caption{Superconducting state (SC) and extrapolated normal
  state (PG) contributions to SIN tunneling characteristics $dI/dV$
  (upper panels) and thermodynamics $S/T$ and $C_v/T$ (lower panels). We
  compare theoretical calculations (left) with experiments (right) on
  tunneling for BSCCO from \cite{Renner} and on specific heat for
  Y$_{0.8}$Ca$_{0.2}$Ba$_2$Cu$_3$O$_{7-\delta}$ (YBCO) from
  \cite{LoramPhysicaC}. The theoretical SIN curve is calculated for $T =
  T_c/2$, while the experimental curves are measured outside (dashed
  line) and inside (solid line) a vortex core.  The \textit{dashed and
    solid} curves in the upper left panel are for SC and PG state,
  respectively.  In both lower panels, the \textit{solid and dashed}
  curves indicate the SC and PG state, respectively.  }
\label{fig:25}
\end{figure*}

To address these coherence effects one has to introduce a distinction
between the self energy \cite{Chen4} associated with noncondensed and
condensed pairs. This distinction is blurred by the approximation of
Eq.~(\ref{eq:sigma3}).  Above, but near $T_c$, or at any temperature
below we now use an improved approximation \cite{Maly1,Maly2}
\begin{equation}
\Sigma_{pg} \approx \frac{\Delta^2}{\omega +\xi_{\bf k} +i\gamma}
\end{equation}
This is to be distinguished from $\Sigma_{sc}$ where the condensed pairs
are infinitely long-lived and there is no counterpart for $\gamma$.  The
value of this parameter, and even its $T$-dependence is not particularly
important, as long as it is nonzero.

Figure \ref{fig:24} plots the fermionic spectral function at $\xi_{\bf
  k} =0$, called $A(\omega)$, as the system passes from above to below
$T_c$. One can see in this figure that just below $T_c$, $A(\omega)$ is
zero at a point $\omega =0$, and that as temperature further decreases
the spectral function evolves smoothly into approximately two slightly
broadened delta functions, which are just like their counterparts in
BCS.  In this way there is a clear signature associated with
superconducting coherence. To compare with experimental data is somewhat
complicated, since measurements of the spectral function
\cite{arpesstanford,arpesanl} in the cuprates also reveal other higher
energy features (``dip, hump and kink"), not specifically associated
with the effects of phase coherence.  Nevertheless, this Figure, like
its experimental counterpart, illustrates that sharp gap features can be
seen in the spectral function, but only below $T_c$.

\begin{figure}
\centerline{\includegraphics[width=2.8in,clip]{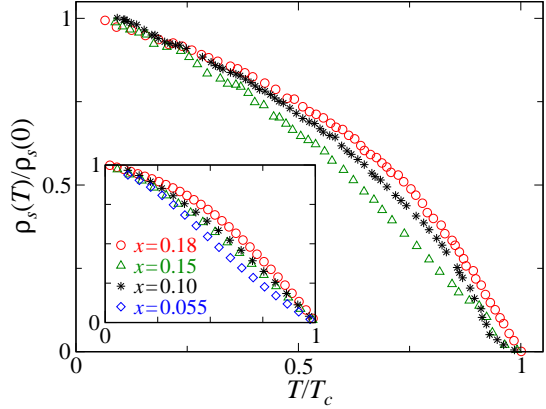}}
\caption{Comparison of normalized superfluid density $\rho_s/\rho_s(0)$
  versus $T/T_c$ for experiment (main figure) and theory (inset) in 
  high $T_c$ superconductors with variable doping concentration $x$. Both
  experiment and theory show a quasi-universal behavior with respect to
  doping. }
\label{fig:26}
\end{figure}

Analogous plots of superconducting coherence effects are presented in
Fig.~\ref{fig:25} in the context of more direct comparison with
experiment.  Shown are the results of specific heat and tunneling
calculations and their experimental counterparts
\cite{LoramPhysicaC,Renner}.  The latter measures, effectively, the
density of fermionic states.  Here the label ``PG" corresponds to an
extrapolated normal state in which we set the superconducting order
parameter $\Delta_{sc}$ to zero, but maintain the the total excitation
gap $\Delta$ to be the same as in a phase coherent, superconducting
state.  Agreement between theory and experiment is satisfactory. We
present this artificial normal state extrapolation in discussing
thermodynamics in order to make contact with its experimental
counterpart. However, it should be stressed that in zero magnetic field,
there is no coexistent non-superconducting phase. BCS theory is a rather
special case in which there are two possible phases below $T_c$, and one
can, thereby, use this coexistence to make a reasonable estimate of the
condensation energy. When Bose condensation needs to be accommodated,
there seems to be no alternative ``normal" phase below $T_c$.

\begin{figure}
\centerline{\includegraphics[width=2.9in,clip]{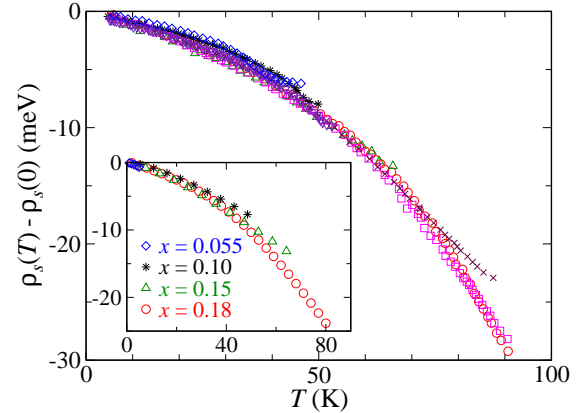}}
\caption{Offset plot of superfluid density $\rho_s$ comparing experiment
  (main figure) with theory (inset). Both experiment and theory show the
  same quasi-universal slope $d\rho_s/dT$ for different hole 
  concentrations $x$.}
\label{fig:27}
\end{figure}

\subsection{Electrodynamics and Thermal Conductivity}
\label{sec:6F}

In some ways the subtleties of phase coherent pairing are even more
perplexing in the context of electrodynamics.  Figure \ref{fig:10}
presents a paradox in which the excitation gap for fermions appears to
have little to do with the behavior of the superfluid density.  To
address these data \cite{JS1} one may use the formalism of
Sec.~\ref{sec:3A}.  One has to introduce the variable $x$ (which
accounts for Mott physics) and this is done via a fit to $T^*(x)$ in the
phase diagram.  Here for Fig.~\ref{fig:27} it is also necessary to fit
$\rho_s(T=0,x)$ to experiment, although this is not important in
Fig.~\ref{fig:26}.  The figures show a reasonable correspondence
\cite{JS1} with experiment. The paradox raised by Fig.~\ref{fig:10} is
resolved by noting that there are bosonic excitations of the condensate,
as in Eq.~(\ref{Lambda_BCS_Eq2}) and that these become more marked with
underdoping, as pseudogap effects increase.  In this way $\rho_s$ does
not exclusively reflect the fermionic gap, but rather vanishes
``prematurely" before this gap is zero, as a result of pair excitations.

The optical conductivity $\sigma(\omega)$ is similarly
modified \cite{Timusk,Iyengar}.  Indeed, there is an intimate relation
between $\rho_s$ and $\sigma(\omega)$ known as the f-sum rule.  Optical
conductivity studies in the literature, both theory and experiment, have
concentrated on the low $\omega ,T$ regime and the interplay between
impurity scattering and $d$ wave superconductivity \cite{Berlinsky}.
Also of interest are unusually high $\omega$
tails \cite{vanderMarel,Bontemps2} in the real part of $\sigma(\omega)$
which can be inferred from sum-rule arguments and experiment.  Figure
\ref{fig:10} raises a third set of questions which pertain to the more
global behavior of $\sigma$.  In the strong PG regime, where $\Delta$
has virtually no $T$ dependence below $T_c$, the BCS-computed
$\sigma(\omega)$ will be similarly $T$-independent. This is in contrast
to what is observed experimentally where $\sigma(\omega)$ reflects the
same $T$ dependence as in $\rho_s(T)$, as dictated by the f-sum rule.

One may deduce a consequence of this sum rule, based on
Eq.~(\ref{Lambda_BCS_Eq2}). If we associate the fermionic contributions
with the first term in square brackets in this equation, and the bosonic
contributions with the second.  Note that the fermionic transport terms
reflect the full $\Delta$ just as do the single particle properties.
(This may be seen by combining the superconducting and Maki Thompson or
PG diagrams in Appendix \ref{App:Ward}).  We may infer a sum rule
constraint on the bosonic contributions, which vanish as expected in the
BCS regime where $\Delta_{pg}$ vanishes.  We write
\begin{eqnarray}
\frac{2}{\pi} \int_{0}^{\infty}\:d\Omega\:
\sigma^{bosons}(\Omega,T)
 = \frac{\Delta_{pg}^2}{\Delta^2}
 \left(\frac{n_s}{m}\right)^{BCS}\!\!\!\!(T)\:.
 \label{boson_weight}
 \end{eqnarray}
 The bosonic contributions can be determined most readily from a
 framework such as time dependent Ginzburg Landau theory, which
 represents rather well the contributions from the Aslamazov-Larkin
 diagrams, discussed in Appendix \ref{App:Ward}.  
 The bosons make a
 maximum contribution at $T_c$.  

\begin{figure}
\includegraphics[width=3.2in,clip]{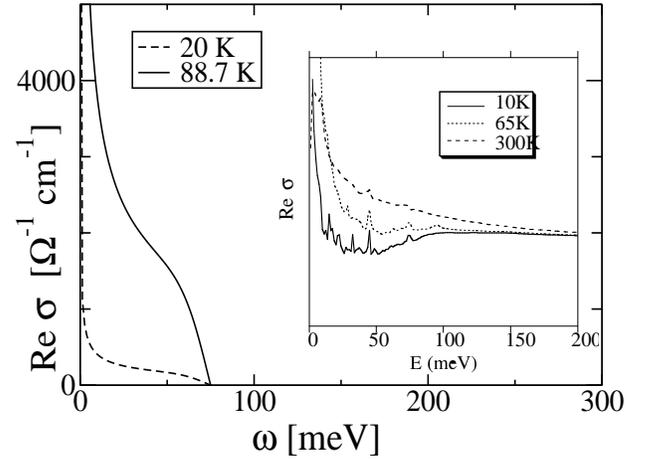}
\caption{Real part of ac conductivity at low and high T below $T_c$.
  Here pseudogap effects, which insure that $\Delta(T) \approx constant$
  below $T_c$, require that one go beyond BCS theory to explain the data
  shown in the inset \cite{Puchkov}. In the theory, inelastic and other
  effects which presumably lead to the large constant background term
  observed in the data have been ignored. }
\label{fig:28}
\end{figure}

The resulting optical conductivity \cite{Iyengar} in the superconducting
state is plotted in Fig.~\ref{fig:28} below.  The key feature to note is
the decrease in width of the so-called Drude peak with decreasing
temperature. In the present picture this pseudogap effect can be
understood as coming from the contribution of bosonic excitations of the
condensate, which are present at $T_c$, but which disappear at low $T$.
This behavior is reasonably consistent with what is reported
experimentally \cite{Timusk}, where ``the effect of the opening of a
pseudogap is a narrowing of the coherent Drude peak at low frequency".
The calculations leading to Fig.~\ref{fig:28}, indicate that by 20~K
most of the conductivity is fermionic except for a high frequency tail
associated with the bosons. This tail may be responsible in part for the
anomalously high $\omega$ contributions to $\sigma$ required to satisfy
the sum rule \cite{vanderMarel,Bontemps2}.
 
Although the normal state calculations have not been done in detail,
once the temperature is less than $T^*$, pseudogap effects associated
with the fermionic contribution will also lead to a progressive
narrowing in the Drude peak with decreasing $T$, down to $T_c$, due to
the opening of an excitation gap. This mechanism (which is widely
ascribed to \cite{Timusk} in the literature) reflects longer lived
electronic states, associated with the opening of this gap. However, it
cannot be applied to explain the progressive narrowing of the
conductivity peaks, once $\Delta(T)$ becomes temperature independent, as
is the case at low $T$.  In this way, one can see that pseudogap effects
show up in the optical conductivity of underdoped samples, both from the
fermionic contributions (near and below $T^*$) and the bosonic
contributions (below $T_c$) as a narrowing of the Drude peak.
 
 The data shown in the inset of Fig.~\ref{fig:28} reveal that the
 condensate is made from frequency contributions primarily up to
 $\approx 0.15$ eV.  These data pose a challenge for applying strict BCS
 theory to this problem, since one consequence of a temperature
 independent $\Delta$ in the underdoped phase is the absence of $T$
 dependence in the optical conductivity. This is the same challenge as
 posed by Fig.~\ref{fig:10}.  Strikingly there is no explicit ``gap
 feature" in the optical conductivity in the normal state.  Such a gap
 feature, that is a near vanishing of $\sigma$ for an extended range of
 frequencies, will occur in models where the pseudogap is unrelated to
 superconductivity; it is not seen experimentally, thus far

By contrast with the electrical conductivity, the thermal conductivity
is dominated by the fermionic contributions at essentially all $T$.
This is because the bosonic contribution to the heat current (as in
standard TDGL theory \cite{Larkinreview}), is negligibly small,
reflecting the low energy scales of the bosons.  Thermal conductivity,
experiments \cite{Taillefer} in the high $T_c$ superconductors provide
some of the best evidence for the presence of fermionic $d$-wave
quasiparticles below $T_c$. In contrast to the ac conductivity, here one
sees a universal low $T$ limit \cite{Lee1993}, and there is little to
suggest that something other than conventional $d$-wave BCS physics is
going on here. This cannot quite be the case however, since in the
pseudogap regime, the temperature dependence of the fermionic excitation
gap is highly anomalous, as shown in Fig.~\ref{fig:Delta_Deltasc},
compared to the BCS analogue.

\subsection{Thermodynamics and Pair-breaking Effects}
\label{sec:6B}

The existence of noncondensed pair states below $T_c$ affects
thermodynamics, in the same way that electrodynamics is affected, as
discussed above. Moreover, one can predict \cite{Chen3} that the $q^2$
dispersion will lead to ideal Bose gas power laws in thermodynamical and
transport properties.  These will be present in addition to the usual
power laws or (for $s$-wave) exponential temperature dependences
associated with the fermionic quasiparticles.  Note that the $q^2$
dependence reflects the spatial extent $\xi_{pg}$, of the composite
pairs , and this size effect has no natural counterpart in true Bose
systems.  It should be stressed that numerical calculations show that
these pair masses, as well as the residue $Z$ are roughly $T$
independent constants at low $T$. As a result, Eq.~(\ref{eq:81}) implies
that $\Delta_{pg}^2 = \Delta^2(T) -\Delta_{sc}^2(T) \propto T^{3/2}$.

The consequences of these observations can now be listed  \cite{Chen3}.
For a quasi-two dimensional system, $C_v/T$ will appear roughly constant
at the lowest temperatures, although it vanishes strictly at $T=0$ as
$T^{1/2}$. The superfluid density $\rho_s(T)$ will acquire a $T^{3/2}$
contribution in addition to the usual fermionic terms. By contrast, for
spin singlet states, there is no explicit pair contribution to the
Knight shift. In this way the low $T$ Knight shift reflects only the
fermions and exhibits a scaling with $T/\Delta(0)$ at low temperatures.
Experimentally, in the cuprates, one usually sees a finite low $T$
contribution to $C_v/T$.  A Knight shift scaling is seen.  Finally, also
observed is a deviation from the predicted $d$-wave linear in $T$ power
law in $\rho_s$.  The new power laws in $C_v$ and $\rho_s$ are
conventionally attributed to impurity effects, where $\rho_s \propto
T^2$, and $C_v/T \propto const$. Experiments are not yet at a stage to
clearly distinguish between these two alternative explanations.

Pair-breaking effects are viewed as providing important insight into the
origin of the cuprate pseudogap.  Indeed, the different pair-breaking
sensitivities of $T^*$ and $T_c$ are usually proposed to support the
notion that the pseudogap has nothing to do with superconductivity.  To
counter this inference, a detailed set of studies was
conducted, (based on the BEC-BCS scenario), of pair-breaking in the
presence of impurities \cite{Chen-Schrieffer,Kao4} and of magnetic
fields \cite{Kao3}. These studies make it clear that the superconducting
coherence temperature $T_c$ is far more sensitive to pair-breaking than
is the pseudogap onset temperature $T^*$. Indeed, the phase diagram of
Fig.~\ref{fig:23} which mirrors its experimental counterpart, shows the
very different, even competing nature of $T^*$ and $T_c$, despite the
fact that both arise from the same pairing correlations.

\subsection{Anomalous Normal State Transport: Nernst
and Other Precursor Transport Contributions}
\label{sec:6c}

Much attention is given to the anomalous behavior of the Nernst
coefficient in the cuprates \cite{Nernst}.  This coefficient is rather
simply related to the transverse thermoelectric coefficient
$\alpha_{xy}$ which is plotted in Fig.~\ref{fig:11c}.  In large part,
the origin of the excitement in the literature stems from the fact that
the Nernst coefficient behaves smoothly through the superconducting
transition. Below $T_c$ it is understood to be associated with
superconducting vortices. Above $T_c$ if the system were a Fermi liquid,
there are arguments to prove that the Nernst coefficient should be
essentially zero.  Hence the observation of a non-negligible Nernst
contribution has led to the picture of normal state vortices. 

Another way to view these data is that there are bosonic degrees of
freedom present in the normal state, as in the BCS-BEC crossover
approach. To describe this dynamics we begin with the TDGL scheme
outlined in Sec.~\ref{sec:TDGL}. This formalism must be modified somewhat
however.  When ``pre-formed" pairs are present at temperatures $T^*$
high compared to $T_c$, the classical bosonic fields of conventional
TDGL must be replaced by their quantum counterparts \cite{Tan}.  We
investigate this fairly straightforward extension of TDGL in the next
sub-section, and subsequently explore the resulting implications for
transport.

\subsubsection{Quantum Extension of TDGL}

To make progress on the quantum analogue of TDGL we study a Hamiltonian
describing bosons coupled to a particle reservoir of fermion pairs. The
latter are addressed in a quantum mechanical fashion. Our treatment of
the reservoir has strong similarities to the approach of Caldeira and
Leggett \cite{Caldeira}.  These bosons are in the presence of an
electromagnetic field, which interacts with the fermion pairs of the
reservoir as well, since they have charge $e^* = 2e$.  For simplicity we
assume that the fermions are dispersionless.  This Hamiltonian is given
by
\begin{multline} \label{eq:H}
 H=\sum_{\vect{l}\vect{m}}\varepsilon_{\vect l\vect m}\psi_{\vect{l}}
 ^\dagger(t)\exp
 \bigl(-ie^*C_{\vect l\vect m}(t)\bigr)\psi_{\vect{m}}(t)\\
 +\sum_{\vect l}e^* \phi\psi^\dagger\psi
 +\sum_{i\vect l}\Bigl\{(a_i+e^*\phi)w_i^\dagger w_i+\eta_i\psi^\dagger w_i
 +\eta_i^*w_i^\dagger\psi\Bigr\}\\
+\sum_{i\vect l}\Bigl\{(b_i-e^*\phi)v_i^\dagger v_i+\zeta_i\psi^\dagger v_i
^\dagger
+\zeta_i^*v_i\psi\Bigr\}.
\end{multline}

Here $\varepsilon_{\vect l\vect m}$ is the hopping matrix element for
the bosons.  $C_{\vect l\vect m}(t)$
is the usual phase factor associated with the vector potential.
Annihilation operators for the reservoir, $w_i$ and $v_i$ (with
infinitesimal coupling constants $\eta_i$ and $\zeta_i$), represent
positive and negative frequencies respectively and need to be treated
separately.  The energies $a_i$ and $b_i$ of the reservoir are both
positive.

The bosonic propagator $T(\kw)$ can be exactly computed from the
equations of motion. Here $A(\kw)=\realpart 2iT(\kw)$ is the boson
spectral function, and
\begin{equation}\label{eq:vertex}
T(\kw) \equiv \Bigl(\omega-\varepsilon(\vect k)-\Sigma_1(\omega)
+\frac{i}{2}\Sigma_2(\omega)\Bigr)^{-1}
\end{equation}
We have
\begin{equation}\label{eq:Sigma2}
\Sigma_2(\omega)\equiv 2\pi\sum_i\lvert\eta_i\rvert^2\delta(\omega-a_i)
 -2\pi\sum_i\lvert\zeta_i\rvert^2\delta(\omega+b_i),
  \end{equation}
with $\Sigma_1(\omega)$ defined by a Kramers Kronig transform.

The reservoir parameters $a_i$, $b_i$, $\eta_i$ and $\zeta_i$ which are
of no particular interest, are all subsumed into the boson self energy
$\Sigma_2(\omega)$. From this point forward we may ignore these
quantities in favor of the boson self energy.  We reiterate that this
theory makes an important simplification, that the reservoir pairs have
no dispersion. \textit{For this case one can solve for the exact
  transport coefficients}.

For small, but constant magnetic and electric fields and thermal
gradients we obtain the linearized response.  For notational
convenience, we define $v_{ab}(\vect k)\equiv \frac{\partial^2}{\partial
  k_a \partial k_b}\varepsilon(\vect k)$.
Then we can write a few of the in-plane transport coefficients which
appear in Eqs.~(\ref{eq:tan1}) and (\ref{eq:tan2}) as
\begin{equation}
\label{eq:tan3}
\sigma =-\frac{e^{*2}}{2}\int\measure v_xv_xA^2({\bf k}, \omega)
\frac{\partial b(\omega)}{\partial\omega}
 \end{equation}
and
\begin{multline}
\label{eq:tan4}
 \alpha_{xy }=
  -\frac{e^{*2}B_{z}}{6T}\int\measure v_xv_{x}v_{yy}
  \omega A^3({\bf k} , \omega)
  \frac{\partial b(\omega)}{\partial\omega}\,.
 \end{multline}

These equations naturally correspond to their TDGL
counterparts (except for different phenomenological
coefficients) when $T \approx T_c$. 

\begin{figure}
\includegraphics[width=3in]{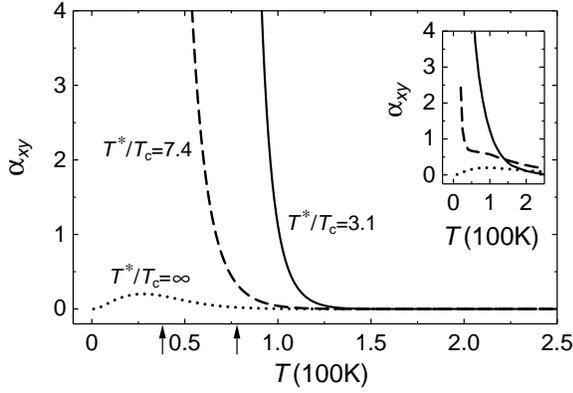}
\caption{Calculated transverse thermoelectric response,
  which appears in the Nernst coefficient. Inset plots experimental
  \cite{Nernst} data.  Both the theoretical and experimental values of
  $\alpha_{xy}$ are normalized by the value of $\alpha_{xy}$ for
  an optimally doped sample at $2T_c$, $0.0232
  (B/\mathrm{T})\;\mathrm{V}/(\mathrm{K\Omega m})$.}
\label{fig:33}
\end{figure}

\subsubsection{Comparison with Transport Data}

The formalism of the previous sub-section and in particular,
Eqs.~(\ref{eq:tan3}) and (\ref{eq:tan4}), can be used to address
transport data within the framework of BCS-BEC crossover.  The results
for $\alpha_{xy}$, which relates to the important Nernst effect, are
plotted in Fig.~\ref{fig:33} with a subset of the data plotted in the
upper right inset.  It can be seen that the agreement is reasonable. In
this way a ``pre-formed pair" picture is, at present, a viable
alternative to ``normal state vortices".

\begin{figure}
\centerline{\includegraphics[angle=0,width=3.3in,height=2.0in,clip]{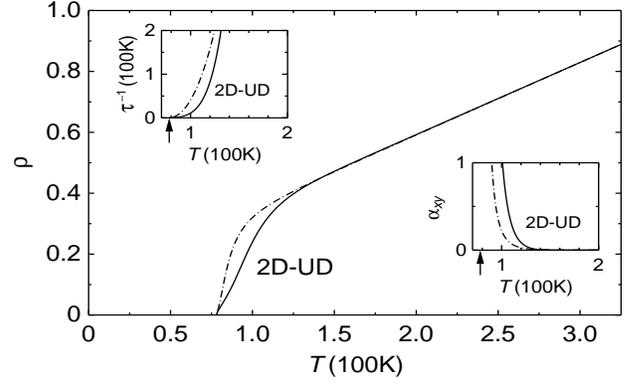}}
\caption{ 
  dc resistivity $\rho$ (normalized by $\rho_Q\equiv s/e^2$) vs. $T$,
  for quasi-2D underdoped (UD) samples with two different forms for
  $\mu_\text{pair}$. These are shown plotted in the left inset as
  $\tau^{-1} \equiv -\mu_\text{pair}/\mbox{Im} \gamma$.  The right inset
  shows corresponding $\alpha_{xy}$, and arrows indicate $T_c$.}
\label{fig:Tan4}
\end{figure}

Within the crossover scenario, just as in TDGL, there are strong
constraints on other precursor transport effects: paraconductivity,
diamagnetism and optical conductivity are all indirectly or directly
connected to the Nernst coefficient, in the sense that they all derive
from the same dynamical equation of motion for the bosons. Analogous
inferences have been made \cite{Larkin}, based on similar bosonic
generalizations of TDGL.

The calculated behavior of a number of these other properties is
summarized \cite{Tan} in Figs.~\ref{fig:Tan4} and \ref{fig:Tan5} below,
which indicate both quasi-two dimensional and more three dimensional
behavior.  In the plots for the resistivity, $\rho$, a fairly generic
\cite{Leridon} linear-in-$T$ contribution \footnote{In reality, one
  should include an additional temperature dependence from pseudogapped
  fermions which presumably has a small upturn, relative to a linear
  curve, as $T$ decreases.} from the fermions has been added, as a
background term.  A central theme of concern in the literature is
whether one can understand the behavior of $\rho$ simultaneously with
that of $\alpha_{xy}$.  Experiments \cite{Leridon,Watanabe} indicate
that there is a small feature corresponding to $T^*$ at which $\rho$
deviates from the background linear fermionic term, but that the
deviation is not noticeably large except in the immediate vicinity of
$T_c$. This is in contrast to the behavior of $\alpha_{xy}$ which shows
a significant high temperature onset.  A key observation from the
theoretical calculations shown in the figures is that the onset
temperatures for the Nernst coefficient and the resistivity are rather
different with the former being considerably higher than the latter.
Moreover all onsets appear to be significantly less than $T^*$.
Qualitative agreement with experiment is not unreasonable.  However, a
more quantitative comparison with experiment has not been established,
since it depends in detail on a phenomenological input: the temperature
dependence of the pair chemical potential $\mu_\text{pair}$.

Another important feature of these plots is the fact that there is
virtually no diamagnetic precursor, as was also found to be the case in
simple TDGL. To see this, note that the magnetization can be written
\cite{Schmid,Tan}
\begin{equation}
\frac{\mu_0 M}{B}=-\frac{4}{3}\frac{\alpha k_B T}
{\hbar c}\frac{\xi_{ab}^2}
{s\sqrt{1+(2\xi_c/s)^2} }, \nonumber
\end{equation}
where $\alpha$ is the fine structure constant, $c$ is the speed of
light, $s$ is the interlayer spacing, and $\xi_{ab}$ ($\xi_c$) are
coherence lengths in the direction parallel (perpendicular) to the
planes.  From this result we can estimate the order of magnitude of the
diamagnetic susceptibility due to preformed pairs. For a typical high
$T_c$ superconductor, $T\sim 50\mathrm{K}$, so that
$\lambda\equiv\frac{hc}{k_BT} \sim 3\times 10^{-4}\mathrm{m}$, and
$\lambda/\alpha\sim 4\times 10^{-2}\mathrm{m}$. Because the coherence
lengths and the interlayer spacing are on the nanometer length scale,
precursor effects in the diamagnetic susceptibility are of the order
$10^{-6}$, and thus extremely small, as is consistent with experiment
\cite{Ong2}.

In order to make the BCS-BEC scenario more convincing it will be
necessary to take these transport calculations below $T_c$.  This is a
project for future research and in this context it will ultimately be
important to establish in this picture how superconducting state
vortices are affected by the persistent pseudogap. Magnetic field
induced magnetic order which has been associated with vortices
\cite{Lake} will, moreover, need to be addressed.  The analogous
interplay of vortices and pseudogap will also be of interest in the
neutral superfluids.

\begin{figure}
\centerline{\includegraphics[angle=0,width=3.3in,height=2.0in]{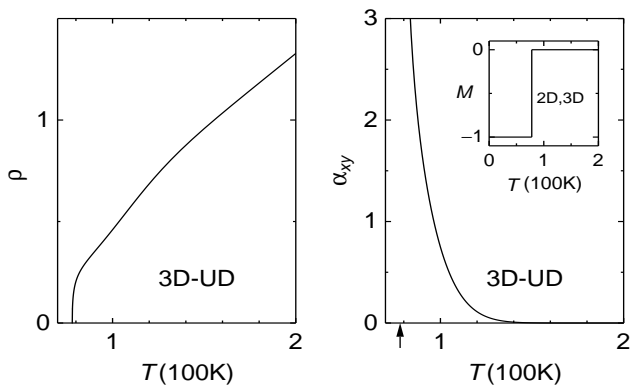}}
\caption{ 
  dc resistivity and $\alpha_{xy}$, for an underdoped, more
  three-dimensional, system like YBCO. The arrow indicates $T_c$. The
  magnetization $M$ is plotted as the inset, and is independent of
  dimensionality.}
\label{fig:Tan5}
\end{figure}

\section{Conclusions}

In this Review we have summarized a large body of work on the subject of
the BCS-BEC crossover scenario. In this context, we have explored the
intersection of two very different fields: high $T_c$ superconductivity
and cold atom superfluidity.  Theories of cuprate superconductivity can
be crudely classified as focusing on ``Mott physics" which reflects the
anomalously small zero temperature superfluid density and ``crossover
physics", which reflects the anomalously short coherence length. Both
schools are currently very interested in explaining the origin of the
mysterious pseudogap phase. In this Review we have presented a case for
its origin in crossover physics.  The pseudogap in the normal state can
be associated with meta-stable pairs of fermions; a (pseudogap) energy
must be supplied to break these pairs apart into their separate
components.  The pseudogap also persists below $T_c$ in the sense that
there are non condensed fermion pair excitations of the condensate.
These concepts have a natural analogue in self consistent theories of
superconducting fluctuations, but for the crossover problem the width of
the ``critical region" is extremely large. This reflects the much
stronger-than-BCS attractive interaction.

It was not our intent to shortchange the role of Mott physics which will
obviously be of importance in our ultimate understanding of the
superconducting cuprates. There is, however, much in this regard which
is still uncertain associated with establishing the simultaneous
relevance and existence of spin-charge separation \cite{Anderson},
stripes \cite{KivelsonRMP}, and hidden order parameters \cite{Laughlin}.
What we do have in hand, though, is a very clear experimental picture of
an extremely unusual superconductor in which superconductivity seems to
evolve gradually from above $T_c$ to below. We have in this Review tried
to emphasize the common ground between high $T_c$ superconductors and
ultracold superfluids. These Mott issues may nevertheless, set the
agenda for future cold atom studies of fermions in optical lattices
\cite{Demler2}.

The recent discovery of superfluidity in cold fermion gases opens the
door to a new set of fascinating problems in condensed matter physics.
Unlike the bosonic system, there is no ready-made counterpart of
Gross-Pitaevskii theory. A new mean field theory which goes beyond BCS
and encompasses BEC in some form or another will have to be developed in
concert with experiment.  \textit{As of this writing, there are four
  experiments where the simple mean field theory discussed in this
  review is in reasonable agreement with the data}.  These are the
collective mode studies over the entire range of accessible magnetic
fields \cite{Tosi,Heiselberg}.  In addition in the unitary regime, RF
spectroscopy-based pairing gap studies \cite{Torma2}, as well as density
profile \cite{JS5} and thermodynamic studies \cite{ThermoScience} all
appear to be compatible with this theory.

The material in this Review is viewed as the first of many steps in a
long process.  It was felt, however, that some continuity should be
provided from another community which has addressed and helped to
develop the BCS-BEC crossover, since the early 1990's when the early
signs of the cuprate pseudogap were beginning to appear.  As of this
writing, it appears likely that the latest experiments on cold atoms
probe a counterpart to this pseudogap regime. That is, on both sides,
but near the resonance, fermions form in long lived metastable pair
states at higher temperatures than those at which they Bose condense.

\acknowledgments

We gratefully acknowledge the help of our many close collaborators over
the years: J. Maly, B. Jank\'o, I. Kosztin, Y.-J. Kao and A. Iyengar.
We also thank our colleagues T.~R. Lemberger, B.~R.  Boyce, J.~N.
Milstein, M.~L. Chiofalo and M.J. Holland, and most recently, J.~E.
Thomas, J. Kinast, and A. Turlapov.  This work was supported by
NSF-MRSEC Grant No.~DMR-0213765 (JS,ST and KL), NSF Grant No.~DMR0094981
and JHU-TIPAC (QC).

\appendix
\section{Derivation of the $T=0$ variational conditions}
\label{App:Var}

The ground state wavefunction is given by
\begin{equation}
\bar{\Psi}^0=e^{-\lambda^2/2+\lambda b_0^{\dagger}}|0\rangle \otimes
 \Pi_{\bf k}(u_{\bf k}+v_{\bf k}a^{\dagger}_{\bf k}a^{\dagger}_{\bf -k})
 |0\rangle
\end{equation}
Thus one has three variational parameters $\lambda,u_{\bf k},v_{\bf k}$
with normalization condition $u_{\bf k}^2+v_{\bf k}^2=1$. We can define
$\theta_{\bf k}$ such that $\uk=\sin \theta_{\bf k}$ and $\vk=\cos
\theta_{\bf k}$, so that we have only two variational parameters,
$\lambda$ and $\theta_{\bf k}$.

We then have
\begin{eqnarray}
\langle \bar{\Psi}^0|H-\mu N|\bar{\Psi}^0\rangle &=& (\nu -2 
\mu)\lambda^2+
2 \sum_{\bf k}(\epsilon_{\bf k}-\mu)v_{\bf k}^2  \nonumber \\
&+& 2U \sum_{\bf k,k'}\uk \vk
u_{\bf k'}v_{\bf k'}
+2 g \lambda \sum_{\bf k}\uk \vk\;,
\nonumber
\end{eqnarray}
where we have chosen $\phik=1$ in Eq.~(\ref{eq:0c}) for simplicity.

Let us define
\begin{equation}
 \Delta_{sc}=-2 U \sum_{\bf k}\uk\vk.
\end{equation}
The variational conditions are
\begin{equation}
\frac{\delta \langle H-\mu N\rangle}{\delta \lambda}=0
\end{equation}
and
\begin{equation}
\frac{\delta \langle H-\mu N\rangle}{\delta \theta_{\bf k}}=0
\end{equation}
These conditions yield two equations: a relationship between the
two condensates
\begin{equation}
\frac{g
\lambda}{\Delta_{sc}}=\frac{1}{U}\frac{g^2}{\nu-2 \mu}
\end{equation}
and the $T=0$ gap equation:
\begin{equation}
1+(U+\frac{g^2}{2 \mu-\nu})\sum_{\bf k}\frac{1}{2
\Ek}=0\;.
\end{equation}

Similarly, we can write an equation for the total number of fermions
\begin{eqnarray}
n^{tot}&=&\langle\bar{\Psi}^0|2 \sum_{\bf k}b^{\dagger}_{\bf k}b_{\bf 
k}+
2 \sum_{\bf k}a^{\dagger}_{\bf k}a_{\bf k}|\bar{\Psi}^0\rangle=2
\lambda^2  \nonumber \\
&+&\sum_{\bf k}\left[1-\frac{\ek-\mu}{\Ek}\right]
\nonumber
\end{eqnarray}
where
\begin{equation}
\Ek=\sqrt{(\ek-\mu)^2+\tilde{\Delta}_{sc}^2},
\end{equation}
with $\tilde{\Delta}_{sc}=\Delta_{sc}-g\lambda$.  Thus we have obtained
the three equations which characterize the system at $T=0$.  It should
be clear that the variational parameter $\lambda$ is the same as the
condensate weight so that $\lambda = \phi _m \equiv\langle b_{{\bf q}=0}
\rangle$.

\section{Proof of Ward Identity Below $T_c$}
\label{App:Ward}

\begin{figure}
 \includegraphics[width=3in, clip]{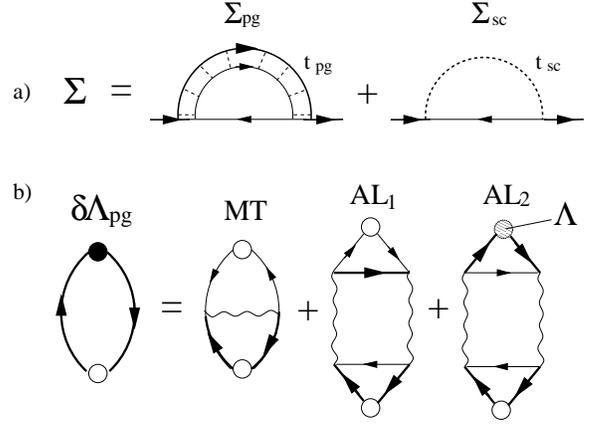} 
\caption{Self energy contributions (a) and response diagrams for 
the vertex correction corresponding to $\Sigma_{pg}$ (b).
   Heavy lines are for dressed, while light lines
  are for bare Green's functions. Wavy lines indicate $t_{pg}$. See text
   for details.}
\label{fig:full_jj}
\end{figure}

We now want to establish that the generalized Ward identity in
Eq.~(\ref{eq:Ward}) is correct for the superconducting state as well.
The vertex $\delta \Lambdak_{pg}$ may be decomposed into Maki-Thompson
($MT$) and two types of Aslamazov-Larkin ($AL_1$, $AL_2$) diagrams,
whose contribution to the response is shown here in Fig.
\ref{fig:full_jj}b.  We write
\begin{equation}
  \delta \Lambdak_{pg} \equiv \delta \Lambdak_{MT} + \delta
  \Lambdak_{AL}^1 + \delta \Lambdak_{AL}^2 (\bm{\Lambda})\:.
\end{equation}
Using conventional diagrammatic rules one can see that the $MT$ term has
the same sign reversal as the anomalous superconducting self energy
diagram. This provides insight into Eq.~(\ref{eq:79}).  Here, however,
the pairs in question are noncondensed and their internal dynamics (via
$t_{pg}$ as distinguished from $t_{sc}$) requires additional $AL_1$ and
$AL_2$ terms as well, which will ultimately be responsible for the
absence of a Meissner contribution from this normal state self energy
effect.

Note that the $AL_2$ diagram is specific to our $GG_0$ scheme, in which
the field couples to the full $G$ appearing in the $T$-matrix through a
vertex $\bm{\Lambda}$.  It is important to distinguish the vertex
$\bm{\Lambda}$ which appears in the $AL_2$ diagram from the full EM
vertex $\Lambda^{EM}$.

In particular, we have
\begin{equation}
\bm{\Lambda} = \bm{\lambda} + \delta\bm{\Lambda}_{pg} -
\delta\bm{\Lambda}_{sc} \:,
\label{Lambda_Eq2}
\end{equation}
where the sign change of the superconducting term (relative to
$\bm{\Lambda}^{EM}$) is a direct reflection of the sign 
in Eq.~(\ref{eq:79}).

We now show that there is a precise cancellation between the $MT$ and
$AL_1$ pseudogap diagrams at $Q=0$.  This cancellation follows directly
from the normal state Ward identity
\begin{equation}
Q\cdot\lambda(K,K+Q) = G_0^{-1}(K) - G_0^{-1}(K+Q)\:, 
\end{equation}
which implies 
\begin{equation}
Q\cdot[\delta \Lambda_{AL}^1 (K,K+Q)  + \delta \Lambda_{MT}(K,K+Q)] = 0
\label{AL1_cancellation}
\end{equation}
so that $\delta \Lambdak_{AL}^1(K,K) = - \delta \Lambdak_{MT}(K,K)$ is
obtained exactly from the $Q \rightarrow 0$ limit.  

Similarly, it can be shown that 
\begin{equation}
Q\cdot\Lambda(K,K+Q) = G^{-1}(K) - G^{-1}(K+Q)
\label{GWI}
\end{equation}
The above result can be used to infer a relation analogous to
Eq.~(\ref{AL1_cancellation}) for the $AL_2$ diagram, leading to
\begin{equation}
\delta \Lambdak_{pg}(K,K) = - \delta \Lambdak_{MT} (K,K)\:,
\end{equation}
which expresses this pseudogap contribution to the vertex entirely in
terms of the Maki-Thompson diagram shown in Fig.~\ref{fig:full_jj}b.
It is evident that $\delta\bm{\Lambda}_{MT}$ is simply the pseudogap
counterpart of $\delta\bm{\Lambda}_{sc}$, satisfying
\begin{equation}
- \delta\bm{\Lambda}_{MT}(K,K) = \frac{\partial\Sigma_{pg}(K)}{\partial
  \mb{k}} \:.
\label{Lambda_MT_Eq}
\end{equation}
Therefore, one observes that for $ T \le T_c$ 
\begin{equation} 
\delta\bm{\Lambda}_{pg}(K,K) = \frac{\partial\Sigma_{pg}(K)}{\partial
  \mb{k}} \:,
\label{Lambda_PG_Eq2}
\end{equation}
which establishes that Eq.~(\ref{eq:Ward}) applies to the
superconducting phase as well.  As expected, there is no direct Meissner
contribution associated with the pseudogap self-energy.

\section{Quantitative
Relation between BCS-BEC crossover and Hartree-TDGL}
\label{App:2}

In this appendix we make more precise the relation between
Hartree-approximated TDGL theory and the $T$-matrix of our $GG_0$ theory.
The Ginzburg-Landau (GL) free energy functional in momentum space is
given by \cite{Wilkins}
\begin{eqnarray}
F[\Psi]&=&N(0) \sum_{\bf q} |\Psi_{\bf q}|^2 (\epsilon +\xi
_1^2q^2)\nonumber \\  
&&{}+\frac{b_0}{2} \sum_{{\bf q}_i}
\Psi^*_{{\bf q}_1}\Psi^*_{{\bf q}_2}\Psi_{{\bf q}_3}\Psi_{{\bf q}_1+{\bf
    q}_2-{\bf q}_3} \:,\hspace{4ex} 
\label{free_energy}
\end{eqnarray} 
where $\Psi_{\bf q}$ are the Fourier components of the order parameter
$\Psi({\mathbf r},t)$, $\epsilon=\frac{T-T^*}{T^*}$, $\xi _1$ is the GL
coherence length, $b_0=[1/T^2 \pi^2]\frac{7}{8} \zeta(3)$, $T^*$ is the
critical temperature when feedback effects from the quartic term are
neglected, and $N(0)$ is the density of states at the Fermi level in the
normal state.  The quantity $\langle |\Psi_{{\bf q}}|^2\rangle$ is
determined self-consistently via
\begin{equation}
\langle |\Psi_{{\bf q}}|^2\rangle=\frac{\int D\Psi e^{-\beta F[\Psi]}
  |\Psi_{{\bf q}}|^2}{\int D\Psi e^{-\beta F[\Psi]}} \;, 
\label{psi_q}
\end{equation}
where $F[\Psi]$ is evaluated in the Hartree approximation \cite{JS1}.
Our self consistency condition can be then written as
\begin{equation}
\langle |\Psi_{{\bf q}}|^2\rangle=\frac{T}{N(0}\bigg[\epsilon +
b_0 \sum_{{\bf q}'} \langle |\Psi_{{\bf q}'0}|^2\rangle 
+\xi _1^2 q^2\bigg]^{-1}\;. 
\label{self_cons_psi_q}
\end{equation}
If we sum Eq.~(\ref{self_cons_psi_q}) over ${\bf q}$ and identify
$\sum_{\bf q} \langle |\Psi_{{\bf q}0}|^2\rangle$ with $\Delta
^2$, we obtain a self-consistency equation for the energy "gap" (or
pseudogap) $\Delta$ above $T_c$
\begin{equation}
 \Delta ^2=\sum_{\bf q} \frac{T}{N(0)}\bigg[\epsilon + 
b_0 \Delta ^2
+\xi _1^2 q^2\bigg]^{-1},
\label{gap_GL}
\end{equation}
\begin{equation}
\Delta ^2=\sum_{\bf q} \frac{T}{N(0)}\; \frac{1}{-
  \bar{\mu}_{pair}(T) +\xi _1^2 q^2} \;,
\label{gap_GL2}
\end{equation}
where
\begin{equation}
\bar{\mu}_{pair}(T)= -\epsilon- b_0 \Delta^2\;.
\label{mu_def_GL}
\end{equation}
Note that the critical temperature is renormalized downward with respect
to $T^*$ and satisfies
\begin{equation}
\bar{\mu}_{pair}(T_c)=0 \,.
\label{tc_cond_GL}
\end{equation}

To compare with GL theory we expand our $T$-matrix equations to first
order in the self energy correction. The $T$-matrix can be written in
terms of the attractive coupling constant $U$ as
\begin{equation}
t(Q)=  \frac{U}{1+U\chi_0(Q)+U\delta \chi(Q)}\;,
\label{t-matrix2}
\end{equation}
where 
\begin{equation}
\chi _0(Q)=\sum_{K} G_0(Q-K)G_0(K)\;. 
\label{chi2}
\end{equation}
Defining 
\begin{equation}
\Delta^2 = - \mathop{\sum_Q} t(Q)
\label{gap_approx}
\end{equation}
we arrive at the same equation as was derived in 
Section \ref{sec:2C} 
\begin{equation}
\Sigma (K) \approx -G_0 (-K) \Delta^2 
\label{self_energy2_gg0}
\end{equation}
where one can derive a self consistency condition on $\Delta^2$ in terms
of the quantity $\delta \chi (0)$ (which is first order in $\Sigma$),
which satisfies
\begin{equation}
\delta \chi (0)=-b_0 N(0) \Delta^2,
\label{delta_chi2}
\end{equation}
implying that 
\begin{equation}
  \delta \chi(0)=-b_0 T \sum_{\bf q} 
  \;  \frac{1}{\epsilon+\xi_1^2 q^2-\delta \chi(0)/N(0)}\;,
\label{self-cons_gg0}
\end{equation}
which coincides with the condition derived earlier in  Eq.~(\ref{gap_GL}).

\section{Convention and Notation}
\label{sec:Convention}

\subsection{Notation}

We follow standard notations as much as possible. They are summarized below.

$E_F$ --- Fermi energy

$k_F$ --- Fermi momentum

$\hbar$ --- Planck constant

$k_B$ --- Boltzmann constant

$c$ --- Speed of light

\mbox{e} --- Electron charge

$T_c$ --- Critical temperature for (superfluid/superconducting) phase
transition

$T^*$ --- Pair formation or pseudogap onset temperature.

$T$ --- Temperature

$\mu$, $\mu_{pair}$, $\mu_B$ --- Fermionic, pair and bosonic chemical
potential, respectively.

$\Delta$ --- Fermionic excitation gap

$\Delta_{sc}$ --- Superconducting/superfluid order parameter

$\Delta_{pg}$ --- Pseudogap 

Four vector $K \equiv (i\omega_n, \mathbf{k})$, $\sum_K \equiv T\sum_n
\sum_{\bf k} $, where $\omega_n = (2n+1)\pi T$ is the odd (fermionic)
Matsubara frequency.

Four vector $Q \equiv (i\Omega_n, \mathbf{q})$, $\sum_Q \equiv T\sum_n
\sum_{\bf q} $, where $\Omega_n = 2n\pi T$ is the even (bosonic)
Matsubara frequency.

$f(x) = 1/(e^{x/k_BT}+1)$ --- Fermi distribution function

$b(x) = 1/(e^{x/k_BT}-1)$ --- Bose distribution function

$G(K)$, $G_0(K)$ --- Full and bare Green's functions for fermions

$D(Q)$, $D_0(Q)$ --- Full and bare Green's functions for Feshbach bosons.

$\Sigma(K)$, $\Sigma_B(Q)$ --- Self energy of fermions and bosons, respectively.

$\chi(Q)$ --- Pair susceptibility

$t(Q)$ --- $T$-matrix

$\bm{\lambda}$, $\bm{\Lambda}$ --- Bare and full vertex, respectively.

$\nu$, $\nu_0$ --- Magnetic detuning parameter. 

$a_s$, $a_s^*$ --- S-wave inter-fermion scattering length

$U$, $U_0$ --- Renormalized and bare strength of the attractive
interaction between fermions in the open-channel. $U_0 \propto a_s
(\nu=+\infty)$.

$U_c$ --- Critical coupling strength at which the two-body scattering
length diverges.

$A = (\phi, \mathbf{A})$ --- Four vector potential 

$J = (\rho, \mathbf{J})$ --- Four current

$K^{\mu\nu}(Q)$ --- Electromagnetic response kernel

$\ek = k^2/2m$ --- Bare fermion dispersion in free space

$\xi_{\bf k} =\ek -\mu$ --- Bare fermion dispersion measured from
chemical potential.

$\Ek$ --- Bogoliubov quasiparticle dispersion

$\uk^2 = \frac{1}{2}(1+\xi_{\bf k}/\Ek)$, $\vk^2= \frac{1}{2}(1-\xi_{\bf
  k}/\Ek)$ 
--- Coherence factor as given in BCS theory.

$n^{tot}=n + 2(n_b +n_{b0})$ --- Total, fermion, finite momentum boson,
condensed boson density, respectively.

$V(r) = \frac{1}{2}m\omega^2 r^2$ --- Harmonic trap potential

We use alphabetic ordering when multiple authors appear together.

We always refer to the absolute value when we refer to the interaction
parameters $U$, $U_0$, $g$, $g_0$ as increasing or decreasing.

\subsection{Convention for units}

Throughout this Review, we use the convention for units where it is not
explicitly spelled out:

$\hbar = k_B = c = 1$.

$E_F = T_F = k_F = 2m = 1$.

In a harmonic trap, $E_F$, $T_F$ and $k_F$ are defined by the
noninteracting Fermi gas with the same total particle number $N$. $E_F =
\hbar\omega (3N)^{1/3}$. We further take the Thomas-Fermi radius $R_{TF}
= 1$.

In this convention, the fermion density in a homogeneous gas in three
dimensions is $n=1/3\pi^2$, the harmonic trap frequency $\omega = 2$, and
the total fermion number in a trap is $N=1/24$.

In the two-channel model, the units are $E_F/k_F^3$ for $U$ and $U_0$,
and $E_F/k_F^{3/2}$ for $g$ and $g_0$.

Our fermionic chemical potential $\mu$ is measured with respect to the
bottom of the energy band, which leads to (i) $\mu = E_F$ in the
non-interacting limit at $T=0$, and (ii) $\mu$ changes sign when the
system crosses the boundary between fermionic and bosonic regimes.

\subsection{Abbreviations}

BCS --- Bardeen-Cooper-Schrieffer (theory)

BEC --- Bose-Einstein condensation

GL --- Ginzburg-Landau (theory)

TDGL --- Time-dependent Ginzburg-Landau (theory)

RPA --- Random phase approximation 

LDA --- Local density approximation 

GP --- Gross-Pitaevskii

TF --- Thomas-Fermi (approximation)

NSR --- Nozi\'eres and Schmitt-Rink

FLEX --- FLuctuation EXchange

AB --- Anderson-Bogoliubov (mode)

PG --- Pseudogap

SC --- Superconductor

AFM --- Antiferromagnet

FB --- Feshbach boson

LSCO --- La$_{1-x}$Sr$_{x}$CuO$_4$

BSCCO --- Bi$_2$Sr$_2$CaCu$_2$O$_{8+ \delta}$

YBCO --- YBa$_2$Cu$_3$O$_{7-\delta}$

ARPES --- Angle-resolved photoemission spectroscopy

STM --- Scan tunneling microscopy

RF -- Radio frequency (spectroscopy)

SIN --- Superconductor-insulator-normal metal (tunneling junction)

SI --- Superconductor-insulator (transition)

3D --- Three dimensions

\bibliographystyle{apsrmp}

\end{document}